\documentclass[aps,pra,twocolumn,showpacs,superscriptaddress,shortbibliography]{revtex4-1} 

\usepackage{tikz}
\usepackage{upgreek}
\usepackage[implicit=false]{hyperref}
\usepackage{comment}
\usepackage{graphicx}
\usepackage{dcolumn}
\usepackage{bm}
\usepackage[T1]{fontenc}
\usepackage{epstopdf}
\usepackage{amsmath,amssymb}
\usepackage[normalem]{ulem}
\usepackage{comment}
\usepackage{dsfont}
\usepackage{mathptmx}

\usepackage[separate-uncertainty = true]{siunitx} 
\DeclareSIUnit\amu{\text{u}}
\DeclareSIUnit\au{\text{au}}
\DeclareSIUnit\angstrom{\text{\AA}}

\usepackage{hyperref}

\usepackage[sectionbib]{bibunits}
\defaultbibliographystyle{apsrev4-1} 
\defaultbibliography{VibronicQTM}

\newcolumntype{L}[1]{>{\raggedright\let\newline\\\arraybackslash\hspace{0pt}}m{#1}}
\newcolumntype{C}[1]{>{\centering\let\newline\\\arraybackslash\hspace{0pt}}m{#1}}
\newcolumntype{R}[1]{>{\raggedleft\let\newline\\\arraybackslash\hspace{0pt}}m{#1}}

\newcommand{\ket}[1]    { | #1 \rangle }
\newcommand{\bra}[1]    { \langle #1 | }

\newcommand{\avg}[1]    { \langle #1 \rangle }

\newcommand{\ii}{\textrm{i}}
\newcommand{\dd}{\textrm{d}}
\newcommand{\dycp}{$[\text{Dy}(\text{Cp}^\text{ttt})_2]^+$}
\newcommand{\dybbpen}{$[\text{Dy(bbpen)Br}]$}

\newcommand{\nocontentsline}[3]{}
\newcommand{\tocless}[2]{\bgroup\let\addcontentsline=\nocontentsline#1{#2}\egroup}

\begin{document}

\begin{bibunit}

\title{Vibronic Effects on the Quantum Tunnelling of Magnetisation in Kramers Single-Molecule Magnets}

\author{Andrea Mattioni}
\email{andrea.mattioni@manchester.ac.uk}
\affiliation{Department of Chemistry, School of Natural Sciences, The University of Manchester, Oxford Road, Manchester, M13 9PL, UK}

\author{Jakob K. Staab}
\affiliation{Department of Chemistry, School of Natural Sciences, The University of Manchester, Oxford Road, Manchester, M13 9PL, UK}

\author{William J. A. Blackmore}
\affiliation{Department of Chemistry, School of Natural Sciences, The University of Manchester, Oxford Road, Manchester, M13 9PL, UK}

\author{Daniel Reta}
\affiliation{Department of Chemistry, School of Natural Sciences, The University of Manchester, Oxford Road, Manchester, M13 9PL, UK}
\affiliation{Faculty of Chemistry, The University of the Basque Country UPV/EHU, Donostia, 20018, Spain}
\affiliation{Donostia International Physics Center (DIPC), Donostia, 20018, Spain}
\affiliation{IKERBASQUE, Basque Foundation for Science, Bilbao, 48013, Spain}

\author{Jake Iles-Smith}
\affiliation{Department of Physics and Astronomy, School of Natural Sciences, The University of Manchester, Oxford Road, Manchester M13 9PL, UK}

\author{Ahsan Nazir}
\affiliation{Department of Physics and Astronomy, School of Natural Sciences, The University of Manchester, Oxford Road, Manchester M13 9PL, UK}

\author{Nicholas F. Chilton}
\email{nicholas.chilton@manchester.ac.uk}
\affiliation{Department of Chemistry, School of Natural Sciences, The University of Manchester, Oxford Road, Manchester, M13 9PL, UK}

\begin{abstract}

{
\color{black}
Single-molecule magnets are among the most promising platforms for achieving molecular-scale data storage and processing. Their magnetisation dynamics are determined by the interplay between electronic and vibrational degrees of freedom, which can couple coherently, leading to complex vibronic dynamics.
Building on an ab initio description of the electronic and vibrational Hamiltonians, we formulate a non-perturbative vibronic model of the low-energy magnetic degrees of freedom in monometallic single-molecule magnets.
Describing their low-temperature magnetism in terms of magnetic polarons, we are able to quantify the vibronic contribution to the quantum tunnelling of the magnetisation, a process that is commonly assumed to be independent of spin-phonon coupling.
We find that the formation of magnetic polarons lowers the tunnelling probability in both amorphous and crystalline systems by stabilising the low-lying spin states.
This work, thus, shows that spin-phonon coupling subtly influences magnetic relaxation in single-molecule magnets even at extremely low temperatures where no vibrational excitations are present.
}
\end{abstract}

\maketitle

\section{Introduction}

Single-molecule magnets (SMMs) hold the potential for realising high-density data storage and quantum information processing \cite{Leuenberger2001,Sessoli2017,Coronado2020,Chilton2022}.
These molecules exhibit a ground state comprising two states characterised by a large magnetic moment with opposite orientation, which represents an ideal platform for storing digital data.
Slow reorientation of this magnetic moment results in magnetic hysteresis at the single-molecule level at sufficiently low temperatures \cite{Sessoli1993}.
The main obstacle to extending this behaviour to room temperature is the coupling of the magnetic degrees of freedom to molecular and lattice vibrations, often referred to as spin-phonon coupling \cite{kragskow2023}.
Thermal excitation of the molecular vibrations cause transitions between different magnetic states, ultimately leading to a complete loss of magnetisation.
Advances in design, synthesis and characterisation of SMMs have shed light on the microscopic mechanisms underlying their desirable magnetic properties, and have allowed extending the nanomagnet behaviour to increasingly higher temperatures 
\cite{Goodwin2017,Guo2018,Gould2022}.

The mechanism responsible for magnetic relaxation in SMMs strongly depends on temperature.
At higher temperatures, relaxation is driven by one (Orbach) and two (Raman) phonon transitions between magnetic sublevels \cite{GatteschiBook}.
When temperatures approach absolute zero, 
all vibrations are predominantly found in their ground state.
Thus, both Orbach and Raman transitions become negligible and the dominant mechanism is quantum tunnelling of the magnetisation (QTM) \cite{Thomas1996,Garanin1997}.
This mechanism originates from a coherent coupling between the two magnetic ground states, which leads to the opening of a tunnelling gap.
The tunnel coupling allows population to redistribute between states of opposite magnetisation, and thus facilitates magnetic reorientation.

\begin{figure*}[th!]
	\begin{centering}
		\includegraphics[width=\linewidth]{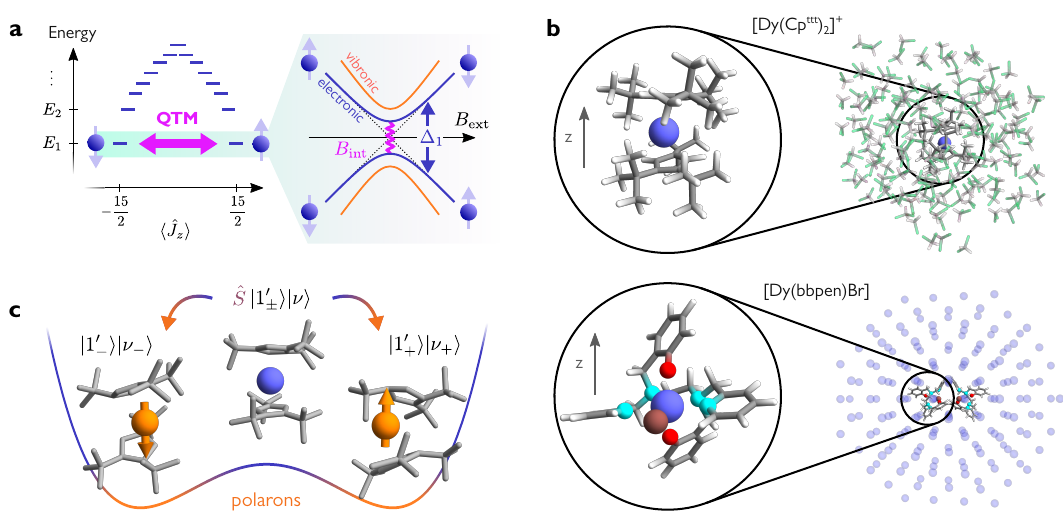}
		\caption{
{\bf Quantum tunnelling in Dy(III) single-molecule magnets.}
{\bf a}, Typical energy level diagram of the lowest-energy $J$ multiplet with angular momentum $J=15/2$ in a Dy(III) single-molecule magnet (SMM), with degenerate doublets at energies $E_1$, $E_2$, etc.
States are organised according to the expectation value of the total angular momentum along the magnetic anisotropy axis $\avg{\hat{J}_z}$. 
Dipolar and hyperfine magnetic fields ($B_\text{int}$) can lift the degeneracy  of the ground doublet and cause quantum tunnelling of the magnetisation (QTM), which results in avoided crossings when sweeping an external magnetic field $B_\text{ext}$.
Molecular vibrations can influence the magnitude of the energy splitting $\Delta_1$.
{\bf b}, Top: Molecular structure of \dycp{} surrounded by a dichloromethane (DCM) bath.
Bottom: Structure of a \dybbpen{} molecular crystal. Only the two SMMs in the primitive unit cell are shown; violet spheres represent Dy atoms at other lattice positions.
Atoms are colour coded as follow: Dy (violet), Br (brown), Cl (green), O (red), N (cyan), C (grey), H (white).  
In both cases, $z$ indicates the direction of the easy axis.
{\bf c}, Idea behind the polaron transformation $\hat{S}$ of Eq.~\eqref{e:polaron_transformation}. 
Each spin state $\ket{1'_\pm}$ is accompanied by a vibrational distortion (greatly exaggerated for visualisation), thus forming a magnetic polaron.
Vibrational states $\ket{\nu}$ are now described in terms of harmonic displacements around the deformed structure, which depends on the state of the spin.
Polarons provide an accurate physical picture when the spin-phonon coupling is strong and mostly modulates the energy of different spin states but not the coupling between them.}
        \label{f:fig_1}
	\end{centering}
\end{figure*}

While the role of vibrations in high-temperature magnetic relaxation is well understood in terms of weak-coupling rate equations for the
electronic populations
\cite{Reta2021,Briganti2021,Staab2022,Lunghi2022}, the connection between QTM and spin-phonon coupling is still largely unexplored.
Some analyses have looked at the influence of vibrations on QTM in integer-spin SMMs, where a model spin system was used to show that spin-phonon coupling could open a tunneling gap \cite{Irlander2020,Irlander2021}.
However, QTM remains more elusive to grasp in half-integer spin systems, such as monometallic Dy(III) SMMs.
In this case, a magnetic field is needed to break the time-reversal symmetry of the molecular Hamiltonian and lift the degeneracy of the ground doublet, as a consequence of Kramers theorem \cite{kramers1930}.
This magnetic field can be provided by hyperfine interaction with nuclear spins or by dipolar coupling to other SMMs;
both these effects have been shown to affect tunnelling behaviour \cite{Ishikawa2005,Moreno2017,Chilton2013, Ortu2019,Pointillart2015,Kishi2017,FloresGonzales2019,Blackmore2023}.
Once the tunnelling gap is opened by a magnetic field, molecular vibrations can in principle affect its magnitude in a nontrivial way (Fig. \ref{f:fig_1}a).
In a recent work, Ortu \textit{et al.} analysed the magnetic hysteresis of a series of Dy(III) SMMs, suggesting that QTM efficiency correlates with molecular flexibility \cite{Ortu2019}.
In another work, hyperfine coupling was proposed to assists QTM by facilitating the interaction between molecular vibrations and spin sublevels \cite{Moreno2019}.
However, a clear and unambiguous demonstration of the influence of the spin-phonon coupling on QTM beyond toy-model approaches is still lacking to this date.
A reason for this shortfall is found in the common wisdom that vibrations only cause transitions between electronic states when thermally excited, and therefore are unable to influence magnetic relaxation when thermal energy is much lower than their frequency.

In this work we present a theoretical analysis of the effect of molecular vibrations on the tunnelling dynamics in two prototypical Dy(III) SMMs,
\dycp{} \cite{Goodwin2017} and \dybbpen{} \cite{Liu2016} (Fig. \ref{f:fig_1}b).
Our approach is based on a fully ab initio description of the SMM vibrational environment and accounts for the spin-phonon coupling in a non perturbative way.
In this aspect, this work represents a step forward compared to previous theoretical analyses, which relied on a simplified description of phonons as small rotational displacements of the magnetic anisotropy axis and on a standard weak-coupling master equation approach \cite{Ho2018}.
After deriving an effective low-energy model for the relevant vibronic degrees of freedom based on a polaron approach \cite{Silbey1984}, we demonstrate that vibrations can either enhance or reduce the quantum tunnelling gap, depending on the orientation of the magnetic field relative to the main anisotropy axis of the SMM.
Lastly, we show that different vibrational modes can have competing effects on QTM;
depending on how vibrations impact the axiality of the lowest energy magnetic doublet, they can lead to either a decrease or an increase of the tunnelling probability.
While identifying vibrations that selectively tune QTM through chemical
design of new SMMs goes beyond the scope of this work, our improved description of vibronic QTM provides a useful framework to articulate further studies in that direction.

\tocless\section{Results}

\subsection{Ab initio simulations}

In this work we investigate two representative examples of Dy(III) SMMs and explore both amorphous and crystalline phonon environments.
The first compound is \dycp{}, shown in Fig. \ref{f:fig_1}b, top \cite{Goodwin2017}.
It consists of a dysprosium ion Dy(III) enclosed between two negatively charged cyclopentadienyl rings with \textit{tert}-butyl groups at positions 1, 2 and 4 (Cp$^\text{ttt}$).
The crystal field generated by the axial ligands makes the states with larger angular momentum be energetically favourable, resulting in the energy level diagram sketched in Fig. \ref{f:fig_1}a.
The energy barrier separating the two degenerate ground states  results in magnetic hysteresis, which was observed up to $T=60\text{~K}$ \cite{Goodwin2017}.

To single out the contribution of molecular vibrations, we focus on a magnetically diluted sample in a frozen solution of dichloromethane (DCM). Thus, our computational model consists of a solvated \dycp{} cation (Fig.~\ref{f:fig_1}b, top), which provides a realistic description of the low-frequency vibrational environment, comprised of pseudo-acoustic vibrational modes (Supplementary Note~1).
These constitute the basis to consider further contributions of dipolar and hyperfine interactions to QTM. 

Once the equilibrium geometry and vibrational modes of the solvated SMM (which are in general combinations of molecular and solvent vibrations) are obtained at the density-functional level of theory, we proceed to determine the equilibrium electronic structure via complete active space self-consistent field spin-orbit (CASSCF-SO) calculations.
The electronic structure is projected onto an effective crystal-field Hamiltonian.
The spin-phonon couplings are obtained from a single CASSCF calculation by computing the analytic derivatives of the molecular Hamiltonian with respect to the nuclear coordinates \cite{Staab2022}. Further details can be found in the Methods section.

The second compound considered in this work is the highly stable \dybbpen{} ($\mathrm{H}_2$bbpen$ = N,N'$-bis(2-hydroxybenzyl)-$N,N'$-bis(2-methylpyridyl)ethylenediamine),  shown in Fig. \ref{f:fig_1}b, bottom  \cite{Liu2016}.
It consists of a Dy(III) ion with pentagonal bipyramidal local geometry, with four N and one Br atom coordinating equatorially.
Two axially coordinating O atoms give rise to strong easy-axis magnetic anisotropy.
The effective barrier for magnetic reversal is around 1,000~K and magnetic hysteresis was observed up to 14~K \cite{Liu2016}.
The small size of the unit cell and the relatively high-symmetry space group ($C222_1$) make this system amenable for spin-phonon coupling calculations in a crystalline environment.
The primitive unit cell, consisting of two symmetry-related replicas of \dybbpen{}, was optimised at the density functional level of theory, and phonons were calculated using a $ 2 \times 2 \times 1 $ supercell expansion.
The electronic structure of the Dy(III) centres was obtained with state-average CASSCF-SO and parametrised with a crystal field Hamiltonian. Spin-phonon couplings were obtained via the linear vibronic coupling model \cite{Staab2022}.
A full account of these methods can be found in ref. \cite{nabi2023}.

\subsection{Polaron model}

The lowest-energy angular momentum multiplet of a Dy(III) SMM ($J=15/2$) can be described by the ab initio vibronic Hamiltonian 
\begin{equation} \label{e:vibronic}
\hat{H} = \sum_{m} E_m \ket{m}\bra{m}
+\sum_j \hat{V}_j \otimes (\hat{b}_j + \hat{b}_j^\dagger)
+\sum_j \omega_j \hat{b}_j^\dagger \hat{b}_j,
\end{equation}
where $E_m$ denotes the energy of the $m$-th eigenstate $\ket{m}$ of the crystal field Hamiltonian and $\hat{V}_j\otimes(\hat{b}_j + \hat{b}_j^\dagger)$ represent the spin-phonon coupling operators.
The harmonic vibrational modes are described in terms of their bosonic annihilation (creation) operators $\hat{b}_j$ ($\hat{b}_j^\dagger$) and frequencies $\omega_j$.

In the absence of magnetic fields, the Hamiltonian \eqref{e:vibronic} is symmetric under time reversal.
This symmetry results in a two-fold degeneracy of the energy levels $E_m$, whose corresponding eigenstates $\ket{m}$ and $\ket{\bar{m}}$ form a time-reversal conjugate Kramers doublet.
The degeneracy is lifted by introducing a magnetic field $\mathbf{B}$, which couples to the electronic degrees of freedom via the Zeeman interaction
$\hat{H}_\text{Zee}=\mu_\text{B} g_J \mathbf{B}\cdot\hat{\mathbf{J}}$, where $g_J$ is the Land\'e $g$-factor and $\hat{\mathbf{J}}$ is the total angular momentum operator.
To linear order in the magnetic field, each Kramers doublet splits into two energy levels $E_m\pm\Delta_m/2$ corresponding to the states
\begin{eqnarray} \label{e:m_plus}
\ket{m_+} &= \cos\frac{\theta_m}{2}\ket{m} + e^{\ii{} \phi_m} \sin\frac{\theta_m}{2}\ket{\bar{m}} \\
\label{e:m_minus}
\ket{m_-} &= -\sin\frac{\theta_m}{2}\ket{m} + e^{\ii{} \phi_m} \cos\frac{\theta_m}{2}\ket{\bar{m}}
\end{eqnarray}
where the energy splitting $\Delta_m$ and the mixing angles $\theta_m$ and $\phi_m$ are determined by the matrix elements 
of the Zeeman Hamiltonian on the subspace $\{\ket{m},\ket{\bar{m}}\}$.
In addition to the intra-doublet mixing described by Eqs. \eqref{e:m_plus} and \eqref{e:m_minus}, the Zeeman interaction also mixes Kramers doublets at different energies.
The ground doublet acquires contributions from higher-lying states
\begin{equation} \label{e:1pm}
\ket{1_\pm'} = \ket{1_\pm} + \sum_{m\neq 1,\bar{1}} \ket{m} \frac{\bra{m}\hat{H}_\text{Zee}\ket{1_\pm}}{E_1-E_m} + \mathcal{O}(B^2).
\end{equation}
These states no longer form a time-reversal conjugate doublet, meaning that the spin-phonon coupling can now contribute to transitions between them.

Since QTM is typically observed at much lower temperatures than the energy gap between the lowest and first excited doublets (which here is $\gtrsim 600$~K \cite{Goodwin2017, Liu2016}) we focus on the perturbed ground doublet $\ket{1'_\pm}$.
Within this subspace, the Hamiltonian $\hat{H}+\hat{H}_\text{Zee}$
takes the form
\begin{eqnarray} \label{e:Heff}
\hat{H}_\text{eff} &=& E_1 + \frac{\Delta_1}{2} \sigma'_z
+ \sum_j \omega_j \hat{b}_j^\dagger\hat{b}_j \\
&+&
\sum_j
\left(\bra{1}\hat{V}_j\ket{1} - w_j^z \sigma'_z \right)
\left(\hat{b}_j + \hat{b}_j^\dagger\right) \nonumber \\
&-&
\sum_j
\left(w^x_j\sigma'_x+w^y_j\sigma'_y\right)
\left(\hat{b}_j + \hat{b}_j^\dagger\right).
\nonumber
\end{eqnarray}
This Hamiltonian describes the interaction between vibrational modes and an effective spin one-half represented by the Pauli matrices $\bm{\sigma}'=(\sigma'_x,\sigma'_y,\sigma'_z)$, where
$\sigma'_z = \ket{1'_+}\bra{1'_+}-\ket{1'_-}\bra{1'_-}$.
The vector
$\mathbf{w}_j=(
\Re\bra{1_-}\hat{W}_j\ket{1_+},
\Im\bra{1_-}\hat{W}_j\ket{1_+},
\bra{1_+}\hat{W}_j\ket{1_+}
)$
is defined in terms of the operator
$\hat{W}_j=\sum_{m\neq 1,\bar{1}} \hat{V}_j \ket{m}\bra{m} \hat{H}_\text{Zee}/(E_m-E_1)+\text{ h.c.}$, describing the effect of the Zeeman interaction on the spin-phonon coupling.
Due to the strong magnetic axiality of the SMM considered here,
the longitudinal component of the spin-phonon coupling $w_j^z$  dominates over the transverse part $w_j^x$, $w_j^y$.
In this case, we can get a better physical picture of the system by transforming the Hamiltonian \eqref{e:Heff} to the polaron frame defined by the unitary operator
\begin{equation}
\label{e:polaron_transformation}
\hat{S} = \exp\left[ \sum_{s=\pm} \ket{1'_s}\bra{1'_s}\ 
\sum_j \xi_j^s
\left(\hat{b}_j^\dagger - \hat{b}_j\right) \right],
\end{equation}
which mixes electronic and vibrational degrees of freedom by displacing the mode operators by $\xi_j^\pm = (\bra{1}\hat{V}_j\ket{1} \mp w_j^z)/\omega_j$ depending on the state of the effective spin one-half  \cite{Silbey1984}.
The idea behind this transformation is to allow nuclei to relax around a new equilibrium geometry, which may be different for every spin state. This lowers the energy of the system and provides a good description of the vibronic eigenstates when the spin-phonon coupling is approximately diagonal in the spin basis (Fig.~\ref{f:fig_1}c).
In the polaron frame, the longitudinal spin-phonon coupling is fully absorbed into the purely electronic part of the Hamiltonian, while the transverse components can be
approximated by their thermal average over vibrations,
neglecting their vanishingly small quantum fluctuations (Supplementary Note~2).
After transforming back to the original frame, we are left with an effective spin one-half Hamiltonian with no residual spin-phonon coupling
$H_\text{eff}\approx \hat{H}_\text{eff}^\text{(pol)} + \sum_j \omega_j \hat{b}_j^\dagger\hat{b}_j$, where
\begin{equation} \label{e:H_eff^pol}
\hat{H}_\text{eff}^\text{(pol)} = E_1 + \frac{\Delta_1}{2} \sigma''_z
+ 2\sum_j \frac{\bra{1}\hat{V}_j\ket{1}}{\omega_j} \mathbf{w}_j \cdot
\bm{\sigma}''.
\end{equation}
The set of Pauli matrices
$\bm{\sigma}''=\hat{S}^\dagger (\bm{\sigma}'\otimes\mathds{1}_\text{vib}) \hat{S}$ 
describe the two-level system formed by the magnetic polarons of the form
$\hat{S}^\dagger \ket{1'_\pm}\ket{\{\nu_j\}}_\text{vib}$,
where $\{\nu_j\}$ is a set of occupation numbers for the vibrational modes of the solvent-SMM system.
These magnetic polarons can be thought as
magnetic electronic states strongly coupled to a distortion of the molecular geometry.
They inherit the magnetic properties of the corresponding electronic states, and can be seen as the molecular equivalent of the magnetic polarons observed in a range of magnetic materials \cite{Yakovlev2010,Schott2019,Godejohann2020}.
Polaron representations of vibronic systems have been employed in a wide variety of settings, ranging from spin-boson models \cite{Silbey1984,Chin2011} to photosynthetic complexes 
\cite{Yang2012,Kolli2011,Pollock2013}, to quantum dots \cite{Wilson2002,McCutcheon2010,Nazir2016}, 
providing a convenient basis to describe the dynamics of quantum systems strongly coupled to a vibrational environment.
These methods are particularly well suited for condensed matter systems where the electron-phonon coupling is strong but causes very slow transitions between different electronic states, allowing exact treatment of the pure-dephasing part of the electron-phonon coupling and renormalising the electronic parameters.
For this reason, the polaron transformation is especially effective for describing  our system (Supplementary Note~3).
The most striking advantage of this approach is that the average effect of the spin-phonon coupling is included non-perturbatively into the electronic part of the Hamiltonian, leaving behind a vanishingly small residual spin-phonon coupling. 

As a last step, we bring the Hamiltonian in Eq. \eqref{e:H_eff^pol} into a more familiar form by expressing it in terms of an effective $g$-matrix.
We recall that the quantities $\Delta_1$ and $\mathbf{w}_j$ depend linearly on the magnetic field $\mathbf{B}$ via the Zeeman Hamiltonian $\hat{H}_\text{Zee}$.
An additional dependence on the orientation of the magnetic field comes from the mixing angles $\theta_1$ and $\phi_1$ introduced in Eqs. \eqref{e:m_plus} and \eqref{e:m_minus}, appearing in the states
$\ket{1_\pm}$ used in the definition of $\mathbf{w}_j$.
This further dependence is removed by transforming the Pauli operators back to the basis $\{\ket{1},\ket{\bar{1}}\}$ via a three-dimensional rotation $\bm{\sigma} = \mathbf{R}_{\theta_1,\phi_1} \cdot \bm{\sigma}''$.
Finally, we obtain
\begin{equation} \label{e:H_eff_g-tensor}
\hat{H}_\text{eff}^\text{(pol)} = E_1 + \mu_\text{B} \mathbf{B} \cdot
\left( \mathbf{g}^\text{el} + \sum_j \mathbf{g}^\text{vib}_{j} \right)
\cdot \frac{\bm{\sigma}}{2},
\end{equation}
for appropriately defined electronic and single-mode vibronic $g$-matrices
$\mathbf{g}^\text{el}$ and $\mathbf{g}^\text{vib}_j$.
These are directly related to the electronic splitting term $\Delta_1$ and to the vibronic corrections described by $\mathbf{w}_j$ in Eq. \eqref{e:H_eff^pol}, respectively 
(see Supplementary Note~2 for a thorough derivation).
The main advantage of representing the ground Kramers doublet with an effective spin one-half Hamiltonian
is that it provides a conceptually simple foundation for studying low-temperature magnetic behaviour of the SMM, confining all microscopic details, including vibronic effects, to an effective $g$-matrix.


\subsection{Vibronic modulation of the ground Kramers doublet}

\begin{figure}[t]
	\begin{centering}
		\includegraphics[width=\linewidth]{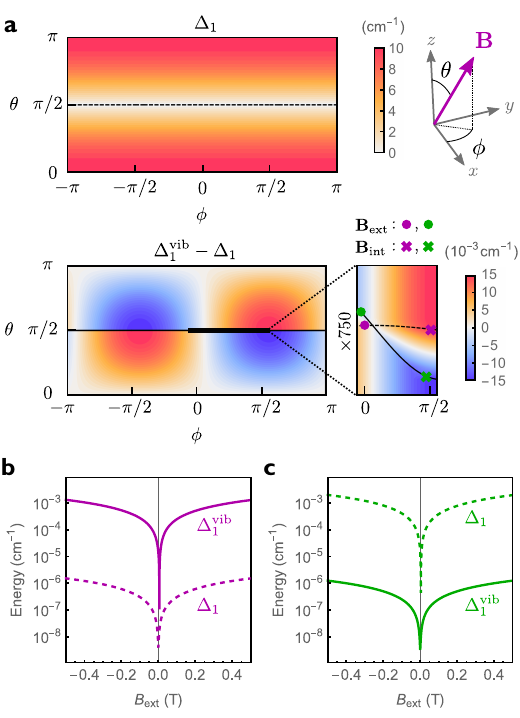}
		\caption{
{\bf Zeeman splitting of the ground Kramers doublet in \dycp{}.} 
\textbf{a}, Electronic ground doublet splitting ($\Delta_1$, top) and vibronic correction ($\Delta_1^\text{vib}-\Delta_1$, bottom) as a function of the orientation of the magnetic field, parametrised in terms of polar and azimuthal angles $\theta$ and $\phi$.
The polar angle $\theta$ is measured with respect to the axis joining the cyclopentadienyl centroids, corresponding approximately to the easy axis.
The dashed (solid) line corresponds to the electronic (vibronic) hard plane.
The magnitude of the magnetic field is fixed to 1~T.
\textbf{b}, \textbf{c}, Electronic (dashed) and vibronic (solid) Zeeman splitting of the ground doublet as a function of the external field magnitude $B_\text{ext}$ in the presence of a transverse internal field $B_\text{int}=1\text{~mT}$ calculated from Eq. \eqref{e:H_eff_g-tensor}.
External and internal fields are perpendicular to each other and were both chosen to lie in the hard plane of either the electronic (\textbf{b}, purple) or vibronic (\textbf{c}, green) $g$-matrix.
The orientation of the external (internal) field is shown for both cases as circles (crosses) in the inset in (\textbf{a}), with colors matching the ones in (\textbf{b}) and (\textbf{c}).
}
		\label{f:fig_2}
	\end{centering}
\end{figure}

We begin by considering the influence of vibrations on the Zeeman splitting of the lowest Kramers doublet.
The Zeeman splitting in absence of vibrations is simply given by $\Delta_1=\mu_\text{B}| \mathbf{B}\cdot\mathbf{g}^\text{el}|$.
In~the presence of vibrations, the electronic $g$-matrix $\mathbf{g}^\text{el}$ is modified by adding the vibronic correction $\sum_j\mathbf{g}^\text{vib}_j$, resulting in the Zeeman splitting  $\Delta_1^\text{vib}$.
In Fig. \ref{f:fig_2}a we show the Zeeman splittings as a function of the orientation of the magnetic field $\mathbf{B}$ for \dycp{}, parametrised in terms of the polar angles $(\theta,\phi)$.
Depending on the field orientation, vibrations can lead to either an increase or decrease of the Zeeman splitting.
These changes seem rather small when compared to the largest electronic splitting, obtained when $\mathbf{B}$ is oriented along the $z$-axis (Fig.~\ref{f:fig_1}b), as expected for a system with easy-axis anisotropy.
However, they become quite significant for field orientations close to the $xy$-plane, where the purely electronic splitting $\Delta_1$ becomes vanishingly small and $\Delta_1^\text{vib}$ can be dominated by the vibronic contribution.
This is clearly shown in Fig. \ref{f:fig_2}b,c where we decompose the total field $\mathbf{B}=\mathbf{B}_\text{int}+\mathbf{B}_\text{ext}$ in a fixed internal component $\mathbf{B}_\text{int}$ originating from dipolar and hyperfine interactions, responsible for opening a tunnelling gap, and an external part $\mathbf{B}_\text{ext}$ which we sweep along a fixed direction across zero.
When these fields lie in the plane perpendicular to the purely electronic easy axis, i.e. the hard plane, the vibronic splitting can be three orders of magnitude larger than the electronic one (Fig. \ref{f:fig_2}b).
The situation is reversed when the fields lie in the hard plane of the vibronic $g$-matrix (Fig. \ref{f:fig_2}c).
We note that this effect is specific to states with easy-axis magnetic anisotropy, however this is the defining feature of SMMs, such that our results should be generally applicable to all Kramers SMMs. In fact, we observe very similar results for \dybbpen{} (Supplementary Note~4).

\subsection{Internal fields and QTM probability}

So far we have seen that spin-phonon coupling can either enhance or reduce the tunnelling gap in the presence of a magnetic field depending on its orientation.
For this reason, it is not immediately clear 
whether its effects survive ensemble averaging in a collection of randomly oriented SMMs, such as for frozen solutions or polycrystalline samples considered in magnetometry experiments.
In order to check this, let us consider an ideal field-dependent magnetisation measurement.
When sweeping a magnetic field $B_\text{ext}$ at a constant rate from positive to negative values along a given direction, QTM is typically observed as a sharp step in the magnetisation of the sample when crossing the region around $B_\text{ext}=0$ \cite{Thomas1996, Blackmore2023}.
This sudden change of the magnetisation is due to a non-adiabatic spin-flip transition between the two lowest energy spin states, that occurs when traversing an avoided crossing (see diagram in Fig. \ref{f:fig_1}a, right).
The spin-flip probability is given by the celebrated Landau-Zener expression \cite{Landau1932i,Landau1932ii,Zener1932,Stueckelberg1932,Majorana1932,Ivakhnenko2023}, which in our case takes the form
\begin{equation} \label{e:P_LZ}
P_\text{LZ} = 1-\exp{ \left(-\frac{\pi |\bm{\Delta}_\perp|^2}{2 |{\bf v}|} \right)},
\end{equation}
where we have defined
$\mathbf{v} = \mu_\text{B} \dd{\mathbf{B}_\text{ext}}/\dd{t} \cdot \mathbf{g}$,
and
$\bm{\Delta}_\perp$ is the component of
$\bm{\Delta} = \mu_\text{B} \mathbf{B}_\text{int} \cdot \mathbf{g}$ perpendicular to $\mathbf{v}$, while $\mathbf{g}$ denotes the total electronic-vibrational  $g$-matrix appearing in Eq.~\eqref{e:H_eff_g-tensor}
(see Supplementary Note~2 for a derivation of Eq. \eqref{e:P_LZ}).
%

\begin{figure}[t!]
	\begin{centering}
		\includegraphics[width=\linewidth]{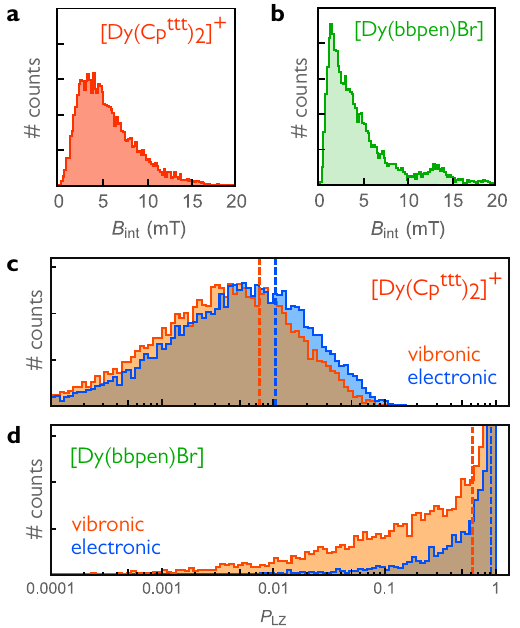}
		\caption{
{\bf Internal fields and spin-flip probability.} 
\textbf{a}, \textbf{b}, Distribution of internal field magnitudes $B_\mathrm{int}$ experienced by a Dy centre due to the dipolar fields produced by surrounding Dy centres, magnetised by a randomly oriented external field $\mathbf{B}_\mathrm{ext}$.
For \dycp{} (\textbf{a}), a uniform spatial distribution of 1,000 randomly oriented single-molecule magnets (SMMs) around a central Dy(III) was assumed, corresponding to a 170~mM solution in dichloromethane.
For the \dybbpen{} molecular crystal (\textbf{b}), we considered the total dipolar field arising from all Dy centres within a 100~\AA{} radius from a central Dy assuming 5\% diamagnetic dilution.
\textbf{c}, \textbf{d}, Distribution of electronic (blue) and vibronic (orange) Landau-Zener spin-flip probabilities $P_\mathrm{LZ}$, calculated for a randomly oriented SMM subjected to the dipolar fields shown above, assuming an external field sweep rate of 10~Oe/s.  Average values are shown as dashed lines: (\textbf{c}) 0.0104 (blue) and 0.0074 (orange); (\textbf{d}) 0.903 (blue) and 0.618 (orange).
All histograms are obtained from an ensemble of 10,000 random external field orientations and dipole arrangements.
}
		\label{f:fig_3}
	\end{centering}
\end{figure}

In order to fully characterise the spin-flip process, we need to quantify the internal fields that cause QTM in Kramers SMMs, which originate from either dipolar or hyperfine interactions.
In the following we focus on dipolar fields, since their effects can be observed at much higher temperatures than those required to witness hyperfine interactions
(Supplementary Note~5).
Samples studied in magnetometry experiments typically contain a macroscopic number of SMMs, each of which produces a microscopic dipole field.
We estimate the combined effect of these microscopic dipoles in a \dycp{} DCM frozen solution of by generating random spatial configurations of SMMs and calculating the resulting field at a specific point in space corresponding to a randomly selected SMM.
We repeat this process 10,000 times to obtain the internal field distribution $\mathbf{B}_\text{int}$, as shown in Fig.~\ref{f:fig_3}a.
The orientation of this field is random and its magnitude averages to 5.5~mT for a SMM concentration of 170~mM \cite{Goodwin2017}
(Supplementary Note~5).

In the case of the \dybbpen{} molecular crystal, the effect of all Dy atoms 
within a 100~\AA{} radius of a central magnetic centre was considered in a 5\% Dy in Y diamagnetically diluted crystallite \cite{Liu2016}.
Random Dy/Y subsitutions at different sites and random orientations of the magnetising field $\mathbf{B}_\mathrm{ext}$ were considered to mimic a powder sample, leading to the distribution shown in Fig.~\ref{f:fig_3}b with average magnitude 4.9~mT.

We then sample the distribution of internal fields to calculate the corresponding spin-flip probabilities for a randomly oriented SMM using Eq.~\eqref{e:P_LZ}.
The effect of spin-phonon coupling
on the spin-flip dynamics of an ensemble of SMMs is shown in Fig. \ref{f:fig_3}c,d.
The vibronic correction to the ground doublet $g$-matrix leads to a suppression of spin-flip events (orange) compared to a purely electronic model (blue).
Despite the significant overlap between the two distributions, spin-phonon coupling results in a $\sim$30\% drop of average spin-flip probabilities, represented by the dashed lines in Fig.~\ref{f:fig_3}c,d.
The vibronic suppression of QTM can be intuitively understood in terms of the polaron energy landscape sketched in Fig.~\ref{f:fig_1}c: strong coupling between spin degrees of freedom and molecular distortions can stabilise spin states, introducing a vibrational energy cost for spin reversal; i.e. flipping a spin requires reorganisation of the molecular structure.

From Fig.~\ref{f:fig_3}c,d, we also note that crystalline \dybbpen{} exhibits much larger QTM than \dycp{}{} in frozen solution.
This can be understood in terms of the different microscopic dipole fields in the two systems.
In
Supplementary Note~5
we show that  $\mathbf{B}_\mathrm{int}$ is perfectly isotropic in a frozen solution.
On the contrary, due to the symmetry of the \dybbpen{} molecular crystal, the component of the internal field along the intra-unit cell Dy-Dy direction survives orientational averaging, resulting in an average transverse component of 1.2~mT
(Supplementary Note~5).

\section{Discussion}

\begin{figure*}[t!]
	\begin{centering}
        \includegraphics[width=\linewidth]{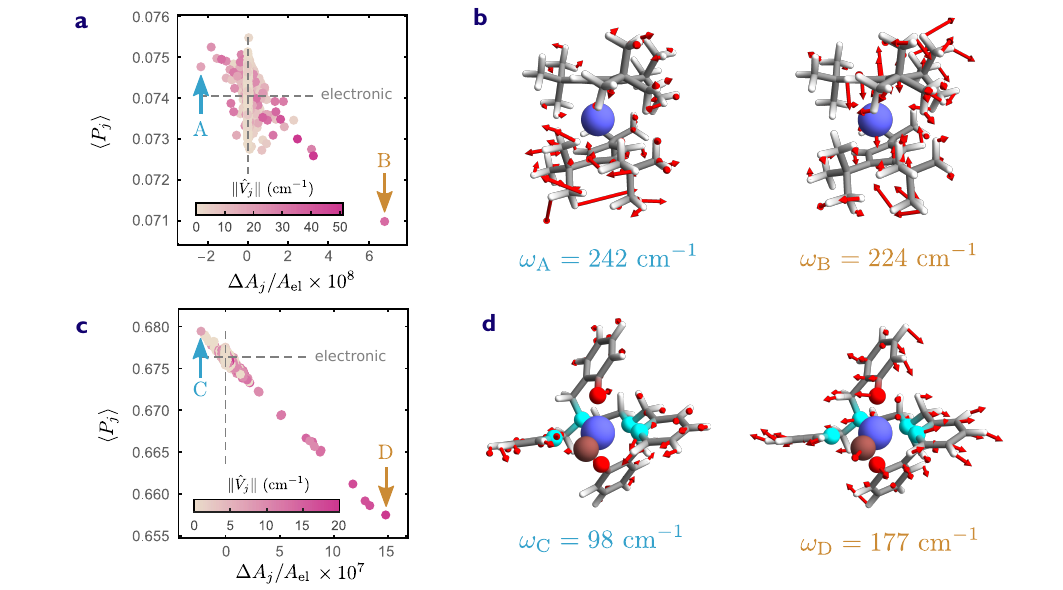}
	\caption{
{\bf Single-mode contributions to tunnelling of the magnetisation.}
\textbf{a}, \textbf{c}, Single-mode vibronic spin-flip probabilities plotted for each vibrational mode, shown as a function of the mode axiality $\Delta A_j = A_j - A_\text{el}$ relative to the electronic axiality $A_\text{el}$.
The magnitude of the internal field is fixed to $B_\text{int}=1\text{~mT}$ and the external field sweep rate is 10~Oe/s. The probabilities $\avg{P_j}$ are obtained by averaging over random orientations of external and internal fields. The color coding represents the spin-phonon coupling strength $\|\hat{V}_j\|$. Grey dashed lines corresponds to a purely electronic model.
(\textbf{a}) and (\textbf{c}) correspond to amorphous \dycp{} and crystalline \dybbpen{}. 
\textbf{b}, \textbf{d} Visual representation of the displacements induced by the vibrational modes indicated by arrows in (\textbf{a}) and (\textbf{c}) denoted by A, B, C, D; the corresponding vibrational frequencies are denoted by $\omega_\mathrm{A}$,$\omega_\mathrm{B}$,$\omega_\mathrm{C}$,$\omega_\mathrm{D}$.
}
		\label{f:fig_4}
	\end{centering}
\end{figure*}

As shown above, the combined effect of all vibrations in a randomly oriented ensemble of SMMs is to reduce QTM.
However, not all vibrations contribute to the same extent.
Based on the polaron model introduced above, vibrations with large spin-phonon coupling
and low frequency
have a larger impact on the magnetic properties of the ground Kramers doublet.
This can be seen from Eq. \eqref{e:H_eff^pol}, where 
the vibronic correction to the effective ground Kramers Hamiltonian 
is weighted by the factor $\bra{1}\hat{V}_j\ket{1}/\omega_j$.
Another property of vibrations that can influence QTM is their symmetry.
In monometallic SMMs, QTM has generally been correlated with a reduction of axial symmetry, either by the presence of flexible ligands or by transverse magnetic fields.
Since we are interested in symmetry only as long as it influences magnetism, it is useful to introduce a measure of axiality on the $g$-matrix, such as
\begin{equation} \label{e:axiality}
A(\mathbf{g}) =
\frac{
\left\lVert \mathbf{g} - \frac{1}{3} \text{Tr }\mathbf{g} \right\rVert }
{ \sqrt{\frac{2}{3}} \text{Tr }\mathbf{g} },
\end{equation}
where $\|\cdot\|$ denotes the Frobenius norm.
This measure yields 1 for perfect easy-axis anisotropy, 1/2 for an easy-plane system, and 0 for the perfectly isotropic case.
The axiality of an individual vibrational mode can be quantified as
$A_j = A(\mathbf{g}^\text{el}+\mathbf{g}^\text{vib}_j)$
by building a single-mode vibronic $g$-matrix, analogous to the multi-mode one introduced in Eq. \eqref{e:H_eff_g-tensor}.
We might be tempted to intuitively conclude that polaron formation always increases the axiality with respect to its electronic value $A_\text{el} = A(\mathbf{g}^\text{el})$, given that the collective effect of the spin-phonon coupling is to reduce QTM.
However, when considered individually, some vibrations can have the opposite effect 
of effectively reducing the magnetic axiality.

In order to see how axiality correlates to QTM, we calculate the single-mode spin-flip probabilities $\avg{P_j}$.
These are obtained by
replacing the multi-mode vibronic $g$-matrix in Eq.~\eqref{e:H_eff_g-tensor} with the single-mode one $\mathbf{g}^\text{el}+\mathbf{g}^\text{vib}_j$, 
and following the same procedure detailed in
Supplementary Note~2.
The single-mode contribution to the spin-flip probability unambiguously correlates with mode axiality, as shown in Fig.~\ref{f:fig_4}a for \dycp{}; the correlation is even starker for crystalline \dybbpen{} (Fig.~\ref{f:fig_4}c).
Vibrational modes that lead to a larger QTM probability are likely to reduce the magnetic axiality (top-left sector).
Vice versa, those vibrational modes that enhance axiality also suppress QTM (bottom-right sector).

As a first step towards uncovering the microscopic basis of this unexpected behaviour, we single out the vibrational modes that have the largest impact on magnetic axiality in both directions.
These vibrational modes, labelled  A, B for \dycp{} and C, D for \dybbpen{}, represent a range of qualitatively distinct vibrations, as can be observed in Fig. \ref{f:fig_4}b,d.
In the case of \dycp{}, mode A is mainly localised on one of the $\mathrm{Cp}^\mathrm{ttt}$ ligands and features atomic displacements predominantly perpendicular to the easy axis.
Mode B, on the other hand, involves axial distortions of the Cp rings and, to a lesser extent, rotations of the methyl groups.
Thus, it makes sense intuitively that A would lead to an increased QTM probability, while the opposite is true for B, as observed in Fig.~\ref{f:fig_4}a.

However, the connection between the magnetic axiality defined in Eq.~\eqref{e:axiality}
and vibrational motion is not always straightforward.
In the case of \dybbpen{}, mode C mainly involves a tilt of the two equatorial pyridyl groups. This movement disrupts axiality and enhances QTM.
On the other hand, mode D features equatorial motion of the first coordination sphere of the Dy(III) ion, involving movement of Br and Dy itself in the hard plane.
However, this vibrational mode induces a suppression of QTM, as seen in Fig.~\ref{f:fig_4}c, rather than in increase, as would be expected based on the above symmetry arguments. This shows that $\Delta A_j$ does not necessarily correlate to atomic motions, but can be a useful proxy for determining a given vibration's contribution to the QTM probability.
In fact, the correlation between the two quantities can be rationalised with the help of the simple toy model presented in Supplementary Note~6.
Nonetheless, we note that the out-of-phase motion of the equatorial pyridyl groups in D preserves axiality and could contribute to its efficiency at suppressing QTM.
It is also worth noting that Briganti \textit{et al.} recently demonstrated that motion of atoms beyond the first coordination sphere of the central Dy(III) ion can greatly influence spin dynamics in the Raman regime through bond polarisation effects \cite{Briganti2021}.
Performing a similar electrostatic analysis in the context of our polaron model is beyond the scope of this work; however, it represents an interesting direction for further investigations elucidating the role of vibrations on QTM.


In conclusion, we have presented a detailed description of the effect of molecular and solvent vibrations on the quantum tunnelling between low-energy spin states in two different single-ion Dy(III) SMMs, 
corresponding to amorphous and crystalline environments.
Our theoretical results, based on an ab initio approach, are complemented by a polaron treatment of the relevant vibronic degrees of freedom, which does not suffer from any weak spin-phonon coupling assumption and is therefore well-suited to other strong coupling scenarios.
We have been able to derive a non-perturbative vibronic correction to the effective $g$-matrix of the lowest-energy Kramers doublet, 
which we have used as a basis to determine the tunnelling dynamics in an idealised magnetic field sweep experiment, building on Landau-Zener theory.
This has allowed us to formulate the observation that spin-phonon coupling does have an  influence on QTM, albeit a subtle one ($\sim 30$\%), as opposed to the widespread belief that magnetic tunnelling is not influenced by vibrations since it only becomes effective at low temperatures.
This effect is rooted in the formation of magnetic polarons, which results in a redefinition of the magnetic anisotropy of the ground Kramers doublet.
Our theoretical treatment is fully ab initio and represents a significant improvement over other theoretical descriptions of QTM which rely on weak coupling assumptions.
Lastly, we observe that specific vibrational modes can either enhance or suppress QTM.
This behaviour correlates to the magnetic axiality of each mode, which can be used as a proxy for determining whether a specific vibration enhances or hinders tunnelling.
Our analysis suggests that there may be a positive side to spin-phonon coupling in QTM. Enhancing the coupling to specific vibrations via appropriate chemical design while keeping detrimental vibrations under control, could in principle increase magnetic axiality and thus suppress QTM even further.
However, translating this observation into clear-cut chemical design guidelines remains an open question, that requires the analysis of other molecular systems.
As ab initio spin-phonon coupling calculations become more accessible, the approach presented here can be applied to the study of vibronic QTM in other SMMs, and thus represents a valuable tool for understanding the role of vibrations in low-temperature magnetic relaxation.

%
%

\section*{Methods}

The ab initio model of the DCM-solvated \dycp{} molecule is constructed using a multi-layer approach.
During geometry optimisation and frequency calculation the system is partitioned into two layers following the ONIOM scheme~\cite{svensson1996}. The high-level layer, consisting of the SMM itself and the first solvation shell of 26 DCM molecules, is described by Density Functional Theory (DFT) while the outer bulk of the DCM ball constitutes the low-level layer modelled by the semi-empirical PM6 method.
All DFT calculations are carried out using the pure PBE exchange-correlation functional \cite{perdew1996} with Grimme's D3 dispersion correction.
Dysprosium is replaced by its diamagnetic analogue yttrium for which the Stuttgart RSC 1997 ECP basis is employed~\cite{andrae1990}.
Cp ring carbons directly coordinated to the central ion are equipped with Dunning's correlation consistent triple-zeta polarised cc-pVTZ basis set and all remaining atoms with its double-zeta analogue cc-pVDZ~\cite{dunning1989}.
Subsequently, the electronic spin states and spin-phonon coupling parameters are calculated at the CASSCF-SO level explicitly accounting for the strong static correlation present in the f-shell of Dy(III) ions. At this level, environmental effects are treated using an electrostatic point charge representation of all DCM atoms.
All DFT/PM6 calculations are carried out with \textsc{Gaussian} version 9 revision D.01~\cite{g09} and the CASSCF calculations are carried out with \textsc{OpenMolcas} version 21.06~\cite{omolcas2019}.

The starting \dycp{} solvated system was obtained using the solvate program belonging to the AmberTool suite of packages, with box as method and CHCL3BOX as solvent model. Chloroform molecules were subsequently converted to DCM. From this large system, only molecules falling within 9~\AA{} from the central metal atom are considered from now on.
The initial disordered system of 160 DCM molecules packed around the \dycp{} crystal structure \cite{Goodwin2017} is pre-optimised in steps, starting by only optimising the high-level layer atoms and freezing the rest of the system. The low-layer atoms are pre-optimised along the same lines starting with DCM molecules closest to the SMM and working in shells towards the outside. Subsequently, the whole system is geometry optimised until RMS (maximum) values in force and displacement corresponding to \SI{0.00045}{\au} (\SI{0.0003}{\au}) and \SI{0.0018}{\au} (\SI{0.0012}{\au}) are reached, respectively.
After adjusting the isotopic mass of yttrium to that of dysprosium $m_{\mathrm{Dy}} = \SI{162.5}{\amu}$, vibrational normal modes and frequencies of the entire molecular aggregate are computed within the harmonic approximation.

Electrostatic atomic point charge representations of the environment DCM molecules are evaluated for each isolated solvent molecule independently at the DFT level of theory employing the CHarges from ELectrostatic Potentials using a Grid-based (ChelpG) method~\cite{breneman1990}, which serve as a classical model of environmental effects in the subsequent CASSCF calculations.

The evaluation of equilibrium electronic states and spin-phonon coupling parameters is carried out at the CASSCF level including scalar relativistic effects using the second-order Douglas-Kroll Hamiltonian and spin-orbit coupling through the atomic mean field approximation implemented in the restricted active space state interaction approach~\cite{malmqvist1989,malmqvist2002}.
The dysprosium atom is equipped with the ANO-RCC-VTZP, the Cp ring carbons with the ANO-RCC-VDZP and the remaining atoms with the
ANO-RCC-VDZ basis set~\cite{widmark1990}. The resolution of the identity approximation with an on-the-fly acCD auxiliary basis is employed to handle the two-electron integrals~\cite{aquilante2007}.
The active space of 9 electrons in 7 orbitals, spanned by 4f atomic orbitals, is employed in a state-average CASSCF calculation including the 18 lowest lying sextet roots which span the $^6\mathrm{H}$ and $^6\mathrm{F}$ atomic terms.
%
%

We use our own implementation of spin Hamiltonian parameter projection
to obtain the crystal field parameters $B_k^q$ entering the Hamiltonian
\begin{equation} \label{se:HCF}
\hat{H}_\text{CF} = \sum_{k=2,4,6}\sum_{q=-k}^{k} \theta_k B_k^q
O_k^q(\hat{\mathbf{J}}),
\end{equation}
describing the $^6\text{H}_{15/2}$ ground state multiplet.
Operator equivalent factors and Stevens operators are denoted by $\theta_k$ and  $O_k^q(\hat{\mathbf{J}})$, where 
$\hat{\mathbf{J}}=(\hat{J}_x,\hat{J}_y,\hat{J}_z)$
are the angular momentum components.
Spin-phonon coupling arises from changes to the Hamiltonian \eqref{se:HCF} due to 
slight distortions of the molecular geometry, parametrised as 
\begin{equation}
B_k^q(\{X_j\}) = B_k^q + \sum_{j=1}^M \frac{\partial B_k^q}{\partial X_j} X_j + \dots,
\end{equation}
where $X_j$ denotes the dimensionless $j$-th normal coordinate of the molecular aggregate.
The derivatives $\partial B_k^q/\partial X_j$ are calculated using the Linear Vibronic Coupling (LVC) approach described in Ref. \cite{Staab2022} based on the state-average CASSCF density-fitting gradients and non-adiabatic coupling involving all 18 sextet roots.
Finally, we express the dimensionless normal coordinates in terms of bosonic creation and annihilation operators as $\hat{X}_j = (\hat{b}_j+\hat{b}_j^\dagger)/\sqrt{2}$, which defines the 
system part of the spin-phonon coupling operators in Eq. \eqref{e:vibronic} as
\begin{equation}
\hat{V}_j = \frac{1}{\sqrt{2}}  \sum_{k,q} \theta_k
\frac{\partial B_k^q}{\partial X_j}
O_k^q(\hat{\mathbf{J}}).
\end{equation}

\section*{Data availability}
The data generated in this study have been deposited in the Figshare database and can be accessed at
\url{http://doi.org/10.48420/21892887} \cite{data}.
Source data for all figures are provided with this paper.

\section*{Code availability}
The code used to calculate ab initio spin-phonon couplings is part of our in-house Python packages \texttt{spin\_phonon\_suite} and \texttt{angmom\_suite},
freely available from the
PyPI repository at \url{https://pypi.org/project/spin-phonon-suite/}
and \url{https://pypi.org/project/angmom-suite/}.



%


\section*{Acknowledgements}
This work was made possible thanks to the ERC grant 2019-STG-851504 and Royal Society fellowship URF191320 (N.F.C.).
The authors acknowledge support from the Computational Shared Facility at the University of Manchester.

\section*{Author contributions}
A.M. formulated and implemented the effective polaron model with input from J.I.-S. and A.N.;
J.K.S. and D.R. performed the ab initio calculations with guidance from N.F.C.;
A.M. estimated dipolar fields with input from W.J.A.B.;
N.F.C. supervised the work.
All authors contributed towards analysis, discussions and preparation of the manuscript.

\section*{Competing interests}
The authors declare no competing interests.

\section*{Published version}
This version of the article has been accepted for publication, after peer review
but is not the Version of Record and does not reflect post-acceptance improvements, or any
corrections. The Version of Record is available online at: \url{http://dx.doi.org/10.1038/s41467-023-44486-3}.

\end{bibunit}

\begin{onecolumngrid}

\clearpage

\newcommand{\beginsupplement}{
	\setcounter{table}{0}
	\setcounter{figure}{0}
	\setcounter{equation}{0}
	\setcounter{page}{1}
	\setcounter{section}{0}
	\renewcommand{\thetable}{\arabic{table}}
	\renewcommand{\thefigure}{\arabic{figure}}
	\renewcommand{\theequation}{S\arabic{equation}}	
	\renewcommand{\thesection}{Supplementary Note \arabic{section}}

    \renewcommand{\figurename}{Supplementary Figure}
    \renewcommand{\tablename}{Supplementary Table}

}

\beginsupplement\setcounter{tocdepth}{1}

\begin{center}
{\bf\Large Supplementary Information: \vspace{5mm} \\
Vibronic Effects on the Quantum Tunnelling of Magnetisation \\
in Kramers Single-Molecule Magnets \\}
\vspace{5mm}
{Andrea Mattioni,$^{1,*}$
Jakob K. Staab,$^{1}$
William J. A. Blackmore,$^{1}$
Daniel Reta,$^{1,2,3,4}$
\\ Jake Iles-Smith,$^{5}$
Ahsan Nazir,$^{5}$
and Nicholas F. Chilton$^{1,\dagger}$ \\ }
\vspace{5mm}
$^1$\textit{Department of Chemistry, School of Natural Sciences, \\ The University of Manchester, Oxford Road, Manchester, M13 9PL, UK}
\\
\vspace{1mm}
$^2$\textit{Faculty of Chemistry, The University of the Basque Country UPV/EHU, Donostia, 20018, Spain}
\\
\vspace{1mm}
$^3$\textit{Donostia International Physics Center (DIPC), Donostia, 20018, Spain}
\\
\vspace{1mm}
$^4$\textit{IKERBASQUE, Basque Foundation for Science, Bilbao, 48013, Spain}
\\
\vspace{1mm}
$^5$\textit{Department of Physics and Astronomy, School of Natural Sciences, \\ The University of Manchester, Oxford Road, Manchester M13 9PL, UK}
\end{center}
\vspace{10mm}


\vfill
\noindent {\footnotesize $^*$\ andrea.mattioni@manchester.ac.uk}\\
\noindent {\footnotesize $^\dagger$\ nicholas.chilton@manchester.ac.uk}

\newpage

\begin{bibunit}

\newpage
\section{Spin-phonon couplings and phonon density of states}
\label{ss:dos}

\begin{figure}[htb]
	\begin{centering}
		\includegraphics[width=0.7\linewidth]{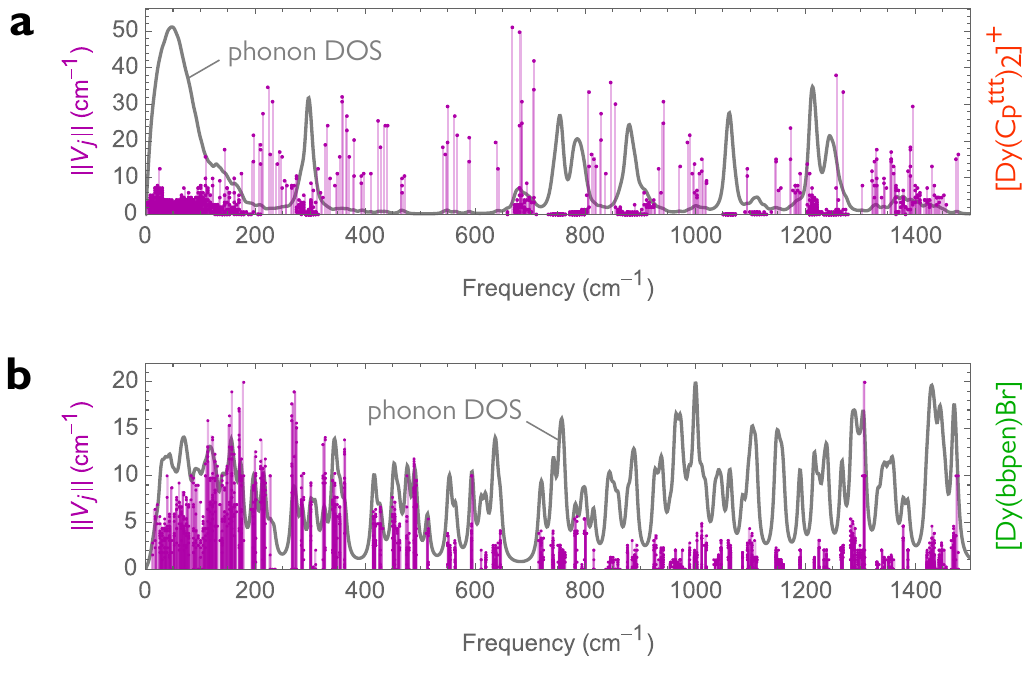}
		\caption{{\bf Spin-phonon couplings and phonon density of states.}
		Spin phonon-couplings (purple sticks) are quantified as the Frobenius norm of the electronic part of the spin-phonon coupling operators $\hat{V}_j$. The phonon density of states (DOS), shown as a grey line, is obtained by broadening each mode with an anti-symmetrised Lorentzian lineshape with full width at half maximum of 10~cm$^{-1}$ \cite{kragskow2023}. \textbf{a}, \dycp{} in dichloromethane solvent ball; \textbf{b}, \dybbpen{} molecular crystal.}
		\label{sf:dos}
	\end{centering}
\end{figure}

\newpage
\section{Derivation of the effective vibronic doublet Hamiltonian}
\label{ss:hamiltonian_derivation}

\subsection{Electronic perturbation theory}

The starting point for our analysis of vibronic effects on QTM is the vibronic Hamiltonian 
\begin{equation} \label{s:e:vibronic+zeeman}
\hat{H} = \sum_{m>0} E_m (\ket{m}\bra{m}+\ket{\bar{m}}\bra{\bar{m}}) + \hat{H}_\textrm{Zee}
+\sum_j \hat{V}_j \otimes (\hat{b}_j + \hat{b}_j^\dagger)
+\sum_j \omega_j \hat{b}_j^\dagger \hat{b}_j,
\end{equation}
where
$\hat{H}_\textrm{Zee} = \mu_B g_J \mathbf{B}\cdot\hat{\mathbf{J}}$.
is the Zeeman interaction with a magnetic field $\bf B$.
The doubly degenerate eigenstates of the crystal field Hamiltonian
$H_\text{CF}=\sum_{m>0} E_m (\ket{m}\bra{m}+\ket{\bar{m}}\bra{\bar{m}})$
are related by time-reversal symmetry, i.e.
$\hat{\Theta}\ket{m} \propto \ket{\bar{m}}$
with $\hat{\Theta}^2 \ket{m} = -\ket{m}$,
where $\hat{\Theta}$ is the time-reversal operator.
In the case of Dy(III), the total electronic angular momentum is $J=15/2$, leading to $2J+1=16$ electronic states.
We label these states in ascending energy with integers $m=\pm 1, \dots, \pm 8$, using the compact notation $\ket{-m}=\ket{\bar{m}}$.

We momentarily neglect the spin-phonon coupling and focus on the purely electronic Hamiltonian $H_\text{el}=H_\text{CF}+H_\text{Zee}$.
Within each degenerate subspace, the Zeeman  term selects a specific electronic basis and lifts its degeneracy.
This can be seen by projecting the electronic Hamitonian onto the $m$-th subspace and diagonalising the $2\times 2$ matrix 
\begin{equation}
H_\text{el}^{(m)} = E_m + \mu_\text{B} g_J
\left( \begin{array}{cc}
\bra{m}       \mathbf{B} \cdot \hat{\mathbf{J}} \ket{m}       &
\bra{m}       \mathbf{B} \cdot \hat{\mathbf{J}} \ket{\bar{m}} \\
\bra{\bar{m}} \mathbf{B} \cdot \hat{\mathbf{J}} \ket{m}       &
\bra{\bar{m}} \mathbf{B} \cdot \hat{\mathbf{J}} \ket{\bar{m}}
\end{array} \right).
\end{equation}
For each individual cartesian component of the angular momentum, we decompose the corresponding $2\times 2$ matrix in terms of Pauli spin operators, which allows to rewrite the Hamiltonian of the $m$-th doublet as $H^{(m)}_\text{el} = E_m + \mu_\text{B} \mathbf{B}\cdot \mathbf{g}_\text{el}^{(m)} \cdot \bm{\sigma}^{(m)}/2$,
where
\begin{equation} \label{se:g_el}
\mathbf{g}_\text{el}^{(m)} = 
2 g_J \left( \begin{array}{ccc}
\Re\bra{\bar{m}}\hat{J}_x\ket{m} & \Im\bra{\bar{m}}\hat{J}_x\ket{m} & \bra{m}\hat{J}_x\ket{m} \\
\Re\bra{\bar{m}}\hat{J}_y\ket{m} & \Im\bra{\bar{m}}\hat{J}_y\ket{m} & \bra{m}\hat{J}_y\ket{m} \\
\Re\bra{\bar{m}}\hat{J}_z\ket{m} & \Im\bra{\bar{m}}\hat{J}_z\ket{m} & \bra{m}\hat{J}_z\ket{m}
\end{array} \right)
\end{equation}
is the $g$-matrix for an effective spin 1/2
and
${\bm{\sigma}}^{(m)}=(\sigma^{(m)}_x,\sigma^{(m)}_y,\sigma^{(m)}_z)$, with
$\sigma^{(m)}_z=\ket{m}\bra{m}-\ket{\bar{m}}\bra{\bar{m}}$.
We note that in general the $g$-matrix in Eq. \eqref{se:g_el} is not hermitean, but can be brought to such form by transforming the spin operators $\bm{\sigma}^{(m)}$ to an appropriate basis \cite{Chibotaru2008}. An easier prescription to find the hermitean form af any 
 $g$-matrix $\mathbf{g}$ is to redefine it as $\sqrt{\mathbf{g}\mathbf{g}^\dagger}$.

To lowest order in the magnetic field, the Zeeman interaction lifts the two-fold degeneracy by selecting the basis
\begin{eqnarray} \label{se:m_plus}
\ket{m_+} &= \cos\frac{\theta_m}{2}\ket{m} + e^{\ii{} \phi_m} \sin\frac{\theta_m}{2}\ket{\bar{m}} \\
\label{se:m_minus}
\ket{m_-} &= -\sin\frac{\theta_m}{2}\ket{m} + e^{\ii{} \phi_m} \cos\frac{\theta_m}{2}\ket{\bar{m}}
\end{eqnarray}
and shifting the energies according to
$
E_{m,\pm} = E_m \pm {\Delta_m}/{2},
$
where the gap
\begin{eqnarray} \label{se:Delta_m}
\Delta_m &=&
\bra{m_+}\hat{H}_\text{Zee}\ket{m_+} - 
\bra{m_-}\hat{H}_\text{Zee}\ket{m_-} \\
&=& 2 \mu_\text{B} g_J \sqrt{ \bra{m} \mathbf{B} \cdot \hat{\mathbf{J}} \ket{m}^2 + |\bra{m} \mathbf{B} \cdot \hat{\mathbf{J}} \ket{\bar{m}}|^2 } \nonumber
\end{eqnarray}
can be obtained as the norm of the vector $\mathbf{j}_m = \mu_\text{B} \mathbf{B} \cdot \mathbf{g}_\text{el}^{(m)}$
and the phase and mixing angles are defined as
\begin{equation}
e^{\ii \phi_m} = \frac{\bra{\bar{m}} \mathbf{B} \cdot \hat{\mathbf{J}} \ket{m}} {|\bra{\bar{m}} \mathbf{B} \cdot \hat{\mathbf{J}} \ket{m}|},
\qquad
\tan\theta_m = \frac{ |\bra{\bar{m}} \mathbf{B} \cdot \hat{\mathbf{J}} \ket{m}| }{ \bra{m} \mathbf{B} \cdot \hat{\mathbf{J}} \ket{m}},
\end{equation}
or equivalently as the azimuthal and polar angles determining the direction of $\mathbf{j}_m$.

Besides selecting a preferred basis and lifting the degeneracy of each doublet, the Zeeman interaction also causes mixing between different doublets.
In particular, the lowest doublet will change according to
\begin{equation} \label{s:e:1pm}
\ket{1_\pm'} = \ket{1_\pm} + \sum_{m\neq 1,\bar{1}} \ket{m} \frac{\bra{m}\hat{H}_\text{Zee}\ket{1_\pm}}{E_1-E_m} + \mathcal{O}(B^2)
\approx \left( 1 - \hat{Q}_1 \hat{H}_\text{Zee} \right) \ket{1_\pm},
\end{equation}
with
\begin{equation}
\hat{Q}_1=\sum_{m\neq 1,\bar{1}} \ket{m}\frac{1}{E_m-E_1}\bra{m}.
\end{equation}

\subsection{Polaron Hamiltonian for the ground doublet}

Now that we have an approximate expression for the relevant electronic states,
we reintroduce the spin-phonon coupling into the picture.
First, we project the vibronic Hamiltonian \eqref{s:e:vibronic+zeeman} onto the subspace spanned by $\ket{1'_\pm}$, yielding
\begin{equation} \label{se:vibronic_hamiltonian_projected}
\hat{H}_\text{eff} = E_1 + \left(\begin{array}{cc} \frac{\Delta_1}{2} & 0 \\ 0 & -\frac{\Delta_1}{2} \end{array} \right)
+ \sum_j \left( \begin{array}{cc}
\bra{1'_+} \hat{V}_j \ket{1'_+} & \bra{1'_+} \hat{V}_j \ket{1'_-} \\
\bra{1'_-} \hat{V}_j \ket{1'_+} & \bra{1'_-} \hat{V}_j \ket{1'_-}
\end{array}  \right) \otimes (\hat{b}_j + \hat{b}_j^\dagger)
+ \sum_j \omega_j \hat{b}_j^\dagger\hat{b}_j.
\end{equation}
On this basis, the purely electronic part $\hat{H}_\text{CF}+\hat{H}_\text{Zee}$ is diagonal with eigenvalues $E_1\pm\Delta_1/2$, and the purely vibrational part is trivially unaffected.
On the other hand, the spin-phonon couplings can be calculated to lowest order in the magnetic field strength $B$ as 
\begin{eqnarray}
\bra{1'_\pm} \hat{V}_j \ket{1'_\pm} & = &
\bra{1_\pm}\left(1-\hat{H}_\text{Zee}\hat{Q}_1\right) \hat{V}_j
\left(1-\hat{Q}_1\hat{H}_\text{Zee}\right)\ket{1_\pm} + \mathcal{O}(B^2) \\
 & = & \bra{1_\pm} \hat{V}_j \ket{1_\pm} 
 - \bra{1_\pm} \left( \hat{V}_j \hat{Q}_1\hat{H}_\text{Zee} + \hat{H}_\text{Zee}\hat{Q}_1 \hat{V}_j \right) \ket{1_\pm} + \mathcal{O}(B^2) \nonumber \\
 & = & \bra{1} \hat{V}_j \ket{1} 
 - \bra{1_\pm} {\hat{W}_j} \ket{1_\pm}
 + \mathcal{O}(B^2), \nonumber 
\end{eqnarray}
\begin{eqnarray}
\bra{1'_\mp} \hat{V}_j \ket{1'_\pm} & = &
\bra{1_\mp}\left(1-\hat{H}_\text{Zee}\hat{Q}_1\right) \hat{V}_j
\left(1-\hat{Q}_1\hat{H}_\text{Zee}\right)\ket{1_\pm} + \mathcal{O}(B^2) \\
 & = & \bra{1_\mp} \hat{V}_j \ket{1_\pm} 
 - \bra{1_\mp} \left( \hat{V}_j \hat{Q}_1\hat{H}_\text{Zee} + \hat{H}_\text{Zee}\hat{Q}_1 \hat{V}_j \right) \ket{1_\pm} + \mathcal{O}(B^2) \nonumber \\
 & = & - \bra{1_\mp} {\hat{W}_j} \ket{1_\pm}
 + \mathcal{O}(B^2), \nonumber 
\end{eqnarray}
where we have defined
\begin{equation} \label{se:Aj}
\hat{W}_j = \hat{V}_j \hat{Q}_1\hat{H}_\text{Zee} + \hat{H}_\text{Zee}\hat{Q}_1 \hat{V}_j
\end{equation}
and used the time-reversal invariance of the spin-phonon coupling operators to obtain
$\bra{1_\pm} \hat{V}_j \ket{1_\pm} = \bra{1} \hat{V}_j \ket{1}$ and
$\bra{1_\mp} \hat{V}_j \ket{1_\pm} =0$.

%
%

The two states $\ket{1_\pm}$ form a conjugate pair under time reversal, meaning that $\hat{\Theta}\ket{1_\pm}=\mp e^{\ii\alpha} \ket{1_\mp}$ for some $\alpha\in\mathbb{R}$.
Using the fact that for any two states $\psi$, $\varphi$, and for any operator $\hat{O}$ we have $\bra{\psi} \hat{O} \ket{\varphi} = \bra{\hat{\Theta}\varphi} \hat{\Theta}\hat{O}^\dagger\hat{\Theta}^{-1}\ket{\hat{\Theta}\psi}$, and recalling that the angular momentum operator is odd under time reversal, i.e. $\hat{\Theta}\hat{\mathbf{J}}\hat{\Theta}^{-1}=-\hat{\mathbf{J}}$, we can show that
\begin{equation}
\bra{1_-} \hat{W}_j \ket{1_-}  = 
\bra{\hat{\Theta}1_-} \hat{\Theta} \hat{W}_j \hat{\Theta}^{-1} \ket{\hat{\Theta} 1_-} 
 = 
- \bra{1_+} \hat{W}_j \ket{1_+}. \nonumber
\end{equation}
Keeping in mind these observations, and defining the vector
\begin{equation} \label{se:aj_pm}
\mathbf{w}_j = \left(\begin{array}{c}
w_j^x \\ w_j^y \\ w_j^z
\end{array} \right)   
=
\left(\begin{array}{c}
\Re\ \bra{1_-} \hat{W}_j \ket{1_+} \\
\Im\ \bra{1_-} \hat{W}_j \ket{1_+} \\
\bra{1_+} \hat{W}_j \ket{1_+}
\end{array} \right),   
\end{equation}
we can rewrite the spin-phonon coupling operators in Eq.  \eqref{se:vibronic_hamiltonian_projected}  as
\begin{equation} \label{se:spin-phonon_coupling_matrix_projected}
\left( \begin{array}{cc}
\bra{1'_+} \hat{V}_j \ket{1'_+} & \bra{1'_+} \hat{V}_j \ket{1'_-} \\
\bra{1'_-} \hat{V}_j \ket{1'_+} & \bra{1'_-} \hat{V}_j \ket{1'_-}
\end{array}  \right)
 = 
\bra{1}\hat{V}_j\ket{1} -
\left( \begin{array}{cc}
\bra{1_+} \hat{W}_j \ket{1_+} & \bra{1_-} \hat{W}_j \ket{1_+}^* \\
\bra{1_-} \hat{W}_j \ket{1_+} & -\bra{1_+} \hat{W}_j \ket{1_+}
\end{array}  \right) 
 = 
\bra{1}\hat{V}_j\ket{1} -
\mathbf{w}_j \cdot \bm{\sigma}'
\end{equation}
where  $\bm{\sigma}'$
is a vector whose entries are the Pauli matrices in the basis $\ket{1'_\pm}$, i.e. $\sigma'_z=\ket{1'_+}\bra{1'_+}-\ket{1'_-}\bra{1'_-}$.
Plugging this back into 
Eq.  \eqref{se:vibronic_hamiltonian_projected}
and explicitly singling out the diagonal components of $\hat{H}_\text{eff}$  in the basis $\ket{1'_\pm}$, we obtain
\begin{eqnarray} \label{se:vibronic_hamiltonian_projected_pre-polaron}
\hat{H}_\text{eff}
&=&
\ket{1'_+}\bra{1'_+}\left[ E_1 + \frac{\Delta_1}{2}
+\sum_j
\left(\bra{1}\hat{V}_j\ket{1} - w_j^z \right)
\left(\hat{b}_j + \hat{b}_j^\dagger\right)
+ \sum_j \omega_j \hat{b}_j^\dagger\hat{b}_j  \right] \\
&+&
\ket{1'_-}\bra{1'_-}\left[ E_1 - \frac{\Delta_1}{2}
+\sum_j
\left(\bra{1}\hat{V}_j\ket{1} + w_j^z \right)
\left(\hat{b}_j + \hat{b}_j^\dagger\right)
+ \sum_j \omega_j \hat{b}_j^\dagger\hat{b}_j  \right] \nonumber \\
&-&
\sum_j
\left(w^x_j\sigma'_x+w^y_j\sigma'_y\right)
\left(\hat{b}_j + \hat{b}_j^\dagger\right).
\nonumber
\end{eqnarray}

At this point,
we apply a unitary polaron transformation to the Hamiltonian 
\eqref{se:vibronic_hamiltonian_projected_pre-polaron}
\begin{eqnarray} \label{se:polaron_transformation}
\hat{S} &=& \exp\left[ \sum_{s=\pm} \ket{1'_s}\bra{1'_s}\ 
\sum_j
\frac{1}{\omega_j}
\left(\bra{1}\hat{V}_j\ket{1} - s w_j^z \right)
\left(\hat{b}_j^\dagger - \hat{b}_j\right) \right] \\
 &=& \sum_{s=\pm} \ket{1'_s}\bra{1'_s}\ \prod_j \hat{D}_j(\xi_j^s)\nonumber
\end{eqnarray}
where $\xi_j^s = \left(\bra{1}\hat{V}_j\ket{1} - s w_j^z\right)/\omega_j$
and 
\begin{equation}
\hat{D}_j(\xi_j^s)=e^{\xi_j^s \left(\hat{b}_j^\dagger - \hat{b}_j\right)}
\end{equation}
is the bosonic displacement operator acting on mode $j$,
i.e.
$
\hat{D}_j(\xi) \hat{b}_j \hat{D}_j^\dagger(\xi)
= \hat{b}_j - \xi
$.
The Hamiltonian thus becomes
\begin{equation} \label{se:vibronic_hamiltonian_projected_polaron}
\hat{S}\hat{H}_\text{eff}\hat{S}^\dagger
=
\sum_{s=\pm} \ket{1'_s}\bra{1'_s}
\left( E_1 + s\frac{\Delta_1}{2}-\sum_j \omega_j|\xi_j^s|^2 \right)
+ \sum_j \omega_j \hat{b}_j^\dagger\hat{b}_j 
-
\sum_j \hat{S}
\left(w^x_j\sigma'_x+w^y_j\sigma'_y\right)
\left(\hat{b}_j + \hat{b}_j^\dagger\right) \hat{S}^\dagger.
\end{equation}
The polaron transformation reabsorbes the diagonal component of the spin-phonon coupling \eqref{se:spin-phonon_coupling_matrix_projected} proportional to $w_j^z$ into the energy shifts $\omega_j|\xi_j^\pm|^2$, leaving a residual off-diagonal spin-phonon coupling proportional to $w_j^x$ and $w_j^y$.
Note that the polaron transformation exactly diagonalises the Hamiltonian \eqref{se:vibronic_hamiltonian_projected} if $w_j^x=w_j^y=0$.
In \ref{ss:pca}, we argue in detail that in our case $|w_j^x|,|w_j^y|\ll|w_j^z|$ to a very good approximation.
Based on this argument, we could decide to neglect the residual 
spin-phonon
coupling in the polaron frame.
The energies of the states belonging to the lowest doublet are shifted by a vibronic correction
\begin{eqnarray}
E_{1'_\pm} &=& E_1\pm\frac{\Delta_1}{2}-\sum_j \frac{1}{\omega_j}\left( \bra{1}\hat{V}_j\ket{1} \mp w_j^z\right)^2 \\
&=& E_1\pm\frac{\Delta_1}{2}-\sum_j \frac{1}{\omega_j}
\left( \bra{1}\hat{V}_j\ket{1}^2 \mp 2 \bra{1}\hat{V}_j\ket{1} w_j^z +\mathcal{O}(B^2) \right),
\end{eqnarray}
leading to a redefinition of the energy gap
\begin{equation}
E_{1'_+}-E_{1'_-}
= \Delta_1 + 4\sum_j\frac{\bra{1}\hat{V}_j\ket{1}}{\omega_j}w_j^z.
\end{equation}

Although the off-diagonal components of the spin-phonon coupling $w_j^x$ and $w_j^y$ are several orders of magnitude smaller than the diagonal one $w_j^z$ (see \ref{ss:pca}), the sheer number of vibrational modes could still lead to an observable effect on the electronic degrees of freedom.
We can estimate this effect by averaging the residual spin-phonon coupling over a thermal phonon distribution in the polaron frame.
Making use of Eq. \eqref{se:polaron_transformation}, the off-diagonal coupling in Eq. \eqref{se:vibronic_hamiltonian_projected_polaron}  can be written as
\begin{eqnarray} \label{se:residual_coupling}
\hat{H}_\text{sp-ph}^\text{(pol)}
&=& -
\sum_j \hat{S}
\left(w^x_j\sigma'_x+w^y_j\sigma'_y\right)
\left(\hat{b}_j + \hat{b}_j^\dagger\right) \hat{S}^\dagger \\
&=& - \sum_j\ket{1'_-}\bra{1_-}\hat{W}_j\ket{1_+}\bra{1'_+}\ 
\hat{D}_j(\xi_j^-)
\left(\hat{b}_j+\hat{b}_j^\dagger\right)
\hat{D}_j^\dagger(\xi_j^+) + \text{h.c.} \nonumber
\end{eqnarray}
Assuming the vibrations to be in a thermal state at temperature $T$ in the polaron frame
\begin{equation}
\rho_\text{ph}^\text{(th)} = \prod_j \rho_j^\text{(th)} = \prod_j \frac{e^{-\omega_j \hat{b}_j^\dagger\hat{b}_j/k_\text{B}T}}{\text{Tr}\left[e^{-\omega_j \hat{b}_j^\dagger\hat{b}_j/k_\text{B}T}\right]},
\end{equation}
%
%
obtaining the average of Eq. \eqref{se:residual_coupling} reduces to calculating the dimensionless quantity
\begin{eqnarray}
\kappa_j &=& - \text{Tr}\left[
\hat{D}_j(\xi_j^-)
\left(\hat{b}_j+\hat{b}_j^\dagger\right)
\hat{D}_j^\dagger(\xi_j^+)
\rho_j^\text{(th)}
\right] \\
&=& 
\left(\xi_j^+ + \xi_j^-\right) e^{-\frac{1}{2}\left(\xi_j^+ - \xi_j^-\right)^2 \coth\left(\frac{\omega_j}{2 k_\text{B} T}\right)}
\nonumber \\
&=& 
2\frac{\bra{1}\hat{V}_j\ket{1}}{\omega_j} e^{-2\frac{(w_j^z)^2}{\omega_j^2}\coth\left(\frac{\omega_j}{2 k_\text{B} T}\right)}
\nonumber \\
&=& 
2\frac{\bra{1}\hat{V}_j\ket{1}}{\omega_j} \left(
1+\mathcal{O}(B^2),
\right) \nonumber
\end{eqnarray}
which appears as a multiplicative rescaling factor for the off-diagonal couplings $\bra{1_\mp}\hat{W}_j\ket{1_\pm}$.
Note that, when neglecting second and higher order terms in the magnetic field, $\kappa_j$ does not show any dependence on temperature or on the magnetic field orientation via $\theta_1$ and $\phi_1$.

After thermal averaging, the effective electronic Hamiltonian for the lowest energy doublet becomes
\begin{equation} \label{se:H_pol_avg}
\hat{H}_\text{el} = \text{Tr}_\text{ph}\left[
\hat{S} \hat{H}_\text{eff} \hat{S}^\dagger \rho_\text{ph}^\text{(th)}\right] = E_1 + \delta E_1 +  
\left(
2\sum_j \frac{\bra{1}\hat{V}_j\ket{1}}{\omega_j} w_j^x,
2\sum_j \frac{\bra{1}\hat{V}_j\ket{1}}{\omega_j} w_j^y,
\frac{\Delta_1}{2}+
2\sum_j \frac{\bra{1}\hat{V}_j\ket{1}}{\omega_j} w_j^z
\right) \cdot
\left( \begin{array}{c}
\sigma'_x\\
\sigma'_y\\
\sigma'_z
\end{array}\right)
\end{equation}
where the energy of the lowest doublet is shifted by
\begin{equation}
\delta E_1 = - \sum_j \frac{\bra{1}\hat{V}_j\ket{1}^2}{\omega_j}
+\sum_j\frac{\omega_j}{e^{\omega_j/k_\text{B}T}-1}
\end{equation}
due to the spin-phonon coupling and to the thermal phonon energy.
Eq. \eqref{se:H_pol_avg} thus represents a refined description of the lowest effective spin-1/2 doublet in the presence of 
spin-phonon coupling.

We can finally recast the Hamiltonian
\eqref{se:H_pol_avg} in terms of a $g$-matrix for an effective spin 1/2, similarly to what we did earlier in the case of no spin-phonon coupling.
In order to do so, we first recall from Eq. \eqref{se:Delta_m} and \eqref{se:aj_pm} that the quantities $\Delta_1$ and
$(w_j^x,w_j^y,w_j^z)$ appearing in Eq. \eqref{se:H_pol_avg} depend on the magnetic field orientation via the states $\ket{1_\pm}$, and on both orientation and intensity via $\hat{H}_\text{Zee}$.
We can get rid of the first dependence by expressing the Zeeman eigenstates $\ket{1_\pm}$ in terms of the original crystal field eigenstates $\ket{1}$, $\ket{\bar{1}}$.
For the spin-phonon 
coupling vector $\mathbf{w}_j$, we obtain
\begin{equation} \label{se:aj_tilde}
\mathbf{w}_j
=
\left( \begin{array}{c}
\Re \bra{1_-}\hat{W}_j \ket{1_+} \\
\Im \bra{1_-}\hat{W}_j \ket{1_+} \\
\bra{1_+}\hat{W}_j \ket{1_+}
\end{array} \right)
= 
\left( \begin{array}{ccc}
\cos\theta_1\cos\phi_1 & \cos\theta_1\sin\phi_1 & -\sin\theta_1 \\
-\sin\phi_1 & \cos\phi_1 & 0 \\
\sin\theta_1\cos\phi_1 & \sin\theta_1\sin\phi_1 & \cos\theta_1
\end{array} \right)
\left( \begin{array}{c}
\Re \bra{\bar{1}}\hat{W}_j \ket{1} \\
\Im \bra{\bar{1}}\hat{W}_j \ket{1} \\
\bra{1}\hat{W}_j \ket{1}
\end{array} \right)
=
\mathbf{R}(\theta_1,\phi_1) \cdot \tilde{\mathbf{w}}_j.
\end{equation}
where
$\mathbf{R}(\theta_1,\phi_1)$ is a rotation matrix. 
Similarly, the elctronic contribution $\Delta_1$ transforms as
%
%
\begin{equation}
\left( 0 , 0 , \Delta_1 \right)
= \mathbf{j}_1 \cdot \mathbf{R}(\theta_1,\phi_1)^T,
= \mu_\text{B} \mathbf{B} \cdot \mathbf{g}^{(1)}_\text{el}
 \cdot \mathbf{R}(\theta_1,\phi_1)^T.
\end{equation}
The Pauli spin operators need to be changed accordingly to
$
\tilde{\bm{\sigma}} = \mathbf{R}(\theta_1,\phi_1)^T \cdot\bm{\sigma}'
$.
Lastly, we single out explicitly the magnetic field dependence of $\hat{W}_j$, defined in Eq. \eqref{se:Aj}, by introducing
a three-component operator
$\hat{\mathbf{K}}_j=(\hat{K}_j^x,\hat{K}_j^y,\hat{K}_j^z)$, such that
\begin{eqnarray} \label{se:Aj_Kj}
\hat{W}_j &=& \mu_\text{B} g_J \mathbf{B} \cdot \left( 
\hat{V}_j \hat{Q}_1 \hat{\mathbf{J}} + \hat{\mathbf{J}}\hat{Q}_1 \hat{V}_j
\right) \\
&=& \mu_\text{B} g_J \mathbf{B} \cdot \hat{\mathbf{K}}_j. \nonumber
\end{eqnarray}
Thus, the effective electronic Hamiltonian in Eq.
\eqref{se:H_pol_avg} can be finally rewritten as
\begin{equation} \label{se:H_el_g_vib}
\hat{H}_\text{el} = E_1 + \delta E_1 + 
\mu_\text{B} \mathbf{B} \cdot
\left(
\mathbf{g}_\text{el}^{(1)} + \mathbf{g}_\text{vib}
\right)
\cdot \tilde{\bm{\sigma}}/2
\end{equation}
where $\mathbf{g}_\text{el}^{(1)}$ is the electronic $g$-matrix
defined in Eq. \eqref{se:g_el}, and
\begin{equation} \label{se:g_vib}
\mathbf{g}_\text{vib} = 
4 g_J \sum_j \frac{\bra{1}\hat{V}_j\ket{1}}{\omega_j}
\left( \begin{array}{ccc}
\Re\bra{\bar{1}}\hat{K}_j^x\ket{1} & 
\Im\bra{\bar{1}}\hat{K}_j^x\ket{1} & 
        \bra{1} \hat{K}_j^x\ket{1} \\
\Re\bra{\bar{1}}\hat{K}_j^y\ket{1} & 
\Im\bra{\bar{1}}\hat{K}_j^y\ket{1} & 
        \bra{1} \hat{K}_j^y\ket{1} \\
\Re\bra{\bar{1}}\hat{K}_j^z\ket{1} & 
\Im\bra{\bar{1}}\hat{K}_j^z\ket{1} & 
        \bra{1} \hat{K}_j^z\ket{1}
\end{array} \right)
\end{equation}
is a vibronic correction.

Note that this correction is non-perturbative in the spin-phonon coupling,
despite only containing quadratic terms in $\hat{V}_j$ (recall that $\hat{\mathbf{K}}_j$ depends linearly on $\hat{V}_j$).
The only approximations leading to Eq. \eqref{se:H_el_g_vib} are a linear perturbative expansion in the magnetic field $\mathbf{B}$ and neglecting quantum fluctuations of the off-diagonal spin-phonon coupling in the polaron frame, which is accounted for only via its thermal expectation value.
This approximation relies on the fact that the off-diagonal couplings are much smaller than the diagonal spin-phonon 
coupling that is treated exactly by the polaron transformation (see \ref{ss:pca}).

\subsection{Landau-Zener probability}

Let us consider a situation in which the magnetic field comprises a time-independent contribution arising from internal dipolar or hyperfine fields $\mathbf{B}_\text{int}$ and a time dependent external field $\mathbf{B}_\text{ext}(t)$.
Let us fix the orientation of the external field and vary its magnitude at a constant rate, such that the field switches direction at $t=0$.
Under these circumstances, the Hamiltonian of Eq. \eqref{se:H_el_g_vib} becomes
\begin{equation}
    \hat{H}_\text{el}(t) = E_1 + \delta E_1 +
    \mu_\text{B} \left( \mathbf{B}_\text{int} + \frac{\dd{\mathbf{B}_\text{ext}}}{\dd{t}} t \right) \cdot\mathbf{g}\cdot\frac{\tilde{\bm{\sigma}}}{2},
\end{equation}
where $\mathbf{g}=\mathbf{g}_\text{el}^{(1)} + \mathbf{g}_\text{vib}$.
Neglecting the constant energy shift and 
introducing the vectors
\begin{eqnarray}
    \bm{\Delta} &= \mu_\text{B} \mathbf{B}_\text{int}\cdot\mathbf{g}, \\
    \mathbf{v} &= \mu_\text{B} {\dd{\mathbf{B}_\text{ext}}}/{\dd{t}} \cdot\mathbf{g}, 
\end{eqnarray}
the Hamiltonian then becomes
\begin{equation}
    \hat{H}_\text{el}(t)=
    \frac{\bm{\Delta}}{2}\cdot\tilde{\bm{\sigma}} +
    \frac{\mathbf{v}t}{2}\cdot\tilde{\bm{\sigma}}
    = \frac{\bm{\Delta}_\perp}{2}\cdot\tilde{\bm{\sigma}} +
    \frac{\mathbf{v}t + \bm{\Delta}_\parallel}{2}\cdot\tilde{\bm{\sigma}}.
\end{equation}
In the second equality, we have split the vector $\bm{\Delta}=\bm{\Delta}_\perp+\bm{\Delta}_\parallel$ into a perpendicular and a parallel component to $\mathbf{v}$.
Choosing an appropriate reference frame, we can write
\begin{equation} \label{se:H_LZ}
    \hat{H}_\text{el}(t') =
    \frac{\Delta_\perp}{2} \tilde{\sigma}_x +
    \frac{v t'}{2}\tilde{\sigma}_z,
\end{equation}
in terms of the new time variable $t'=t+\Delta_\parallel/v$.
Assuming that the spin is initialised in its ground state at $t'\to-\infty$, the probability of observing a spin flip at $t'\to+\infty$ is given by the Landau-Zener formula \cite{Landau1932i,Landau1932ii,Zener1932,Stueckelberg1932,Majorana1932,Ivakhnenko2023}
\begin{equation}
    P_\text{LZ} = 1-\exp\left(-\frac{\pi\Delta_\perp^2}{2v}\right).
\end{equation}

We remark that tunnelling is only made possible by the presence of $\Delta_\perp$, which stems from internal fields that have a perpendicular component to the externally applied field.
We also observe that a perfectly axial system would not exhibit tunnelling behaviour, since in that case the direction of $\mathbf{B}\cdot \mathbf{g}$ would always point along the easy axis (i.e. along the only eigenvector of $\mathbf{g}$ with a non-vanishing eigenvalue), and therefore $\mathbf{v}$ and $\bm{\Delta}$ would always be parallel.
Thus, deviations from axiality and the presence of transverse fields are both required for QTM to occur.

\newpage
\section{Distribution of spin-phonon coupling vectors} \label{ss:pca}

The effective polaron Hamiltonian presented in the main text and  derived in the previous section provides a good description of the ground doublet only if the spin-phonon coupling operators are approximately diagonal in the electronic eigenbasis.
This is equivalent to requiring that the components of the vectors $\mathbf{w}_j$ defined in Eq. \eqref{se:aj_pm} satisfy
\begin{equation} \label{se:a_condition}
|w_j^x|,|w_j^y| \ll |w_j^z|.
\end{equation}
Thus, a value of $\| \mathbf{w}^\perp_j \| = \sqrt{(w_j^x)^2 + (w_j^y)^2}$ much smaller than $\| \mathbf{w}_j \|$ ensures that a polaron model is well justified.
However, we stress that, even when this condition is not met, the polaron Hamiltonian of Eq. \eqref{se:H_pol_avg} still accounts for the transverse spin-phonon couplings $w_j^x$ and $w_j^y$ in an effective way by considering their thermal average.


Supplementary Fig. \ref{sf:pca} shows the
values $\| \mathbf{w}^\perp_j \| / \| \mathbf{w}_j \|$, which determine the validity of the polaron approximation, 
for all modes $\{\mathbf{w}_j, j=1,\dots,M \}$ (where $M$ is the number of vibrational modes) under the effect of a magnetic field applied in the hard plane, $\mathbf{B}=(1,0,0)$, or along the easy axis, $\mathbf{B}=(0,0,1)$.
The polaron approximation is well justified by the observation that, for most vibrational modes,  $\| \mathbf{w}^\perp_j \| / \| \mathbf{w}_j \|$ is below 0.01 for \dycp{} and below 0.1 for \dybbpen{}.

This observation is confirmed by comparing the variance of the set of vectors $\{\mathbf{w}_j\}$ in the $xy$-plane, $\sigma^2_x + \sigma^2_y$,  to the total variance, $\sigma^2 = \sigma^2_x + \sigma^2_y + \sigma^2_z$, where
\begin{equation}
\sigma^2_\alpha = \text{var}(w_j^\alpha) =
\frac{1}{M} \sum_{j=1}^M \left( w_j^\alpha - \mu_\alpha \right)^2,
\end{equation}
with $\alpha=x,y,z$ and $\mu_\alpha=\frac{1}{M}\sum_{j=1}^M w_j^\alpha$.
For \dycp{}, the variance in $xy$-plane only accounts for around $10^{-6}$ of the total variance, whereas for \dybbpen{} the fraction goes up to $10^{-3}$.
Therefore, we conclude that the approach followed in \ref{ss:hamiltonian_derivation} is fully justified.

\begin{figure}[tbh]
	\begin{centering}
		\includegraphics[width=\linewidth]{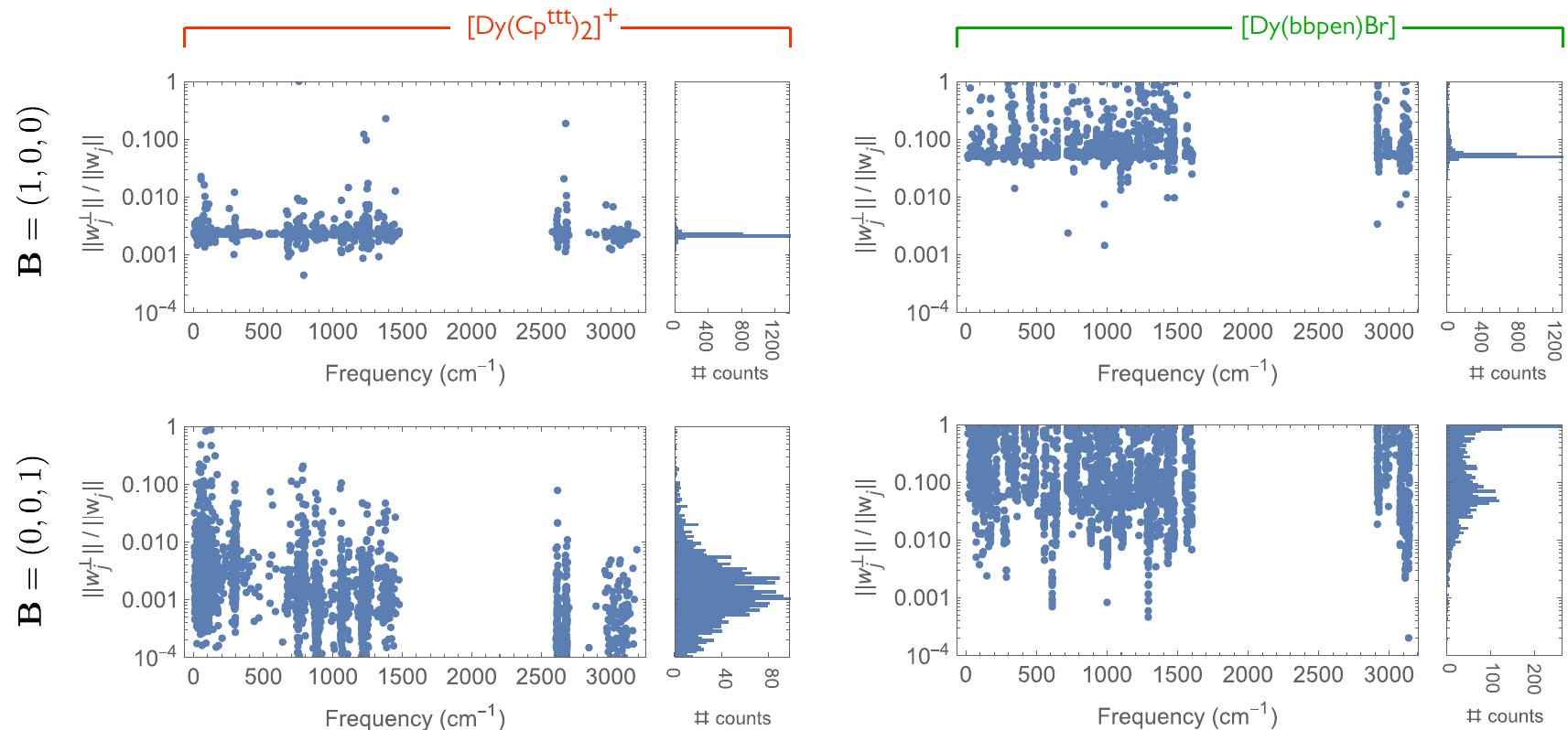}
		\caption{{\bf Distribution of transverse spin-phonon coupling strength $\| \mathbf{w}^\perp_j \| / \| \mathbf{w}_j \| $.}
            The transverse spin-phonon coupling vector $\mathbf{w}^\perp_j$ is the projection onto the hard plane of the spin-phonon coupling vector $\mathbf{w}_j$.
            Left: \dycp{}; right: \dybbpen{}; top: magnetic field $\mathbf{B}$ oriented along $x$ (hard plane); bottom: magnetic field $\mathbf{B}$ oriented along $z$ (easy axis). The field magnitude is fixed to 1~T.}
		\label{sf:pca}
	\end{centering}
\end{figure}

\newpage
\section{Ground Zeeman splitting for \dybbpen{}}

\begin{figure}[htb]
	\begin{centering}
		\includegraphics[width=0.5\linewidth]{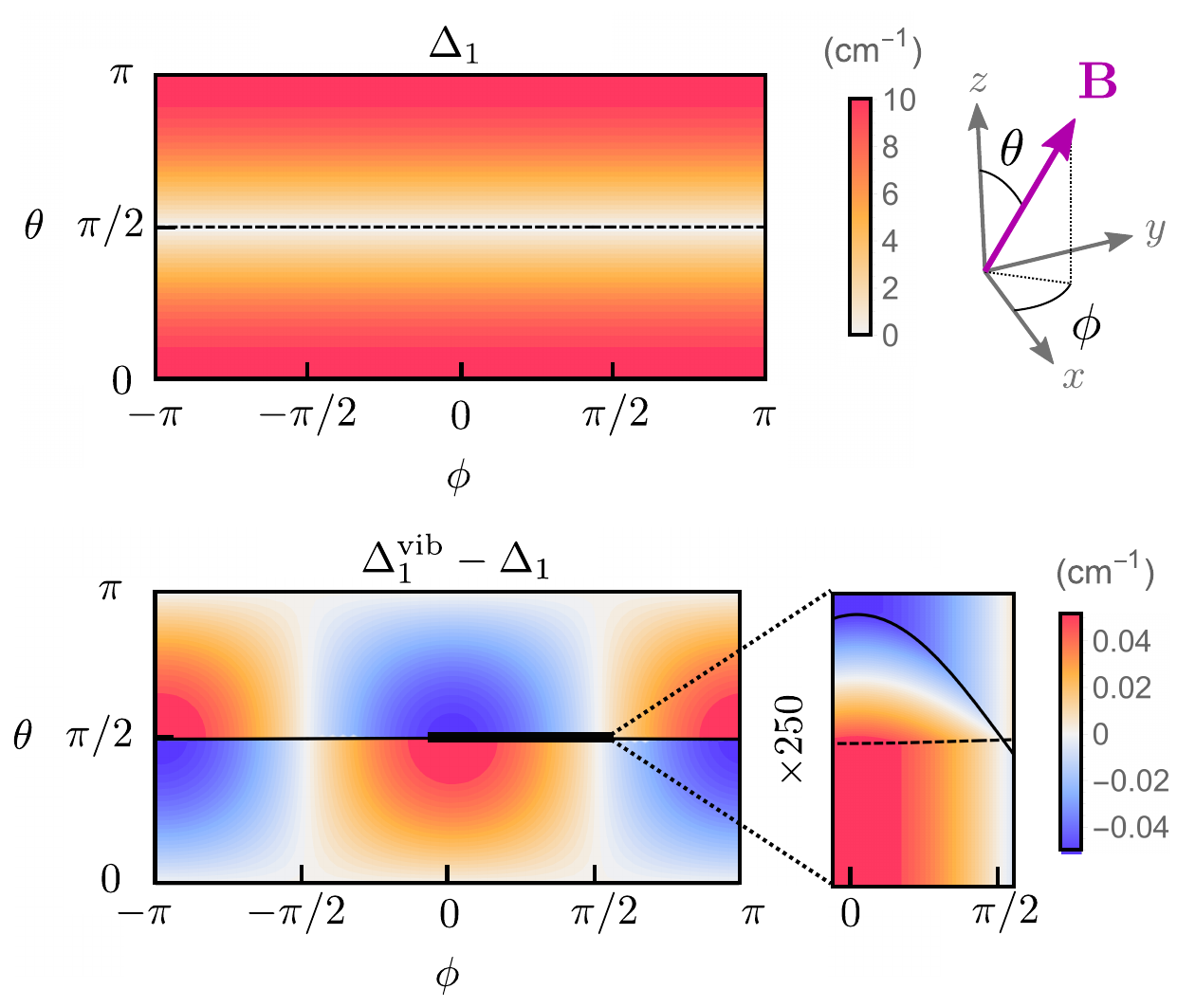}
		\caption{
{\bf Zeeman splitting of the ground Kramers doublet in \dybbpen{}.} 
\textbf{a}, Electronic ground doublet splitting ($\Delta_1$, top) and vibronic correction ($\Delta_1^\text{vib}-\Delta_1$, bottom) as a function of the orientation of the magnetic field, parametrised in terms of polar and azimuthal angles $\theta$ and $\phi$.
The polar angle $\theta$ is measured with respect to the axis joining the two oxygen atoms, corresponding approximately to the easy axis.
The dashed (solid) line corresponds to the electronic (vibronic) hard plane.
The magnitude of the magnetic field is fixed to 1~T.  }
		\label{sf:fig_SI_zeeman_dybbpen}
	\end{centering}
\end{figure}

\newpage
\section{Estimate of the internal fields}
\label{ss:Bint}

\subsection{Dipolar fields}

In this section we provide an estimate of the internal fields $B_\text{int}$ in a disordered ensemble of SMMs.
When a SMM with strongly axial magnetic anisotropy is placed in a strong external magnetic field $\mathbf{B}_\text{ext}$, it gains a non-zero magnetic dipole moment along its easy axis.
Once the external field is removed, the SMM partially retains its magnetisation $\bm{\mu} = \mu \hat{\bm{\mu}}$, which produces a microscopic dipolar field 
\begin{equation} \label{se:Bdip}
    \mathbf{B}_\text{dip}(\mathbf{r}) = \frac{\mu_0 \mu}{4\pi r^3}
    \left[ 3 \hat{\mathbf{r}} (\hat{\bm{\mu}}\cdot\hat{\mathbf{r}}) - \hat{\bm{\mu}} \right]
\end{equation}
at a point $\mathbf{r}=r \hat{\mathbf{r}}$ in space.
This field can then cause a tunnelling gap to open in neighboring SMMs, depending on their relative distance and orientation.

Knowing the spatial distribution and orientation of an ensemble of SMMs, either amorphous or crystalline, we can estimate the internal field experienced by a randomly selected SMM in the ensemble due to all other members of the ensemble.

{\it Frozen solution} --- In the case of a \dycp{} frozen solution, we consider a uniform distribution of randomly oriented SMMs in a sphere of radius $R$ around a central SMM placed at $\mathbf{r}=0$.
We choose a 170~mM SMM concentration to mimic typical experimental conditions \cite{Goodwin2017}.
The ground magnetic moment of the Dy centres can be determined by reading the saturation value of the magnetisation $M_\mathrm{sat}$ of a frozen solution sample of known volume $V$ and concentration $c$, containing $N=c V$ magnetic centres.
Using data from 
ref. \cite{Goodwin2017}, we obtain 
an average magnetic moment per molecule 
\begin{equation} \label{se:mu_z}
\avg{\mu_\parallel} = \frac{M}{N} \approx 4.07 \mu_\text{B}
\end{equation}
along the direction of the external field $\mathbf{B}_\text{ext}$, where $\avg{\cdot}$ denotes the average over the ensemble of SMMs.
Since the orientation of SMMs in a frozen solution is random, the component of the magnetisation $\bm{\mu}$ perpendicular to the applied field averages to zero, i.e. $\avg{\mu_\perp} = 0$.
However, it still contributes to the formation of the microscopic dipolar field \eqref{se:Bdip}, which depends on $\bm{\mu}=\bm{\mu}_\parallel+\bm{\mu}_\perp$.
Since the sample consists of many randomly oriented SMMs, the average magnetisation in Eq.~\eqref{se:mu_z} can also be expressed in terms of $\mu=|\bm{\mu}|$
via the orientational average
\begin{equation} \label{se:mu_z_avg}
    \avg{\mu_\parallel} = \int_0^{\pi/2}\dd{\theta}\ \sin\theta\ \mu_\parallel(\theta)
    =\frac{\mu}{2},
\end{equation}
where $ \mu_\parallel(\theta) = \mu \cos{\theta} $ is the component of the magnetisation of a SMM along the direction of the external field $\mathbf{B}_\text{ext}$.
Thus, the magnetic moment responsible for the microscopic dipolar field is twice as big as the measured value \eqref{se:mu_z}.
We enforce a minimum distance of 10~\AA{} between dipoles, corresponding to approximately twice the RMS distance of ligand atoms from Dy.
Although the dipoles are randomly oriented, the orientation of the dipole is chosen such that the $z$-component is always positive to simulate the presence of an external field $\mathbf{B}_\text{ext}$ along $z$.
We repeat this process 10,000 times in order to sample the full distribution of fields and spin-flip probabilities.
The resulting dipolar field is randomly oriented and has an average magnitude of 5.54~mT, as shown in Supplementary Fig. \ref{sf:fig_SI_DIPOLAR}a.
The corresponding spin-flip probabilities are calculated via Landau-Zener theory (\ref{ss:hamiltonian_derivation}) and are shown in Fig.~3b.
We checked convergence with respect to the solvent sphere radius $R$ and see no significant changes for average number of dipoles ranging from 125 to 1000 (Table~\ref{st:montecarlo}).

\begin{table}[h!]
\begin{centering}
	\begin{tabular}{|c|c|cc|cc|cc|}
        \hline
        $\avg{N}$ & $R$ (\AA{}) & $\avg{B_\text{dip}}$ (mT) & SD & $\avg{P_\text{LZ}^\text{(el)}}$ & SD & $\avg{P_\text{LZ}^\text{(vib)}}$ & SD \\ \hline 
        125 & 66  & 5.49 & 3.22 & 0.0104 & 0.0147 & 0.244 & 0.234 \\
        250 & 84  & 5.47 & 3.24 & 0.0105 & 0.0147 & 0.245 & 0.235 \\
        500 & 105 & 5.48 & 3.23 & 0.0105 & 0.0147 & 0.247 & 0.234 \\
       1000 & 133 & 5.54 & 3.25 & 0.0107 & 0.0148 & 0.250 & 0.238 \\
       \hline
       \end{tabular}
 \caption{{\bf Monte Carlo dipolar field and spin-flip probability of different sized solvent balls.} Average values of dipolar field magnitude $\avg{B_\mathrm{dip}}$, electronic and vibronic spin-flip probabilities $\avg{P_\text{LZ}^\text{(el)}}$, $\avg{P_\text{LZ}^\text{(vib)}}$, and their standard deviations (SD) are reported side by side. The average number of magnetic dipoles included in the calculation is denoted by $\avg{N}$, corresponding to a sphere of radius $R$, assuming magnetic dipoles have a concentration of 170~mM.}
	\label{st:montecarlo}
\end{centering}
\end{table}

{\it Molecular crystal} --- In the case of \dybbpen{}, the spatial distribution of dipoles is fully determined by the crystal structure.
In order to account for polycrystalline samples, we sample random orientations of the magnetising field $\mathbf{B}_\mathrm{ext}$ with respect to the crystal orientation.
Another source of randomness in this molecular crystal comes from diamagnetic dilution of Dy in Y. This is mimicked by setting to zero the dipole moments at the Dy lattice positions with 95\% probability \cite{Liu2016}.
The ground magnetic moment was fixed to $10\mu_\mathrm{B}$, owing to the observation of fully saturated magnetisation at 1~T \cite{Liu2016}.
We consider the field produced by all magnetic dipoles within a sphere of radius $R=100$~\AA{} centred on a Dy atom and repeat this process 10,000 times in order to sample the full distribution of fields shown in Supplementary Fig.~\ref{sf:fig_SI_DIPOLAR}b.
While the average magnitude of the dipolar field is similar to the one obtained for the frozen solution, its orientation is not isotropic.
The component the direction joining two Dy centres belonging to the same unit cell ($z$ in Supplementary Fig.~\ref{sf:fig_SI_DIPOLAR}b) averages to 2.93~mT.
Since the principal anisotropy axis of Dy forms a $23^\circ$ angle 
with respect to that direction, this field results in an average
1.15~mT transverse component.
The presence of this no-vanishing transverse field explains the much higher QTM probabilities obtained in this case. 

\begin{figure}[h!]
	\begin{centering}
		\includegraphics[width=.5\linewidth]{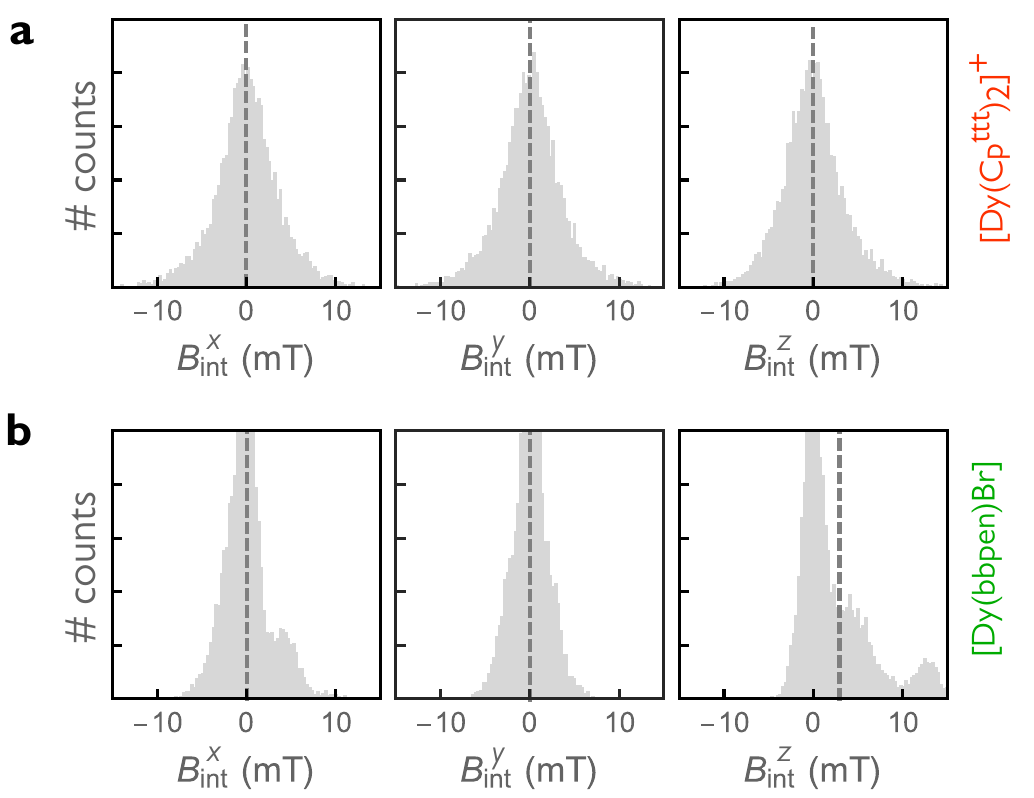}
		\caption{
{\bf Dipolar field distribution.} 
Distributions of the cartesian components of $\mathbf{B}_\mathrm{int} = ({B}^x_\mathrm{int},{B}^y_\mathrm{int},{B}^z_\mathrm{int})^T$.
Average values are indicated by vertical dashed lines.
\textbf{a}, \dycp{} in dichloromethane solvent ball; \textbf{b}, \dybbpen{} molecular crystal.
}
		\label{sf:fig_SI_DIPOLAR}
	\end{centering}
\end{figure}

\subsection{Hyperfine coupling}

Another possible source of microscopic magnetic fields are nuclear spins.
Among the different isotopes of dysprosium, only $^{161}$Dy and $^{163}$Dy have non-zero nuclear spin ($I=5/2$), making up for approximately 44~\% of naturally occurring dysprosium.
The nucear spin degrees of freedom are described by the Hamiltonian
\begin{equation} \label{se:H_nuc}
\hat{H}_\text{nuc} =
\hat{H}_\text{Q} + \hat{H}_\text{HF} =
\hat{\mathbf{I}}\cdot \mathbf{P} \cdot\hat{\mathbf{I}} + \hat{\mathbf{I}}\cdot \mathbf{A} \cdot\hat{\mathbf{J}},
\end{equation}
where the first term is the quadrupole Hamiltonian
$\hat{H}_\text{Q}=\hat{\mathbf{I}}\cdot \mathbf{P} \cdot\hat{\mathbf{I}}$,
accounting for the zero-field splitting of the nuclear spin states, and the second term
$\hat{H}_\text{HF}=\hat{\mathbf{I}}\cdot \mathbf{A} \cdot\hat{\mathbf{J}}$
accounts for the hyperfine coupling between nuclear spin $\hat{\mathbf{I}}$ and electronic angular momentum $\hat{\mathbf{J}}$ operators.
In analogy with the electronic Zeeman Hamiltonian $\hat{H}_\text{Zee} = \mu_\text{B}g_J \mathbf{B}\cdot \hat{\mathbf{J}}$,
we define the effective nuclear magnetic field operator
\begin{equation}
\mu_\text{B} g_J \hat{\mathbf{B}}_\text{nuc} = \mathbf{A}^T \cdot \hat{\mathbf{I}},
\end{equation}
so that the hyperfine coupling Hamiltonian takes the form of a Zeeman interaction $\hat{H}_\text{HF}=\mu_\text{B} g_J \hat{\mathbf{B}}_\text{nuc}^\dagger \cdot \hat{\mathbf{J}}$.
If we consider the nuclear spin to be in a thermal state at temperature $T$ with respect to the quadrupole Hamiltonian $\hat{H}_\text{Q}$, the resulting expectation  value of the nuclear magnetic field vanishes, since the nuclear spin is completely unpolarised.
However, the external field $\mathbf{B}_\text{ext}$ will tend to polarise the nuclear spin via the nuclear Zeeman Hamiltonian
\begin{equation}
\hat{H}_\text{nuc, Zee} = \mu_\text{N} g_I\ \mathbf{B}_\text{ext}\cdot\hat{\mathbf{I}},
\end{equation}
where $\mu_\text{N}$ is the nuclear magneton and $g_I$ is the nuclear $g$-factor of a Dy nucleus.
In this case,
the nuclear spin is described by the thermal state
\begin{equation}
\rho_\text{nuc}^\text{(th)} = \frac{e^{-(\hat{H}_\text{Q}+\hat{H}_\text{nuc, Zee})/k_\text{B}T} }{\text{Tr}\left[e^{-(\hat{H}_\text{Q}+\hat{H}_\text{nuc, Zee})/k_\text{B}T}\right]}
\end{equation}
and the effective nuclear magnetic field can be calculated as
\begin{equation} \label{se:Bnuc}
\mathbf{B}_\text{nuc} = \text{Tr} \left[ \hat{\mathbf{B}}_\text{nuc} \rho_\text{nuc}^\text{(th)}\right].
\end{equation}

To the best of our knowledge, quadrupole and hyperfine coupling tensors for Dy in \dycp{} and \dybbpen{} have not been reported in the literature. However, ab initio calculations of hyperfine coupling tensors have been performed on DyPc$_2$ \cite{Wysocki2020}.
Although the dysprosium atom in DyPc$_2$ and \dycp{} interacts with different ligands, the crystal field is qualitatively similar for these two complexes, therefore we expect the nuclear spin Hamiltonian to be sufficiently close to the one for \dycp{}, at least for the purpose of obtaining an approximate estimate.
Using the quadrupolar and hyperfine tensors determined for DyPc$_2$ \cite{Wysocki2020} and the nuclear $g$-factors measured for $^{161}$Dy  and $^{163}$Dy \cite{Ferch1974}, 
we can compute ${B}_\text{nuc}=|\mathbf{B}_\text{nuc}|$ from Eq. \eqref{se:Bnuc}  for different orientations of the external magnetic field.
As shown in Supplementary Table \ref{st:Bnuc},
the effective nuclear magnetic fields at $T=2\text{~K}$
are at least one order of magnitude smaller than the dipolar fields calculated in the previous section, regardless of the orientation of the external field.

\begin{table}[hb]
\begin{centering}
	\begin{tabular}{lcccc}
	   & \phantom{tttt} & $^{161}$Dy & \phantom{tttt} & $^{163}$Dy \\ \hline 
	$\mathbf{B}_\text{ext}\ ||\ \hat{\mathbf{x}} $ & & 
        $2.82\times 10^{-8}$~T & &
        $5.34\times 10^{-8}$~T \\
	$\mathbf{B}_\text{ext}\ ||\ \hat{\mathbf{y}} $ & & 
        $1.77\times 10^{-8}$~T & & 
        $3.38\times 10^{-8}$~T \\
	$\mathbf{B}_\text{ext}\ ||\ \hat{\mathbf{z}} $ & & 
        $5.51 \times 10^{-5}$~T & & 
        $1.08\times 10^{-4}$~T \\
	\end{tabular}
	\caption{{\bf Effective nuclear magnetic field.} The effective field due to hyperfine coupling is calculated using Eq. \eqref{se:Bnuc}, assuming the nuclear spin to be in a thermal state at temperature $T=2$~K. Different rows correspond to different orientations of the external magnetic field $\mathbf{B}_\mathrm{ext}$, chosen to lie along the three cartesian unit vectors $\hat{\mathbf{x}}$, $\hat{\mathbf{y}}$, $\hat{\mathbf{z}}$. Columns correspond to the two naturally occurring isotopes of Dy.}
	\label{st:Bnuc}
\end{centering}
\end{table}

\newpage
\section{Relation between single-mode axiality and spin-flip probability}
\label{ss:toymodel}

In the following we show that the correlation between single-mode spin-flip probability $\langle P_j \rangle $ and single mode axiality $A_j $ presented in Fig.~4 in the main text can be rationalised in terms of a simple toy model.

Let us work in the reference frame where the electronic $g$-matrix is diagonal.
For a system with strong easy-axis character, this can be approximated as 
\begin{equation}
\mathbf{g}_\mathrm{el} \propto \left( \begin{array}{ccc}
 \lambda  &  &  \\
   & \lambda & \\
   & & 1
\end{array} \right) \mathrm{\quad with \quad} \lambda \ll 1.
\end{equation}
We choose the vibronic correction to the $g$-matrix to have easy-axis anisotropy as well and we only consider its largest $g$-value $\eta$, which is also much less than one.
This approximation is justified by inspection of the $g$-matrices $\mathbf{g}_\mathrm{vib}$ obtained numerically.
The direction of the anisotropy axis corresponding to $\eta$ is determined by $\theta$, the tilt angle  away from the electronic hard plane, as sketched in Supplementary Fig. \ref{sf:toymodel}a.
Thus,
\begin{equation}
\mathbf{g}_\mathrm{vib} \propto \eta \left( \begin{array}{ccc}
 0 & 0                    & 0                    \\
 0 & \cos^2\theta         & \sin\theta\cos\theta \\
 0 & \sin\theta\cos\theta & \sin^2\theta
\end{array} \right).
\end{equation}
Assuming an external field sweep along $z$ and an internal field along $y$, we can calculate axiality and spin-flip probability corresponding to both the electronic $g$-matrix $\mathbf{g}_\mathrm{el}$ and the vibronic one $\mathbf{g}_\mathrm{el}+\mathbf{g}_\mathrm{vib}$.
If no spin-phonon coupling is present ($\eta=0$), we obtain
\begin{eqnarray}
    A_\mathrm{el} =& \frac{1-\lambda}{1+2\lambda}, \\
P_\mathrm{el} =& 1-e^{-C\lambda^2},
\end{eqnarray}
where $C$ is a positive constant determined by sweep rate, internal field and absolute value of the largest electronic $g$-value.
Assuming weak spin-phonon coupling for simplicity, the vibronic analogue of these quantities can be expanded in powers of $\eta$ as 
\begin{eqnarray}
    A_\mathrm{vib} &= A_\mathrm{el} + \alpha(\theta,\lambda) \eta + O(\eta^2), \\
    P_\mathrm{vib} &= P_\mathrm{el} + \beta(\theta,\lambda) \eta + O(\eta^2),
\end{eqnarray}
where
\begin{eqnarray}
\alpha(\theta,\lambda) =& -\frac{3}{4} \frac{1-2\lambda+(1+2\lambda)\cos 2\theta}{(1+2\lambda)^2} \label{se:alpha} \\
\beta(\theta,\lambda) =& \frac{1}{2} C \lambda  e^{-C \lambda^2}
\left( 2-\lambda+(2+\lambda)\cos 2\theta\right). \label{se:beta}
\end{eqnarray}
If the two coefficients $\alpha$ and $\beta$ have opposite signs ($\alpha\beta<0$),
 axiality $A_\mathrm{vib}$ and spin-flip probability $P_\mathrm{vib}$ become anti-correlated: 
switching on the spin-phonon coupling ($\eta\neq 0$) will increase one at the expenses of the other.
In order to satisfy the condition $\alpha\beta<0$, the parameters $\theta$ and $\lambda$ need to be chosen such that
\begin{equation} \label{se:toymodel_bounds}
    \cos 2\theta < -\frac{1-\lambda/2}{1+\lambda/2}
    \mathrm{ \qquad or \qquad }
    \cos 2\theta > -\frac{1-2\lambda}{1+2\lambda}.
\end{equation}
Note that this result only depends on the relative orientation of electronic and vibrational easy axis ($\theta$) and degree of electronic axiality (determined by $\lambda$).
For a given value of $\lambda$, Eq. \eqref{se:toymodel_bounds} is satisfied for all angles $\theta$, except the ones falling in the grey shaded region in  Supplementary Fig. \ref{sf:toymodel}b.
Assuming a uniformly distributed angle $\theta$ across several vibrational modes, spin-phonon coupling will lead to negative correlation between $A_\mathrm{vib}$ and $P_\mathrm{vib}$.
The window of values for $\theta$ that does not lead to this behaviour becomes increasingly smaller upon increasing the axiality of the electronic $g$-matrix, i.e. decreasing $\lambda$.

\begin{figure}[htb]
	\begin{centering}
		\includegraphics[width=0.5\linewidth]{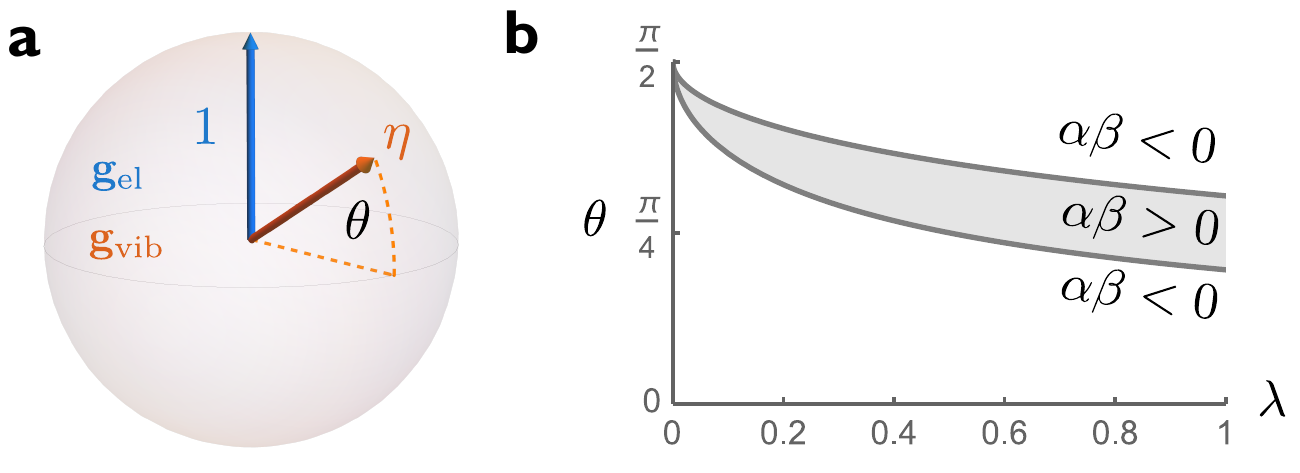}
		\caption{{\bf Toy model for vibronic axiality.}
  \textbf{a}, Main anisotropy axis for the electronic (blue) and vibrational (orange) $g$-matrices $\mathbf{g}_\mathrm{el}$ and $\mathbf{g}_\mathrm{vib}$, with corresponding $g$-values 1 and $\eta$. The angle $\theta$ is defined as the angle between the vibrational easy axis and the electronic hard plane.
  \textbf{b}, Values of angle $\theta$ and electronic hard-plane $g$-value $\lambda$ that lead to the observed anti-correlation between axiality and spin-flip probability. The shaded area between the two gray lines corresponds to the domain where Eq. \eqref{se:toymodel_bounds} is not satisfied. $\alpha$ and $\beta$ are the linear coefficients of the Taylor expansion of vibronic axiality and spin-flip probability with respect to $\eta$, defined in Eq. \eqref{se:alpha} and \eqref{se:beta}. }
		\label{sf:toymodel}
	\end{centering}
\end{figure}

\clearpage


\begin{thebibliography}{60}%
\makeatletter
\providecommand \@ifxundefined [1]{%
 \@ifx{#1\undefined}
}%
\providecommand \@ifnum [1]{%
 \ifnum #1\expandafter \@firstoftwo
 \else \expandafter \@secondoftwo
 \fi
}%
\providecommand \@ifx [1]{%
 \ifx #1\expandafter \@firstoftwo
 \else \expandafter \@secondoftwo
 \fi
}%
\providecommand \natexlab [1]{#1}%
\providecommand \enquote  [1]{``#1''}%
\providecommand \bibnamefont  [1]{#1}%
\providecommand \bibfnamefont [1]{#1}%
\providecommand \citenamefont [1]{#1}%
\providecommand \href@noop [0]{\@secondoftwo}%
\providecommand \href [0]{\begingroup \@sanitize@url \@href}%
\providecommand \@href[1]{\@@startlink{#1}\@@href}%
\providecommand \@@href[1]{\endgroup#1\@@endlink}%
\providecommand \@sanitize@url [0]{\catcode `\\12\catcode `\$12\catcode `\&12\catcode `\#12\catcode `\^12\catcode `\_12\catcode `\%12\relax}%
\providecommand \@@startlink[1]{}%
\providecommand \@@endlink[0]{}%
\providecommand \url  [0]{\begingroup\@sanitize@url \@url }%
\providecommand \@url [1]{\endgroup\@href {#1}{\urlprefix }}%
\providecommand \urlprefix  [0]{URL }%
\providecommand \Eprint [0]{\href }%
\providecommand \doibase [0]{http://dx.doi.org/}%
\providecommand \selectlanguage [0]{\@gobble}%
\providecommand \bibinfo  [0]{\@secondoftwo}%
\providecommand \bibfield  [0]{\@secondoftwo}%
\providecommand \translation [1]{[#1]}%
\providecommand \BibitemOpen [0]{}%
\providecommand \bibitemStop [0]{}%
\providecommand \bibitemNoStop [0]{.\EOS\space}%
\providecommand \EOS [0]{\spacefactor3000\relax}%
\providecommand \BibitemShut  [1]{\csname bibitem#1\endcsname}%
\let\auto@bib@innerbib\@empty
\bibitem [{\citenamefont {Leuenberger}\ and\ \citenamefont {Loss}(2001)}]{Leuenberger2001}%
  \BibitemOpen
  \bibfield  {author} {\bibinfo {author} {\bibfnamefont {M.~N.}\ \bibnamefont {Leuenberger}}\ and\ \bibinfo {author} {\bibfnamefont {D.}~\bibnamefont {Loss}},\ }\bibfield  {title} {\emph {\bibinfo {title} {Quantum computing in molecular magnets},\ }}\href {\doibase 10.1038/35071024} {\bibfield  {journal} {\bibinfo  {journal} {Nature}\ }\textbf {\bibinfo {volume} {410}},\ \bibinfo {pages} {789} (\bibinfo {year} {2001})}\BibitemShut {NoStop}%
\bibitem [{\citenamefont {Sessoli}(2017)}]{Sessoli2017}%
  \BibitemOpen
  \bibfield  {author} {\bibinfo {author} {\bibfnamefont {R.}~\bibnamefont {Sessoli}},\ }\bibfield  {title} {\emph {\bibinfo {title} {Magnetic molecules back in the race},\ }}\href {\doibase 10.1038/548400a} {\bibfield  {journal} {\bibinfo  {journal} {Nature}\ }\textbf {\bibinfo {volume} {548}},\ \bibinfo {pages} {400} (\bibinfo {year} {2017})}\BibitemShut {NoStop}%
\bibitem [{\citenamefont {Coronado}(2020)}]{Coronado2020}%
  \BibitemOpen
  \bibfield  {author} {\bibinfo {author} {\bibfnamefont {E.}~\bibnamefont {Coronado}},\ }\bibfield  {title} {\emph {\bibinfo {title} {Molecular magnetism: from chemical design to spin control in molecules, materials and devices},\ }}\href {\doibase 10.1038/s41578-019-0146-8} {\bibfield  {journal} {\bibinfo  {journal} {Nature Reviews Materials}\ }\textbf {\bibinfo {volume} {5}},\ \bibinfo {pages} {87} (\bibinfo {year} {2020})}\BibitemShut {NoStop}%
\bibitem [{\citenamefont {Chilton}(2022)}]{Chilton2022}%
  \BibitemOpen
  \bibfield  {author} {\bibinfo {author} {\bibfnamefont {N.~F.}\ \bibnamefont {Chilton}},\ }\bibfield  {title} {\emph {\bibinfo {title} {Molecular magnetism},\ }}\href {\doibase 10.1146/annurev-matsci-081420-042553} {\bibfield  {journal} {\bibinfo  {journal} {Annual Review of Materials Research}\ }\textbf {\bibinfo {volume} {52}},\ \bibinfo {pages} {79} (\bibinfo {year} {2022})}\BibitemShut {NoStop}%
\bibitem [{\citenamefont {Sessoli}\ \emph {et~al.}(1993)\citenamefont {Sessoli}, \citenamefont {Gatteschi}, \citenamefont {Caneschi},\ and\ \citenamefont {Novak}}]{Sessoli1993}%
  \BibitemOpen
  \bibfield  {author} {\bibinfo {author} {\bibfnamefont {R.}~\bibnamefont {Sessoli}}, \bibinfo {author} {\bibfnamefont {D.}~\bibnamefont {Gatteschi}}, \bibinfo {author} {\bibfnamefont {A.}~\bibnamefont {Caneschi}}, \ and\ \bibinfo {author} {\bibfnamefont {M.~A.}\ \bibnamefont {Novak}},\ }\bibfield  {title} {\emph {\bibinfo {title} {Magnetic bistability in a metal-ion cluster},\ }}\href {\doibase 10.1038/365141a0} {\bibfield  {journal} {\bibinfo  {journal} {Nature}\ }\textbf {\bibinfo {volume} {365}},\ \bibinfo {pages} {141} (\bibinfo {year} {1993})}\BibitemShut {NoStop}%
\bibitem [{\citenamefont {Kragskow}\ \emph {et~al.}(2023)\citenamefont {Kragskow}, \citenamefont {Mattioni}, \citenamefont {Staab}, \citenamefont {Reta}, \citenamefont {Skelton},\ and\ \citenamefont {Chilton}}]{kragskow2023}%
  \BibitemOpen
  \bibfield  {author} {\bibinfo {author} {\bibfnamefont {J.~G.~C.}\ \bibnamefont {Kragskow}}, \bibinfo {author} {\bibfnamefont {A.}~\bibnamefont {Mattioni}}, \bibinfo {author} {\bibfnamefont {J.~K.}\ \bibnamefont {Staab}}, \bibinfo {author} {\bibfnamefont {D.}~\bibnamefont {Reta}}, \bibinfo {author} {\bibfnamefont {J.~M.}\ \bibnamefont {Skelton}}, \ and\ \bibinfo {author} {\bibfnamefont {N.~F.}\ \bibnamefont {Chilton}},\ }\bibfield  {title} {\emph {\bibinfo {title} {Spin-phonon coupling and magnetic relaxation in single-molecule magnets},\ }}\href {\doibase 10.1039/D2CS00705C} {\bibfield  {journal} {\bibinfo  {journal} {Chem. Soc. Rev.}\ }\textbf {\bibinfo {volume} {52}},\ \bibinfo {pages} {4567} (\bibinfo {year} {2023})}\BibitemShut {NoStop}%
\bibitem [{\citenamefont {Goodwin}\ \emph {et~al.}(2017)\citenamefont {Goodwin}, \citenamefont {Ortu}, \citenamefont {Reta}, \citenamefont {Chilton},\ and\ \citenamefont {Mills}}]{Goodwin2017}%
  \BibitemOpen
  \bibfield  {author} {\bibinfo {author} {\bibfnamefont {C.~A.~P.}\ \bibnamefont {Goodwin}}, \bibinfo {author} {\bibfnamefont {F.}~\bibnamefont {Ortu}}, \bibinfo {author} {\bibfnamefont {D.}~\bibnamefont {Reta}}, \bibinfo {author} {\bibfnamefont {N.~F.}\ \bibnamefont {Chilton}}, \ and\ \bibinfo {author} {\bibfnamefont {D.~P.}\ \bibnamefont {Mills}},\ }\bibfield  {title} {\emph {\bibinfo {title} {Molecular magnetic hysteresis at 60 kelvin in dysprosocenium},\ }}\href {\doibase 10.1038/nature23447} {\bibfield  {journal} {\bibinfo  {journal} {Nature}\ }\textbf {\bibinfo {volume} {548}},\ \bibinfo {pages} {439} (\bibinfo {year} {2017})}\BibitemShut {NoStop}%
\bibitem [{\citenamefont {Guo}\ \emph {et~al.}(2018)\citenamefont {Guo}, \citenamefont {Day}, \citenamefont {Chen}, \citenamefont {Tong}, \citenamefont {Mansikkam{\"a}ki},\ and\ \citenamefont {Layfield}}]{Guo2018}%
  \BibitemOpen
  \bibfield  {author} {\bibinfo {author} {\bibfnamefont {F.-S.}\ \bibnamefont {Guo}}, \bibinfo {author} {\bibfnamefont {B.~M.}\ \bibnamefont {Day}}, \bibinfo {author} {\bibfnamefont {Y.-C.}\ \bibnamefont {Chen}}, \bibinfo {author} {\bibfnamefont {M.-L.}\ \bibnamefont {Tong}}, \bibinfo {author} {\bibfnamefont {A.}~\bibnamefont {Mansikkam{\"a}ki}}, \ and\ \bibinfo {author} {\bibfnamefont {R.~A.}\ \bibnamefont {Layfield}},\ }\bibfield  {title} {\emph {\bibinfo {title} {Magnetic hysteresis up to 80 kelvin in a dysprosium metallocene single-molecule magnet},\ }}\href {\doibase 10.1126/science.aav0652} {\bibfield  {journal} {\bibinfo  {journal} {Science}\ }\textbf {\bibinfo {volume} {362}},\ \bibinfo {pages} {1400} (\bibinfo {year} {2018})}\BibitemShut {NoStop}%
\bibitem [{\citenamefont {Gould}\ \emph {et~al.}(2022)\citenamefont {Gould}, \citenamefont {McClain}, \citenamefont {Reta}, \citenamefont {Kragskow}, \citenamefont {Marchiori}, \citenamefont {Lachman}, \citenamefont {Choi}, \citenamefont {Analytis}, \citenamefont {Britt}, \citenamefont {Chilton}, \citenamefont {Harvey},\ and\ \citenamefont {Long}}]{Gould2022}%
  \BibitemOpen
  \bibfield  {author} {\bibinfo {author} {\bibfnamefont {C.~A.}\ \bibnamefont {Gould}}, \bibinfo {author} {\bibfnamefont {K.~R.}\ \bibnamefont {McClain}}, \bibinfo {author} {\bibfnamefont {D.}~\bibnamefont {Reta}}, \bibinfo {author} {\bibfnamefont {J.~G.~C.}\ \bibnamefont {Kragskow}}, \bibinfo {author} {\bibfnamefont {D.~A.}\ \bibnamefont {Marchiori}}, \bibinfo {author} {\bibfnamefont {E.}~\bibnamefont {Lachman}}, \bibinfo {author} {\bibfnamefont {E.-S.}\ \bibnamefont {Choi}}, \bibinfo {author} {\bibfnamefont {J.~G.}\ \bibnamefont {Analytis}}, \bibinfo {author} {\bibfnamefont {R.~D.}\ \bibnamefont {Britt}}, \bibinfo {author} {\bibfnamefont {N.~F.}\ \bibnamefont {Chilton}}, \bibinfo {author} {\bibfnamefont {B.~G.}\ \bibnamefont {Harvey}}, \ and\ \bibinfo {author} {\bibfnamefont {J.~R.}\ \bibnamefont {Long}},\ }\bibfield  {title} {\emph {\bibinfo {title} {Ultrahard magnetism from mixed-valence dilanthanide complexes with metal-metal bonding},\ }}\href {\doibase 10.1126/science.abl5470} {\bibfield  {journal}
  {\bibinfo  {journal} {Science}\ }\textbf {\bibinfo {volume} {375}},\ \bibinfo {pages} {198} (\bibinfo {year} {2022})}\BibitemShut {NoStop}%
\bibitem [{\citenamefont {Gatteschi}\ \emph {et~al.}(2006)\citenamefont {Gatteschi}, \citenamefont {Sessoli},\ and\ \citenamefont {Villain}}]{GatteschiBook}%
  \BibitemOpen
  \bibfield  {author} {\bibinfo {author} {\bibfnamefont {D.}~\bibnamefont {Gatteschi}}, \bibinfo {author} {\bibfnamefont {R.}~\bibnamefont {Sessoli}}, \ and\ \bibinfo {author} {\bibfnamefont {J.}~\bibnamefont {Villain}},\ }\href {\doibase 10.1093/acprof:oso/9780198567530.001.0001} {\emph {\bibinfo {title} {Molecular Nanomagnets}}}\ (\bibinfo  {publisher} {Oxford University Press},\ \bibinfo {year} {2006})\BibitemShut {NoStop}%
\bibitem [{\citenamefont {Thomas}\ \emph {et~al.}(1996)\citenamefont {Thomas}, \citenamefont {Lionti}, \citenamefont {Ballou}, \citenamefont {Gatteschi}, \citenamefont {Sessoli},\ and\ \citenamefont {Barbara}}]{Thomas1996}%
  \BibitemOpen
  \bibfield  {author} {\bibinfo {author} {\bibfnamefont {L.}~\bibnamefont {Thomas}}, \bibinfo {author} {\bibfnamefont {F.}~\bibnamefont {Lionti}}, \bibinfo {author} {\bibfnamefont {R.}~\bibnamefont {Ballou}}, \bibinfo {author} {\bibfnamefont {D.}~\bibnamefont {Gatteschi}}, \bibinfo {author} {\bibfnamefont {R.}~\bibnamefont {Sessoli}}, \ and\ \bibinfo {author} {\bibfnamefont {B.}~\bibnamefont {Barbara}},\ }\bibfield  {title} {\emph {\bibinfo {title} {Macroscopic quantum tunnelling of magnetization in a single crystal of nanomagnets},\ }}\href {\doibase 10.1038/383145a0} {\bibfield  {journal} {\bibinfo  {journal} {Nature}\ }\textbf {\bibinfo {volume} {383}},\ \bibinfo {pages} {145} (\bibinfo {year} {1996})}\BibitemShut {NoStop}%
\bibitem [{\citenamefont {Garanin}\ and\ \citenamefont {Chudnovsky}(1997)}]{Garanin1997}%
  \BibitemOpen
  \bibfield  {author} {\bibinfo {author} {\bibfnamefont {D.~A.}\ \bibnamefont {Garanin}}\ and\ \bibinfo {author} {\bibfnamefont {E.~M.}\ \bibnamefont {Chudnovsky}},\ }\bibfield  {title} {\emph {\bibinfo {title} {Thermally activated resonant magnetization tunneling in molecular magnets: Mn$_{12}$ac and others},\ }}\href {\doibase 10.1103/PhysRevB.56.11102} {\bibfield  {journal} {\bibinfo  {journal} {Phys. Rev. B}\ }\textbf {\bibinfo {volume} {56}},\ \bibinfo {pages} {11102} (\bibinfo {year} {1997})}\BibitemShut {NoStop}%
\bibitem [{\citenamefont {Reta}\ \emph {et~al.}(2021)\citenamefont {Reta}, \citenamefont {Kragskow},\ and\ \citenamefont {Chilton}}]{Reta2021}%
  \BibitemOpen
  \bibfield  {author} {\bibinfo {author} {\bibfnamefont {D.}~\bibnamefont {Reta}}, \bibinfo {author} {\bibfnamefont {J.~G.~C.}\ \bibnamefont {Kragskow}}, \ and\ \bibinfo {author} {\bibfnamefont {N.~F.}\ \bibnamefont {Chilton}},\ }\bibfield  {title} {\emph {\bibinfo {title} {Ab initio prediction of high-temperature magnetic relaxation rates in single-molecule magnets},\ }}\href {\doibase 10.1021/jacs.1c01410} {\bibfield  {journal} {\bibinfo  {journal} {Journal of the American Chemical Society}\ }\textbf {\bibinfo {volume} {143}},\ \bibinfo {pages} {5943} (\bibinfo {year} {2021})}\BibitemShut {NoStop}%
\bibitem [{\citenamefont {Briganti}\ \emph {et~al.}(2021)\citenamefont {Briganti}, \citenamefont {Santanni}, \citenamefont {Tesi}, \citenamefont {Totti}, \citenamefont {Sessoli},\ and\ \citenamefont {Lunghi}}]{Briganti2021}%
  \BibitemOpen
  \bibfield  {author} {\bibinfo {author} {\bibfnamefont {M.}~\bibnamefont {Briganti}}, \bibinfo {author} {\bibfnamefont {F.}~\bibnamefont {Santanni}}, \bibinfo {author} {\bibfnamefont {L.}~\bibnamefont {Tesi}}, \bibinfo {author} {\bibfnamefont {F.}~\bibnamefont {Totti}}, \bibinfo {author} {\bibfnamefont {R.}~\bibnamefont {Sessoli}}, \ and\ \bibinfo {author} {\bibfnamefont {A.}~\bibnamefont {Lunghi}},\ }\bibfield  {title} {\emph {\bibinfo {title} {A complete ab initio view of {O}rbach and {R}aman spin-lattice relaxation in a dysprosium coordination compound},\ }}\href {\doibase 10.1021/jacs.1c05068} {\bibfield  {journal} {\bibinfo  {journal} {Journal of the American Chemical Society}\ }\textbf {\bibinfo {volume} {143}},\ \bibinfo {pages} {13633} (\bibinfo {year} {2021})}\BibitemShut {NoStop}%
\bibitem [{\citenamefont {Staab}\ and\ \citenamefont {Chilton}(2022)}]{Staab2022}%
  \BibitemOpen
  \bibfield  {author} {\bibinfo {author} {\bibfnamefont {J.~K.}\ \bibnamefont {Staab}}\ and\ \bibinfo {author} {\bibfnamefont {N.~F.}\ \bibnamefont {Chilton}},\ }\bibfield  {title} {\emph {\bibinfo {title} {Analytic linear vibronic coupling method for first-principles spin-dynamics calculations in single-molecule magnets},\ }}\href {\doibase 10.1021/acs.jctc.2c00611} {\bibfield  {journal} {\bibinfo  {journal} {Journal of Chemical Theory and Computation}\ } (\bibinfo {year} {2022}),\ 10.1021/acs.jctc.2c00611}\BibitemShut {NoStop}%
\bibitem [{\citenamefont {Lunghi}(2022)}]{Lunghi2022}%
  \BibitemOpen
  \bibfield  {author} {\bibinfo {author} {\bibfnamefont {A.}~\bibnamefont {Lunghi}},\ }\bibfield  {title} {\emph {\bibinfo {title} {Toward exact predictions of spin-phonon relaxation times: An ab initio implementation of open quantum systems theory},\ }}\href {\doibase 10.1126/sciadv.abn7880} {\bibfield  {journal} {\bibinfo  {journal} {Science Advances}\ }\textbf {\bibinfo {volume} {8}},\ \bibinfo {pages} {eabn7880} (\bibinfo {year} {2022})}\BibitemShut {NoStop}%
\bibitem [{\citenamefont {Irl\"ander}\ and\ \citenamefont {Schnack}(2020)}]{Irlander2020}%
  \BibitemOpen
  \bibfield  {author} {\bibinfo {author} {\bibfnamefont {K.}~\bibnamefont {Irl\"ander}}\ and\ \bibinfo {author} {\bibfnamefont {J.}~\bibnamefont {Schnack}},\ }\bibfield  {title} {\emph {\bibinfo {title} {Spin-phonon interaction induces tunnel splitting in single-molecule magnets},\ }}\href {\doibase 10.1103/PhysRevB.102.054407} {\bibfield  {journal} {\bibinfo  {journal} {Phys. Rev. B}\ }\textbf {\bibinfo {volume} {102}},\ \bibinfo {pages} {054407} (\bibinfo {year} {2020})}\BibitemShut {NoStop}%
\bibitem [{\citenamefont {Irl{\"a}nder}\ \emph {et~al.}(2021)\citenamefont {Irl{\"a}nder}, \citenamefont {Schmidt},\ and\ \citenamefont {Schnack}}]{Irlander2021}%
  \BibitemOpen
  \bibfield  {author} {\bibinfo {author} {\bibfnamefont {K.}~\bibnamefont {Irl{\"a}nder}}, \bibinfo {author} {\bibfnamefont {H.-J.}\ \bibnamefont {Schmidt}}, \ and\ \bibinfo {author} {\bibfnamefont {J.}~\bibnamefont {Schnack}},\ }\bibfield  {title} {\emph {\bibinfo {title} {Supersymmetric spin-phonon coupling prevents odd integer spins from quantum tunneling},\ }}\href {\doibase 10.1140/epjb/s10051-021-00073-3} {\bibfield  {journal} {\bibinfo  {journal} {The European Physical Journal B}\ }\textbf {\bibinfo {volume} {94}},\ \bibinfo {pages} {68} (\bibinfo {year} {2021})}\BibitemShut {NoStop}%
\bibitem [{\citenamefont {Kramers}(1930)}]{kramers1930}%
  \BibitemOpen
  \bibfield  {author} {\bibinfo {author} {\bibfnamefont {A.~H.}\ \bibnamefont {Kramers}},\ }\bibfield  {title} {\emph {\bibinfo {title} {Th{\'e}orie g{\'e}n{\'e}rale de la rotation paramagn{\'e}tique dans les cristaux},\ }}\href@noop {} {\bibfield  {journal} {\bibinfo  {journal} {Proceedings Royal Acad. Amsterdam}\ }\textbf {\bibinfo {volume} {33}},\ \bibinfo {pages} {959} (\bibinfo {year} {1930})}\BibitemShut {NoStop}%
\bibitem [{\citenamefont {Ishikawa}\ \emph {et~al.}(2005)\citenamefont {Ishikawa}, \citenamefont {Sugita},\ and\ \citenamefont {Wernsdorfer}}]{Ishikawa2005}%
  \BibitemOpen
  \bibfield  {author} {\bibinfo {author} {\bibfnamefont {N.}~\bibnamefont {Ishikawa}}, \bibinfo {author} {\bibfnamefont {M.}~\bibnamefont {Sugita}}, \ and\ \bibinfo {author} {\bibfnamefont {W.}~\bibnamefont {Wernsdorfer}},\ }\bibfield  {title} {\emph {\bibinfo {title} {Quantum tunneling of magnetization in lanthanide single-molecule magnets: Bis(phthalocyaninato)terbium and bis(phthalocyaninato)dysprosium anions},\ }}\href {\doibase https://doi.org/10.1002/anie.200462638} {\bibfield  {journal} {\bibinfo  {journal} {Angewandte Chemie International Edition}\ }\textbf {\bibinfo {volume} {44}},\ \bibinfo {pages} {2931} (\bibinfo {year} {2005})}\BibitemShut {NoStop}%
\bibitem [{\citenamefont {Moreno-Pineda}\ \emph {et~al.}(2017)\citenamefont {Moreno-Pineda}, \citenamefont {Damjanovi\'c}, \citenamefont {Fuhr}, \citenamefont {Wernsdorfer},\ and\ \citenamefont {Ruben}}]{Moreno2017}%
  \BibitemOpen
  \bibfield  {author} {\bibinfo {author} {\bibfnamefont {E.}~\bibnamefont {Moreno-Pineda}}, \bibinfo {author} {\bibfnamefont {M.}~\bibnamefont {Damjanovi\'c}}, \bibinfo {author} {\bibfnamefont {O.}~\bibnamefont {Fuhr}}, \bibinfo {author} {\bibfnamefont {W.}~\bibnamefont {Wernsdorfer}}, \ and\ \bibinfo {author} {\bibfnamefont {M.}~\bibnamefont {Ruben}},\ }\bibfield  {title} {\emph {\bibinfo {title} {{N}uclear spin isomers: {E}ngineering a {E}t$_4${N}[{D}y{P}c$_2$] spin qudit},\ }}\href {\doibase https://doi.org/10.1002/anie.201706181} {\bibfield  {journal} {\bibinfo  {journal} {Angewandte Chemie International Edition}\ }\textbf {\bibinfo {volume} {56}},\ \bibinfo {pages} {9915} (\bibinfo {year} {2017})}\BibitemShut {NoStop}%
\bibitem [{\citenamefont {Chilton}\ \emph {et~al.}(2013)\citenamefont {Chilton}, \citenamefont {Langley}, \citenamefont {Moubaraki}, \citenamefont {Soncini}, \citenamefont {Batten},\ and\ \citenamefont {Murray}}]{Chilton2013}%
  \BibitemOpen
  \bibfield  {author} {\bibinfo {author} {\bibfnamefont {N.~F.}\ \bibnamefont {Chilton}}, \bibinfo {author} {\bibfnamefont {S.~K.}\ \bibnamefont {Langley}}, \bibinfo {author} {\bibfnamefont {B.}~\bibnamefont {Moubaraki}}, \bibinfo {author} {\bibfnamefont {A.}~\bibnamefont {Soncini}}, \bibinfo {author} {\bibfnamefont {S.~R.}\ \bibnamefont {Batten}}, \ and\ \bibinfo {author} {\bibfnamefont {K.~S.}\ \bibnamefont {Murray}},\ }\bibfield  {title} {\emph {\bibinfo {title} {Single molecule magnetism in a family of mononuclear $\beta$-diketonate lanthanide({III}) complexes: rationalization of magnetic anisotropy in complexes of low symmetry},\ }}\href {\doibase 10.1039/C3SC22300K} {\bibfield  {journal} {\bibinfo  {journal} {Chem. Sci.}\ }\textbf {\bibinfo {volume} {4}},\ \bibinfo {pages} {1719} (\bibinfo {year} {2013})}\BibitemShut {NoStop}%
\bibitem [{\citenamefont {Ortu}\ \emph {et~al.}(2019)\citenamefont {Ortu}, \citenamefont {Reta}, \citenamefont {Ding}, \citenamefont {Goodwin}, \citenamefont {Gregson}, \citenamefont {McInnes}, \citenamefont {Winpenny}, \citenamefont {Zheng}, \citenamefont {Liddle}, \citenamefont {Mills},\ and\ \citenamefont {Chilton}}]{Ortu2019}%
  \BibitemOpen
  \bibfield  {author} {\bibinfo {author} {\bibfnamefont {F.}~\bibnamefont {Ortu}}, \bibinfo {author} {\bibfnamefont {D.}~\bibnamefont {Reta}}, \bibinfo {author} {\bibfnamefont {Y.-S.}\ \bibnamefont {Ding}}, \bibinfo {author} {\bibfnamefont {C.~A.~P.}\ \bibnamefont {Goodwin}}, \bibinfo {author} {\bibfnamefont {M.~P.}\ \bibnamefont {Gregson}}, \bibinfo {author} {\bibfnamefont {E.~J.~L.}\ \bibnamefont {McInnes}}, \bibinfo {author} {\bibfnamefont {R.~E.~P.}\ \bibnamefont {Winpenny}}, \bibinfo {author} {\bibfnamefont {Y.-Z.}\ \bibnamefont {Zheng}}, \bibinfo {author} {\bibfnamefont {S.~T.}\ \bibnamefont {Liddle}}, \bibinfo {author} {\bibfnamefont {D.~P.}\ \bibnamefont {Mills}}, \ and\ \bibinfo {author} {\bibfnamefont {N.~F.}\ \bibnamefont {Chilton}},\ }\bibfield  {title} {\emph {\bibinfo {title} {{S}tudies of hysteresis and quantum tunnelling of the magnetisation in dysprosium({III}) single molecule magnets},\ }}\href {\doibase 10.1039/C9DT01655D} {\bibfield  {journal} {\bibinfo  {journal} {Dalton Trans.}\ }\textbf
  {\bibinfo {volume} {48}},\ \bibinfo {pages} {8541} (\bibinfo {year} {2019})}\BibitemShut {NoStop}%
\bibitem [{\citenamefont {Pointillart}\ \emph {et~al.}(2015)\citenamefont {Pointillart}, \citenamefont {Bernot}, \citenamefont {Golhen}, \citenamefont {Le~Guennic}, \citenamefont {Guizouarn}, \citenamefont {Ouahab},\ and\ \citenamefont {Cador}}]{Pointillart2015}%
  \BibitemOpen
  \bibfield  {author} {\bibinfo {author} {\bibfnamefont {F.}~\bibnamefont {Pointillart}}, \bibinfo {author} {\bibfnamefont {K.}~\bibnamefont {Bernot}}, \bibinfo {author} {\bibfnamefont {S.}~\bibnamefont {Golhen}}, \bibinfo {author} {\bibfnamefont {B.}~\bibnamefont {Le~Guennic}}, \bibinfo {author} {\bibfnamefont {T.}~\bibnamefont {Guizouarn}}, \bibinfo {author} {\bibfnamefont {L.}~\bibnamefont {Ouahab}}, \ and\ \bibinfo {author} {\bibfnamefont {O.}~\bibnamefont {Cador}},\ }\bibfield  {title} {\emph {\bibinfo {title} {Magnetic memory in an isotopically enriched and magnetically isolated mononuclear dysprosium complex},\ }}\href {\doibase https://doi.org/10.1002/anie.201409887} {\bibfield  {journal} {\bibinfo  {journal} {Angewandte Chemie International Edition}\ }\textbf {\bibinfo {volume} {54}},\ \bibinfo {pages} {1504} (\bibinfo {year} {2015})}\BibitemShut {NoStop}%
\bibitem [{\citenamefont {Kishi}\ \emph {et~al.}(2017)\citenamefont {Kishi}, \citenamefont {Pointillart}, \citenamefont {Lefeuvre}, \citenamefont {Riob{\'e}}, \citenamefont {Le~Guennic}, \citenamefont {Golhen}, \citenamefont {Cador}, \citenamefont {Maury}, \citenamefont {Fujiwara},\ and\ \citenamefont {Ouahab}}]{Kishi2017}%
  \BibitemOpen
  \bibfield  {author} {\bibinfo {author} {\bibfnamefont {Y.}~\bibnamefont {Kishi}}, \bibinfo {author} {\bibfnamefont {F.}~\bibnamefont {Pointillart}}, \bibinfo {author} {\bibfnamefont {B.}~\bibnamefont {Lefeuvre}}, \bibinfo {author} {\bibfnamefont {F.}~\bibnamefont {Riob{\'e}}}, \bibinfo {author} {\bibfnamefont {B.}~\bibnamefont {Le~Guennic}}, \bibinfo {author} {\bibfnamefont {S.}~\bibnamefont {Golhen}}, \bibinfo {author} {\bibfnamefont {O.}~\bibnamefont {Cador}}, \bibinfo {author} {\bibfnamefont {O.}~\bibnamefont {Maury}}, \bibinfo {author} {\bibfnamefont {H.}~\bibnamefont {Fujiwara}}, \ and\ \bibinfo {author} {\bibfnamefont {L.}~\bibnamefont {Ouahab}},\ }\bibfield  {title} {\emph {\bibinfo {title} {Isotopically enriched polymorphs of dysprosium single molecule magnets},\ }}\href {\doibase 10.1039/C7CC00317J} {\bibfield  {journal} {\bibinfo  {journal} {Chem. Commun.}\ }\textbf {\bibinfo {volume} {53}},\ \bibinfo {pages} {3575} (\bibinfo {year} {2017})}\BibitemShut {NoStop}%
\bibitem [{\citenamefont {Flores~Gonzalez}\ \emph {et~al.}(2019)\citenamefont {Flores~Gonzalez}, \citenamefont {Pointillart},\ and\ \citenamefont {Cador}}]{FloresGonzales2019}%
  \BibitemOpen
  \bibfield  {author} {\bibinfo {author} {\bibfnamefont {J.}~\bibnamefont {Flores~Gonzalez}}, \bibinfo {author} {\bibfnamefont {F.}~\bibnamefont {Pointillart}}, \ and\ \bibinfo {author} {\bibfnamefont {O.}~\bibnamefont {Cador}},\ }\bibfield  {title} {\emph {\bibinfo {title} {Hyperfine coupling and slow magnetic relaxation in isotopically enriched {D}y$^\text{III}$ mononuclear single-molecule magnets},\ }}\href {\doibase 10.1039/C8QI01209A} {\bibfield  {journal} {\bibinfo  {journal} {Inorg. Chem. Front.}\ }\textbf {\bibinfo {volume} {6}},\ \bibinfo {pages} {1081} (\bibinfo {year} {2019})}\BibitemShut {NoStop}%
\bibitem [{\citenamefont {Blackmore}\ \emph {et~al.}(2023)\citenamefont {Blackmore}, \citenamefont {Mattioni}, \citenamefont {Corner}, \citenamefont {Evans}, \citenamefont {Gransbury}, \citenamefont {Mills},\ and\ \citenamefont {Chilton}}]{Blackmore2023}%
  \BibitemOpen
  \bibfield  {author} {\bibinfo {author} {\bibfnamefont {W.~J.~A.}\ \bibnamefont {Blackmore}}, \bibinfo {author} {\bibfnamefont {A.}~\bibnamefont {Mattioni}}, \bibinfo {author} {\bibfnamefont {S.~C.}\ \bibnamefont {Corner}}, \bibinfo {author} {\bibfnamefont {P.}~\bibnamefont {Evans}}, \bibinfo {author} {\bibfnamefont {G.~K.}\ \bibnamefont {Gransbury}}, \bibinfo {author} {\bibfnamefont {D.~P.}\ \bibnamefont {Mills}}, \ and\ \bibinfo {author} {\bibfnamefont {N.~F.}\ \bibnamefont {Chilton}},\ }\bibfield  {title} {\emph {\bibinfo {title} {Measurement of the quantum tunneling gap in a dysprosocenium single-molecule magnet},\ }}\href {\doibase 10.1021/acs.jpclett.3c00034} {\bibfield  {journal} {\bibinfo  {journal} {The Journal of Physical Chemistry Letters}\ }\textbf {\bibinfo {volume} {14}},\ \bibinfo {pages} {2193} (\bibinfo {year} {2023})}\BibitemShut {NoStop}%
\bibitem [{\citenamefont {Moreno-Pineda}\ \emph {et~al.}(2019)\citenamefont {Moreno-Pineda}, \citenamefont {Taran}, \citenamefont {Wernsdorfer},\ and\ \citenamefont {Ruben}}]{Moreno2019}%
  \BibitemOpen
  \bibfield  {author} {\bibinfo {author} {\bibfnamefont {E.}~\bibnamefont {Moreno-Pineda}}, \bibinfo {author} {\bibfnamefont {G.}~\bibnamefont {Taran}}, \bibinfo {author} {\bibfnamefont {W.}~\bibnamefont {Wernsdorfer}}, \ and\ \bibinfo {author} {\bibfnamefont {M.}~\bibnamefont {Ruben}},\ }\bibfield  {title} {\emph {\bibinfo {title} {Quantum tunnelling of the magnetisation in single-molecule magnet isotopologue dimers},\ }}\href {\doibase 10.1039/C9SC01062A} {\bibfield  {journal} {\bibinfo  {journal} {Chem. Sci.}\ }\textbf {\bibinfo {volume} {10}},\ \bibinfo {pages} {5138} (\bibinfo {year} {2019})}\BibitemShut {NoStop}%
\bibitem [{\citenamefont {Liu}\ \emph {et~al.}(2016)\citenamefont {Liu}, \citenamefont {Chen}, \citenamefont {Liu}, \citenamefont {Vieru}, \citenamefont {Ungur}, \citenamefont {Jia}, \citenamefont {Chibotaru}, \citenamefont {Lan}, \citenamefont {Wernsdorfer}, \citenamefont {Gao}, \citenamefont {Chen},\ and\ \citenamefont {Tong}}]{Liu2016}%
  \BibitemOpen
  \bibfield  {author} {\bibinfo {author} {\bibfnamefont {J.}~\bibnamefont {Liu}}, \bibinfo {author} {\bibfnamefont {Y.-C.}\ \bibnamefont {Chen}}, \bibinfo {author} {\bibfnamefont {J.-L.}\ \bibnamefont {Liu}}, \bibinfo {author} {\bibfnamefont {V.}~\bibnamefont {Vieru}}, \bibinfo {author} {\bibfnamefont {L.}~\bibnamefont {Ungur}}, \bibinfo {author} {\bibfnamefont {J.-H.}\ \bibnamefont {Jia}}, \bibinfo {author} {\bibfnamefont {L.~F.}\ \bibnamefont {Chibotaru}}, \bibinfo {author} {\bibfnamefont {Y.}~\bibnamefont {Lan}}, \bibinfo {author} {\bibfnamefont {W.}~\bibnamefont {Wernsdorfer}}, \bibinfo {author} {\bibfnamefont {S.}~\bibnamefont {Gao}}, \bibinfo {author} {\bibfnamefont {X.-M.}\ \bibnamefont {Chen}}, \ and\ \bibinfo {author} {\bibfnamefont {M.-L.}\ \bibnamefont {Tong}},\ }\bibfield  {title} {\emph {\bibinfo {title} {A stable pentagonal bipyramidal {D}y({III}) single-ion magnet with a record magnetization reversal barrier over 1000 {K}},\ }}\href {\doibase 10.1021/jacs.6b02638} {\bibfield  {journal}
  {\bibinfo  {journal} {Journal of the American Chemical Society}\ }\textbf {\bibinfo {volume} {138}},\ \bibinfo {pages} {5441} (\bibinfo {year} {2016})}\BibitemShut {NoStop}%
\bibitem [{\citenamefont {Ho}\ and\ \citenamefont {Chibotaru}(2018)}]{Ho2018}%
  \BibitemOpen
  \bibfield  {author} {\bibinfo {author} {\bibfnamefont {L.~T.~A.}\ \bibnamefont {Ho}}\ and\ \bibinfo {author} {\bibfnamefont {L.~F.}\ \bibnamefont {Chibotaru}},\ }\bibfield  {title} {\emph {\bibinfo {title} {Spin-lattice relaxation of magnetic centers in molecular crystals at low temperature},\ }}\href {\doibase 10.1103/PhysRevB.97.024427} {\bibfield  {journal} {\bibinfo  {journal} {Phys. Rev. B}\ }\textbf {\bibinfo {volume} {97}},\ \bibinfo {pages} {024427} (\bibinfo {year} {2018})}\BibitemShut {NoStop}%
\bibitem [{\citenamefont {Silbey}\ and\ \citenamefont {Harris}(1984)}]{Silbey1984}%
  \BibitemOpen
  \bibfield  {author} {\bibinfo {author} {\bibfnamefont {R.}~\bibnamefont {Silbey}}\ and\ \bibinfo {author} {\bibfnamefont {R.~A.}\ \bibnamefont {Harris}},\ }\bibfield  {title} {\emph {\bibinfo {title} {Variational calculation of the dynamics of a two level system interacting with a bath},\ }}\href {\doibase 10.1063/1.447055} {\bibfield  {journal} {\bibinfo  {journal} {The Journal of Chemical Physics}\ }\textbf {\bibinfo {volume} {80}},\ \bibinfo {pages} {2615} (\bibinfo {year} {1984})}\BibitemShut {NoStop}%
\bibitem [{\citenamefont {Nabi}\ \emph {et~al.}(2023)\citenamefont {Nabi}, \citenamefont {Staab}, \citenamefont {Mattioni}, \citenamefont {Kragskow}, \citenamefont {Reta}, \citenamefont {Skelton},\ and\ \citenamefont {Chilton}}]{nabi2023}%
  \BibitemOpen
  \bibfield  {author} {\bibinfo {author} {\bibfnamefont {R.}~\bibnamefont {Nabi}}, \bibinfo {author} {\bibfnamefont {J.~K.}\ \bibnamefont {Staab}}, \bibinfo {author} {\bibfnamefont {A.}~\bibnamefont {Mattioni}}, \bibinfo {author} {\bibfnamefont {J.~G.~C.}\ \bibnamefont {Kragskow}}, \bibinfo {author} {\bibfnamefont {D.}~\bibnamefont {Reta}}, \bibinfo {author} {\bibfnamefont {J.~M.}\ \bibnamefont {Skelton}}, \ and\ \bibinfo {author} {\bibfnamefont {N.~F.}\ \bibnamefont {Chilton}},\ }\bibfield  {title} {\emph {\bibinfo {title} {Accurate and efficient spin--phonon coupling and spin dynamics calculations for molecular solids},\ }}\href {\doibase 10.1021/jacs.3c06015} {\bibfield  {journal} {\bibinfo  {journal} {Journal of the American Chemical Society}\ }\textbf {\bibinfo {volume} {145}},\ \bibinfo {pages} {24558} (\bibinfo {year} {2023})}\BibitemShut {NoStop}%
\bibitem [{\citenamefont {Yakovlev}\ and\ \citenamefont {Ossau}(2010)}]{Yakovlev2010}%
  \BibitemOpen
  \bibfield  {author} {\bibinfo {author} {\bibfnamefont {D.~R.}\ \bibnamefont {Yakovlev}}\ and\ \bibinfo {author} {\bibfnamefont {W.}~\bibnamefont {Ossau}},\ }\bibinfo {title} {Magnetic polarons},\ in\ \href {\doibase 10.1007/978-3-642-15856-8_7} {\emph {\bibinfo {booktitle} {Introduction to the Physics of Diluted Magnetic Semiconductors}}},\ \bibinfo {editor} {edited by\ \bibinfo {editor} {\bibfnamefont {J.~A.}\ \bibnamefont {Gaj}}\ and\ \bibinfo {editor} {\bibfnamefont {J.}~\bibnamefont {Kossut}}}\ (\bibinfo  {publisher} {Springer Berlin Heidelberg},\ \bibinfo {address} {Berlin, Heidelberg},\ \bibinfo {year} {2010})\ pp.\ \bibinfo {pages} {221--262}\BibitemShut {NoStop}%
\bibitem [{\citenamefont {Schott}\ \emph {et~al.}(2019)\citenamefont {Schott}, \citenamefont {Chopra}, \citenamefont {Lemaur}, \citenamefont {Melnyk}, \citenamefont {Olivier}, \citenamefont {Di~Pietro}, \citenamefont {Romanov}, \citenamefont {Carey}, \citenamefont {Jiao}, \citenamefont {Jellett}, \citenamefont {Little}, \citenamefont {Marks}, \citenamefont {McNeill}, \citenamefont {McCulloch}, \citenamefont {McNellis}, \citenamefont {Andrienko}, \citenamefont {Beljonne}, \citenamefont {Sinova},\ and\ \citenamefont {Sirringhaus}}]{Schott2019}%
  \BibitemOpen
  \bibfield  {author} {\bibinfo {author} {\bibfnamefont {S.}~\bibnamefont {Schott}}, \bibinfo {author} {\bibfnamefont {U.}~\bibnamefont {Chopra}}, \bibinfo {author} {\bibfnamefont {V.}~\bibnamefont {Lemaur}}, \bibinfo {author} {\bibfnamefont {A.}~\bibnamefont {Melnyk}}, \bibinfo {author} {\bibfnamefont {Y.}~\bibnamefont {Olivier}}, \bibinfo {author} {\bibfnamefont {R.}~\bibnamefont {Di~Pietro}}, \bibinfo {author} {\bibfnamefont {I.}~\bibnamefont {Romanov}}, \bibinfo {author} {\bibfnamefont {R.~L.}\ \bibnamefont {Carey}}, \bibinfo {author} {\bibfnamefont {X.}~\bibnamefont {Jiao}}, \bibinfo {author} {\bibfnamefont {C.}~\bibnamefont {Jellett}}, \bibinfo {author} {\bibfnamefont {M.}~\bibnamefont {Little}}, \bibinfo {author} {\bibfnamefont {A.}~\bibnamefont {Marks}}, \bibinfo {author} {\bibfnamefont {C.~R.}\ \bibnamefont {McNeill}}, \bibinfo {author} {\bibfnamefont {I.}~\bibnamefont {McCulloch}}, \bibinfo {author} {\bibfnamefont {E.~R.}\ \bibnamefont {McNellis}}, \bibinfo {author} {\bibfnamefont {D.}~\bibnamefont
  {Andrienko}}, \bibinfo {author} {\bibfnamefont {D.}~\bibnamefont {Beljonne}}, \bibinfo {author} {\bibfnamefont {J.}~\bibnamefont {Sinova}}, \ and\ \bibinfo {author} {\bibfnamefont {H.}~\bibnamefont {Sirringhaus}},\ }\bibfield  {title} {\emph {\bibinfo {title} {Polaron spin dynamics in high-mobility polymeric semiconductors},\ }}\href {\doibase 10.1038/s41567-019-0538-0} {\bibfield  {journal} {\bibinfo  {journal} {Nature Physics}\ }\textbf {\bibinfo {volume} {15}},\ \bibinfo {pages} {814} (\bibinfo {year} {2019})}\BibitemShut {NoStop}%
\bibitem [{\citenamefont {Godejohann}\ \emph {et~al.}(2020)\citenamefont {Godejohann}, \citenamefont {Scherbakov}, \citenamefont {Kukhtaruk}, \citenamefont {Poddubny}, \citenamefont {Yaremkevich}, \citenamefont {Wang}, \citenamefont {Nadzeyka}, \citenamefont {Yakovlev}, \citenamefont {Rushforth}, \citenamefont {Akimov},\ and\ \citenamefont {Bayer}}]{Godejohann2020}%
  \BibitemOpen
  \bibfield  {author} {\bibinfo {author} {\bibfnamefont {F.}~\bibnamefont {Godejohann}}, \bibinfo {author} {\bibfnamefont {A.~V.}\ \bibnamefont {Scherbakov}}, \bibinfo {author} {\bibfnamefont {S.~M.}\ \bibnamefont {Kukhtaruk}}, \bibinfo {author} {\bibfnamefont {A.~N.}\ \bibnamefont {Poddubny}}, \bibinfo {author} {\bibfnamefont {D.~D.}\ \bibnamefont {Yaremkevich}}, \bibinfo {author} {\bibfnamefont {M.}~\bibnamefont {Wang}}, \bibinfo {author} {\bibfnamefont {A.}~\bibnamefont {Nadzeyka}}, \bibinfo {author} {\bibfnamefont {D.~R.}\ \bibnamefont {Yakovlev}}, \bibinfo {author} {\bibfnamefont {A.~W.}\ \bibnamefont {Rushforth}}, \bibinfo {author} {\bibfnamefont {A.~V.}\ \bibnamefont {Akimov}}, \ and\ \bibinfo {author} {\bibfnamefont {M.}~\bibnamefont {Bayer}},\ }\bibfield  {title} {\emph {\bibinfo {title} {Magnon polaron formed by selectively coupled coherent magnon and phonon modes of a surface patterned ferromagnet},\ }}\href {\doibase 10.1103/PhysRevB.102.144438} {\bibfield  {journal} {\bibinfo  {journal} {Phys.
  Rev. B}\ }\textbf {\bibinfo {volume} {102}},\ \bibinfo {pages} {144438} (\bibinfo {year} {2020})}\BibitemShut {NoStop}%
\bibitem [{\citenamefont {Chin}\ \emph {et~al.}(2011)\citenamefont {Chin}, \citenamefont {Prior}, \citenamefont {Huelga},\ and\ \citenamefont {Plenio}}]{Chin2011}%
  \BibitemOpen
  \bibfield  {author} {\bibinfo {author} {\bibfnamefont {A.~W.}\ \bibnamefont {Chin}}, \bibinfo {author} {\bibfnamefont {J.}~\bibnamefont {Prior}}, \bibinfo {author} {\bibfnamefont {S.~F.}\ \bibnamefont {Huelga}}, \ and\ \bibinfo {author} {\bibfnamefont {M.~B.}\ \bibnamefont {Plenio}},\ }\bibfield  {title} {\emph {\bibinfo {title} {Generalized polaron ansatz for the ground state of the sub-ohmic spin-boson model: An analytic theory of the localization transition},\ }}\href {\doibase 10.1103/PhysRevLett.107.160601} {\bibfield  {journal} {\bibinfo  {journal} {Phys. Rev. Lett.}\ }\textbf {\bibinfo {volume} {107}},\ \bibinfo {pages} {160601} (\bibinfo {year} {2011})}\BibitemShut {NoStop}%
\bibitem [{\citenamefont {Yang}\ \emph {et~al.}(2012)\citenamefont {Yang}, \citenamefont {Devi},\ and\ \citenamefont {Jang}}]{Yang2012}%
  \BibitemOpen
  \bibfield  {author} {\bibinfo {author} {\bibfnamefont {L.}~\bibnamefont {Yang}}, \bibinfo {author} {\bibfnamefont {M.}~\bibnamefont {Devi}}, \ and\ \bibinfo {author} {\bibfnamefont {S.}~\bibnamefont {Jang}},\ }\bibfield  {title} {\emph {\bibinfo {title} {Polaronic quantum master equation theory of inelastic and coherent resonance energy transfer for soft systems},\ }}\href {\doibase 10.1063/1.4732309} {\bibfield  {journal} {\bibinfo  {journal} {The Journal of Chemical Physics}\ }\textbf {\bibinfo {volume} {137}},\ \bibinfo {pages} {024101} (\bibinfo {year} {2012})}\BibitemShut {NoStop}%
\bibitem [{\citenamefont {Kolli}\ \emph {et~al.}(2011)\citenamefont {Kolli}, \citenamefont {Nazir},\ and\ \citenamefont {Olaya-Castro}}]{Kolli2011}%
  \BibitemOpen
  \bibfield  {author} {\bibinfo {author} {\bibfnamefont {A.}~\bibnamefont {Kolli}}, \bibinfo {author} {\bibfnamefont {A.}~\bibnamefont {Nazir}}, \ and\ \bibinfo {author} {\bibfnamefont {A.}~\bibnamefont {Olaya-Castro}},\ }\bibfield  {title} {\emph {\bibinfo {title} {Electronic excitation dynamics in multichromophoric systems described via a polaron-representation master equation},\ }}\href {\doibase 10.1063/1.3652227} {\bibfield  {journal} {\bibinfo  {journal} {The Journal of Chemical Physics}\ }\textbf {\bibinfo {volume} {135}},\ \bibinfo {pages} {154112} (\bibinfo {year} {2011})}\BibitemShut {NoStop}%
\bibitem [{\citenamefont {Pollock}\ \emph {et~al.}(2013)\citenamefont {Pollock}, \citenamefont {McCutcheon}, \citenamefont {Lovett}, \citenamefont {Gauger},\ and\ \citenamefont {Nazir}}]{Pollock2013}%
  \BibitemOpen
  \bibfield  {author} {\bibinfo {author} {\bibfnamefont {F.~A.}\ \bibnamefont {Pollock}}, \bibinfo {author} {\bibfnamefont {D.~P.~S.}\ \bibnamefont {McCutcheon}}, \bibinfo {author} {\bibfnamefont {B.~W.}\ \bibnamefont {Lovett}}, \bibinfo {author} {\bibfnamefont {E.~M.}\ \bibnamefont {Gauger}}, \ and\ \bibinfo {author} {\bibfnamefont {A.}~\bibnamefont {Nazir}},\ }\bibfield  {title} {\emph {\bibinfo {title} {A multi-site variational master equation approach to dissipative energy transfer},\ }}\href {\doibase 10.1088/1367-2630/15/7/075018} {\bibfield  {journal} {\bibinfo  {journal} {New Journal of Physics}\ }\textbf {\bibinfo {volume} {15}},\ \bibinfo {pages} {075018} (\bibinfo {year} {2013})}\BibitemShut {NoStop}%
\bibitem [{\citenamefont {Wilson-Rae}\ and\ \citenamefont {Imamo\ifmmode~\breve{g}\else \u{g}\fi{}lu}(2002)}]{Wilson2002}%
  \BibitemOpen
  \bibfield  {author} {\bibinfo {author} {\bibfnamefont {I.}~\bibnamefont {Wilson-Rae}}\ and\ \bibinfo {author} {\bibfnamefont {A.}~\bibnamefont {Imamo\ifmmode~\breve{g}\else \u{g}\fi{}lu}},\ }\bibfield  {title} {\emph {\bibinfo {title} {Quantum dot cavity-{QED} in the presence of strong electron-phonon interactions},\ }}\href {\doibase 10.1103/PhysRevB.65.235311} {\bibfield  {journal} {\bibinfo  {journal} {Phys. Rev. B}\ }\textbf {\bibinfo {volume} {65}},\ \bibinfo {pages} {235311} (\bibinfo {year} {2002})}\BibitemShut {NoStop}%
\bibitem [{\citenamefont {McCutcheon}\ and\ \citenamefont {Nazir}(2010)}]{McCutcheon2010}%
  \BibitemOpen
  \bibfield  {author} {\bibinfo {author} {\bibfnamefont {D.~P.~S.}\ \bibnamefont {McCutcheon}}\ and\ \bibinfo {author} {\bibfnamefont {A.}~\bibnamefont {Nazir}},\ }\bibfield  {title} {\emph {\bibinfo {title} {Quantum dot {R}abi rotations beyond the weak exciton-phonon coupling regime},\ }}\href {\doibase 10.1088/1367-2630/12/11/113042} {\bibfield  {journal} {\bibinfo  {journal} {New Journal of Physics}\ }\textbf {\bibinfo {volume} {12}},\ \bibinfo {pages} {113042} (\bibinfo {year} {2010})}\BibitemShut {NoStop}%
\bibitem [{\citenamefont {Nazir}\ and\ \citenamefont {McCutcheon}(2016)}]{Nazir2016}%
  \BibitemOpen
  \bibfield  {author} {\bibinfo {author} {\bibfnamefont {A.}~\bibnamefont {Nazir}}\ and\ \bibinfo {author} {\bibfnamefont {D.~P.~S.}\ \bibnamefont {McCutcheon}},\ }\bibfield  {title} {\emph {\bibinfo {title} {Modelling exciton–phonon interactions in optically driven quantum dots},\ }}\href {\doibase 10.1088/0953-8984/28/10/103002} {\bibfield  {journal} {\bibinfo  {journal} {Journal of Physics: Condensed Matter}\ }\textbf {\bibinfo {volume} {28}},\ \bibinfo {pages} {103002} (\bibinfo {year} {2016})}\BibitemShut {NoStop}%
\bibitem [{\citenamefont {Landau}(1932{\natexlab{a}})}]{Landau1932i}%
  \BibitemOpen
  \bibfield  {author} {\bibinfo {author} {\bibfnamefont {L.~D.}\ \bibnamefont {Landau}},\ }\bibfield  {title} {\emph {\bibinfo {title} {Zur {T}heorie der {E}nergie{\"u}bertragung},\ }}\href@noop {} {\bibfield  {journal} {\bibinfo  {journal} {Phyz. Z. Sowjetunion}\ }\textbf {\bibinfo {volume} {1}},\ \bibinfo {pages} {88} (\bibinfo {year} {1932}{\natexlab{a}})}\BibitemShut {NoStop}%
\bibitem [{\citenamefont {Landau}(1932{\natexlab{b}})}]{Landau1932ii}%
  \BibitemOpen
  \bibfield  {author} {\bibinfo {author} {\bibfnamefont {L.~D.}\ \bibnamefont {Landau}},\ }\bibfield  {title} {\emph {\bibinfo {title} {Zur {T}heorie der {E}nergie{\"u}bertragung {II}},\ }}\href@noop {} {\bibfield  {journal} {\bibinfo  {journal} {Phyz. Z. Sowjetunion}\ }\textbf {\bibinfo {volume} {2}},\ \bibinfo {pages} {46} (\bibinfo {year} {1932}{\natexlab{b}})}\BibitemShut {NoStop}%
\bibitem [{\citenamefont {Zener}\ and\ \citenamefont {Fowler}(1932)}]{Zener1932}%
  \BibitemOpen
  \bibfield  {author} {\bibinfo {author} {\bibfnamefont {C.}~\bibnamefont {Zener}}\ and\ \bibinfo {author} {\bibfnamefont {R.~H.}\ \bibnamefont {Fowler}},\ }\bibfield  {title} {\emph {\bibinfo {title} {Non-adiabatic crossing of energy levels},\ }}\href {\doibase 10.1098/rspa.1932.0165} {\bibfield  {journal} {\bibinfo  {journal} {Proceedings of the Royal Society of London. Series A}\ }\textbf {\bibinfo {volume} {137}},\ \bibinfo {pages} {696} (\bibinfo {year} {1932})}\BibitemShut {NoStop}%
\bibitem [{\citenamefont {St{\"u}ckelberg}(1932)}]{Stueckelberg1932}%
  \BibitemOpen
  \bibfield  {author} {\bibinfo {author} {\bibfnamefont {E.~C.~G.}\ \bibnamefont {St{\"u}ckelberg}},\ }\bibfield  {title} {\emph {\bibinfo {title} {{T}heorie der unelastischen {S}t{\"o}sse zwischen {A}tomen},\ }}\href@noop {} {\bibfield  {journal} {\bibinfo  {journal} {Helv. Phys. Acta}\ }\textbf {\bibinfo {volume} {5}},\ \bibinfo {pages} {369} (\bibinfo {year} {1932})}\BibitemShut {NoStop}%
\bibitem [{\citenamefont {Majorana}(1932)}]{Majorana1932}%
  \BibitemOpen
  \bibfield  {author} {\bibinfo {author} {\bibfnamefont {E.}~\bibnamefont {Majorana}},\ }\bibfield  {title} {\emph {\bibinfo {title} {Atomi orientati in campo magnetico variabile},\ }}\href {\doibase 10.1007/bf02960953} {\bibfield  {journal} {\bibinfo  {journal} {Il Nuovo Cimento}\ }\textbf {\bibinfo {volume} {9}},\ \bibinfo {pages} {43} (\bibinfo {year} {1932})}\BibitemShut {NoStop}%
\bibitem [{\citenamefont {Ivakhnenko}\ \emph {et~al.}(2023)\citenamefont {Ivakhnenko}, \citenamefont {Shevchenko},\ and\ \citenamefont {Nori}}]{Ivakhnenko2023}%
  \BibitemOpen
  \bibfield  {author} {\bibinfo {author} {\bibfnamefont {O.~V.}\ \bibnamefont {Ivakhnenko}}, \bibinfo {author} {\bibfnamefont {S.~N.}\ \bibnamefont {Shevchenko}}, \ and\ \bibinfo {author} {\bibfnamefont {F.}~\bibnamefont {Nori}},\ }\bibfield  {title} {\emph {\bibinfo {title} {Nonadiabatic {L}andau-{Z}ener-{S}t{\"u}ckelberg-{M}ajorana transitions, dynamics, and interference},\ }}\href {\doibase https://doi.org/10.1016/j.physrep.2022.10.002} {\bibfield  {journal} {\bibinfo  {journal} {Physics Reports}\ }\textbf {\bibinfo {volume} {995}},\ \bibinfo {pages} {1} (\bibinfo {year} {2023})}\BibitemShut {NoStop}%
\bibitem [{\citenamefont {Svensson}\ \emph {et~al.}(1996)\citenamefont {Svensson}, \citenamefont {Humbel}, \citenamefont {Froese}, \citenamefont {Matsubara}, \citenamefont {Sieber},\ and\ \citenamefont {Morokuma}}]{svensson1996}%
  \BibitemOpen
  \bibfield  {author} {\bibinfo {author} {\bibfnamefont {M.}~\bibnamefont {Svensson}}, \bibinfo {author} {\bibfnamefont {S.}~\bibnamefont {Humbel}}, \bibinfo {author} {\bibfnamefont {R.~D.~J.}\ \bibnamefont {Froese}}, \bibinfo {author} {\bibfnamefont {T.}~\bibnamefont {Matsubara}}, \bibinfo {author} {\bibfnamefont {S.}~\bibnamefont {Sieber}}, \ and\ \bibinfo {author} {\bibfnamefont {K.}~\bibnamefont {Morokuma}},\ }\bibfield  {title} {\emph {\bibinfo {title} {{ONIOM}:~ a multilayered integrated {MO} + {MM} method for geometry optimizations and single point energy predictions. {A} test for diels-alder reactions and {P}t({P}(t-{B}u)$_3$)$_2$ + {H}$_2$ oxidative addition},\ }}\href {\doibase 10.1021/jp962071j} {\bibfield  {journal} {\bibinfo  {journal} {The Journal of Physical Chemistry}\ }\textbf {\bibinfo {volume} {100}},\ \bibinfo {pages} {19357} (\bibinfo {year} {1996})}\BibitemShut {NoStop}%
\bibitem [{\citenamefont {Perdew}\ \emph {et~al.}(1996)\citenamefont {Perdew}, \citenamefont {Burke},\ and\ \citenamefont {Ernzerhof}}]{perdew1996}%
  \BibitemOpen
  \bibfield  {author} {\bibinfo {author} {\bibfnamefont {J.~P.}\ \bibnamefont {Perdew}}, \bibinfo {author} {\bibfnamefont {K.}~\bibnamefont {Burke}}, \ and\ \bibinfo {author} {\bibfnamefont {M.}~\bibnamefont {Ernzerhof}},\ }\bibfield  {title} {\emph {\bibinfo {title} {Generalized gradient approximation made simple},\ }}\href {\doibase 10.1103/physrevlett.77.3865} {\bibfield  {journal} {\bibinfo  {journal} {Physical Review Letters}\ }\textbf {\bibinfo {volume} {77}},\ \bibinfo {pages} {3865} (\bibinfo {year} {1996})}\BibitemShut {NoStop}%
\bibitem [{\citenamefont {Andrae}\ \emph {et~al.}(1990)\citenamefont {Andrae}, \citenamefont {H{\"a}u{\ss}ermann}, \citenamefont {Dolg}, \citenamefont {Stoll},\ and\ \citenamefont {Preu{\ss}}}]{andrae1990}%
  \BibitemOpen
  \bibfield  {author} {\bibinfo {author} {\bibfnamefont {D.}~\bibnamefont {Andrae}}, \bibinfo {author} {\bibfnamefont {U.}~\bibnamefont {H{\"a}u{\ss}ermann}}, \bibinfo {author} {\bibfnamefont {M.}~\bibnamefont {Dolg}}, \bibinfo {author} {\bibfnamefont {H.}~\bibnamefont {Stoll}}, \ and\ \bibinfo {author} {\bibfnamefont {H.}~\bibnamefont {Preu{\ss}}},\ }\bibfield  {title} {\emph {\bibinfo {title} {Energy-adjusted ab initio pseudopotentials for the second and third row transition elements},\ }}\href {\doibase 10.1007/bf01114537} {\bibfield  {journal} {\bibinfo  {journal} {Theor. Chim. Acta}\ }\textbf {\bibinfo {volume} {77}},\ \bibinfo {pages} {123} (\bibinfo {year} {1990})}\BibitemShut {NoStop}%
\bibitem [{\citenamefont {Dunning}(1989)}]{dunning1989}%
  \BibitemOpen
  \bibfield  {author} {\bibinfo {author} {\bibfnamefont {T.~H.}\ \bibnamefont {Dunning}},\ }\bibfield  {title} {\emph {\bibinfo {title} {Gaussian basis sets for use in correlated molecular calculations. {I}. {T}he atoms boron through neon and hydrogen},\ }}\href {\doibase 10.1063/1.456153} {\bibfield  {journal} {\bibinfo  {journal} {The Journal of Chemical Physics}\ }\textbf {\bibinfo {volume} {90}},\ \bibinfo {pages} {1007} (\bibinfo {year} {1989})}\BibitemShut {NoStop}%
\bibitem [{\citenamefont {Frisch}\ \emph {et~al.}(2009)\citenamefont {Frisch}, \citenamefont {Trucks}, \citenamefont {Schlegel}, \citenamefont {Scuseria}, \citenamefont {Robb}, \citenamefont {Cheeseman}, \citenamefont {Scalmani}, \citenamefont {Barone}, \citenamefont {Mennucci}, \citenamefont {Petersson}, \citenamefont {Nakatsuji}, \citenamefont {Caricato}, \citenamefont {Li}, \citenamefont {Hratchian}, \citenamefont {Izmaylov}, \citenamefont {Bloino}, \citenamefont {Zheng}, \citenamefont {Sonnenberg}, \citenamefont {Hada}, \citenamefont {Ehara}, \citenamefont {Toyota}, \citenamefont {Fukuda}, \citenamefont {Hasegawa}, \citenamefont {Ishida}, \citenamefont {Nakajima}, \citenamefont {Honda}, \citenamefont {Kitao}, \citenamefont {Nakai}, \citenamefont {Vreven}, \citenamefont {Montgomery}, \citenamefont {Jr.}, \citenamefont {Peralta}, \citenamefont {Ogliaro}, \citenamefont {Bearpark}, \citenamefont {Heyd}, \citenamefont {Brothers}, \citenamefont {Kudin}, \citenamefont {Staroverov}, \citenamefont {Kobayashi},
  \citenamefont {Normand}, \citenamefont {Raghavachari}, \citenamefont {Rendell}, \citenamefont {Burant}, \citenamefont {Iyengar}, \citenamefont {Tomasi}, \citenamefont {Cossi}, \citenamefont {Rega}, \citenamefont {Millam}, \citenamefont {Klene}, \citenamefont {Knox}, \citenamefont {Cross}, \citenamefont {Bakken}, \citenamefont {Adamo}, \citenamefont {Jaramillo}, \citenamefont {Gomperts}, \citenamefont {Stratmann}, \citenamefont {Yazyev}, \citenamefont {Austin}, \citenamefont {Cammi}, \citenamefont {Pomelli}, \citenamefont {Ochterski}, \citenamefont {Martin}, \citenamefont {Morokuma}, \citenamefont {Zakrzewski}, \citenamefont {Voth}, \citenamefont {Salvador}, \citenamefont {Dannenberg}, \citenamefont {Dapprich}, \citenamefont {Daniels}, \citenamefont {Farkas}, \citenamefont {Foresman}, \citenamefont {Ortiz}, \citenamefont {Cioslowski},\ and\ \citenamefont {Fox}}]{g09}%
  \BibitemOpen
  \bibfield  {author} {\bibinfo {author} {\bibfnamefont {M.~J.}\ \bibnamefont {Frisch}}, \bibinfo {author} {\bibfnamefont {G.~W.}\ \bibnamefont {Trucks}}, \bibinfo {author} {\bibfnamefont {H.~B.}\ \bibnamefont {Schlegel}}, \bibinfo {author} {\bibfnamefont {G.~E.}\ \bibnamefont {Scuseria}}, \bibinfo {author} {\bibfnamefont {M.~A.}\ \bibnamefont {Robb}}, \bibinfo {author} {\bibfnamefont {J.~R.}\ \bibnamefont {Cheeseman}}, \bibinfo {author} {\bibfnamefont {G.}~\bibnamefont {Scalmani}}, \bibinfo {author} {\bibfnamefont {V.}~\bibnamefont {Barone}}, \bibinfo {author} {\bibfnamefont {B.}~\bibnamefont {Mennucci}}, \bibinfo {author} {\bibfnamefont {G.~A.}\ \bibnamefont {Petersson}}, \bibinfo {author} {\bibfnamefont {H.}~\bibnamefont {Nakatsuji}}, \bibinfo {author} {\bibfnamefont {M.}~\bibnamefont {Caricato}}, \bibinfo {author} {\bibfnamefont {X.}~\bibnamefont {Li}}, \bibinfo {author} {\bibfnamefont {H.~P.}\ \bibnamefont {Hratchian}}, \bibinfo {author} {\bibfnamefont {A.~F.}\ \bibnamefont {Izmaylov}}, \bibinfo {author}
  {\bibfnamefont {J.}~\bibnamefont {Bloino}}, \bibinfo {author} {\bibfnamefont {G.}~\bibnamefont {Zheng}}, \bibinfo {author} {\bibfnamefont {J.~L.}\ \bibnamefont {Sonnenberg}}, \bibinfo {author} {\bibfnamefont {M.}~\bibnamefont {Hada}}, \bibinfo {author} {\bibfnamefont {M.}~\bibnamefont {Ehara}}, \bibinfo {author} {\bibfnamefont {K.}~\bibnamefont {Toyota}}, \bibinfo {author} {\bibfnamefont {R.}~\bibnamefont {Fukuda}}, \bibinfo {author} {\bibfnamefont {J.}~\bibnamefont {Hasegawa}}, \bibinfo {author} {\bibfnamefont {M.}~\bibnamefont {Ishida}}, \bibinfo {author} {\bibfnamefont {T.}~\bibnamefont {Nakajima}}, \bibinfo {author} {\bibfnamefont {Y.}~\bibnamefont {Honda}}, \bibinfo {author} {\bibfnamefont {O.}~\bibnamefont {Kitao}}, \bibinfo {author} {\bibfnamefont {H.}~\bibnamefont {Nakai}}, \bibinfo {author} {\bibfnamefont {T.}~\bibnamefont {Vreven}}, \bibinfo {author} {\bibfnamefont {J.~A.}\ \bibnamefont {Montgomery}}, \bibinfo {author} {\bibnamefont {Jr.}}, \bibinfo {author} {\bibfnamefont {J.~E.}\ \bibnamefont
  {Peralta}}, \bibinfo {author} {\bibfnamefont {F.}~\bibnamefont {Ogliaro}}, \bibinfo {author} {\bibfnamefont {M.}~\bibnamefont {Bearpark}}, \bibinfo {author} {\bibfnamefont {J.~J.}\ \bibnamefont {Heyd}}, \bibinfo {author} {\bibfnamefont {E.}~\bibnamefont {Brothers}}, \bibinfo {author} {\bibfnamefont {K.~N.}\ \bibnamefont {Kudin}}, \bibinfo {author} {\bibfnamefont {V.~N.}\ \bibnamefont {Staroverov}}, \bibinfo {author} {\bibfnamefont {R.}~\bibnamefont {Kobayashi}}, \bibinfo {author} {\bibfnamefont {J.}~\bibnamefont {Normand}}, \bibinfo {author} {\bibfnamefont {K.}~\bibnamefont {Raghavachari}}, \bibinfo {author} {\bibfnamefont {A.}~\bibnamefont {Rendell}}, \bibinfo {author} {\bibfnamefont {J.~C.}\ \bibnamefont {Burant}}, \bibinfo {author} {\bibfnamefont {S.~S.}\ \bibnamefont {Iyengar}}, \bibinfo {author} {\bibfnamefont {J.}~\bibnamefont {Tomasi}}, \bibinfo {author} {\bibfnamefont {M.}~\bibnamefont {Cossi}}, \bibinfo {author} {\bibfnamefont {N.}~\bibnamefont {Rega}}, \bibinfo {author} {\bibfnamefont {J.~M.}\
  \bibnamefont {Millam}}, \bibinfo {author} {\bibfnamefont {M.}~\bibnamefont {Klene}}, \bibinfo {author} {\bibfnamefont {J.~E.}\ \bibnamefont {Knox}}, \bibinfo {author} {\bibfnamefont {J.~B.}\ \bibnamefont {Cross}}, \bibinfo {author} {\bibfnamefont {V.}~\bibnamefont {Bakken}}, \bibinfo {author} {\bibfnamefont {C.}~\bibnamefont {Adamo}}, \bibinfo {author} {\bibfnamefont {J.}~\bibnamefont {Jaramillo}}, \bibinfo {author} {\bibfnamefont {R.}~\bibnamefont {Gomperts}}, \bibinfo {author} {\bibfnamefont {R.~E.}\ \bibnamefont {Stratmann}}, \bibinfo {author} {\bibfnamefont {O.}~\bibnamefont {Yazyev}}, \bibinfo {author} {\bibfnamefont {A.~J.}\ \bibnamefont {Austin}}, \bibinfo {author} {\bibfnamefont {R.}~\bibnamefont {Cammi}}, \bibinfo {author} {\bibfnamefont {C.}~\bibnamefont {Pomelli}}, \bibinfo {author} {\bibfnamefont {J.~W.}\ \bibnamefont {Ochterski}}, \bibinfo {author} {\bibfnamefont {R.~L.}\ \bibnamefont {Martin}}, \bibinfo {author} {\bibfnamefont {K.}~\bibnamefont {Morokuma}}, \bibinfo {author} {\bibfnamefont
  {V.~G.}\ \bibnamefont {Zakrzewski}}, \bibinfo {author} {\bibfnamefont {G.~A.}\ \bibnamefont {Voth}}, \bibinfo {author} {\bibfnamefont {P.}~\bibnamefont {Salvador}}, \bibinfo {author} {\bibfnamefont {J.~J.}\ \bibnamefont {Dannenberg}}, \bibinfo {author} {\bibfnamefont {S.}~\bibnamefont {Dapprich}}, \bibinfo {author} {\bibfnamefont {A.~D.}\ \bibnamefont {Daniels}}, \bibinfo {author} {\bibfnamefont {O.}~\bibnamefont {Farkas}}, \bibinfo {author} {\bibfnamefont {J.~B.}\ \bibnamefont {Foresman}}, \bibinfo {author} {\bibfnamefont {J.~V.}\ \bibnamefont {Ortiz}}, \bibinfo {author} {\bibfnamefont {J.}~\bibnamefont {Cioslowski}}, \ and\ \bibinfo {author} {\bibfnamefont {D.~J.}\ \bibnamefont {Fox}},\ }\href@noop {} {\bibinfo {title} {Gaussian~09 {R}evision {D}.01},\ } (\bibinfo {year} {2009}),\ \bibinfo {note} {{G}aussian Inc. Wallingford CT}\BibitemShut {NoStop}%
\bibitem [{\citenamefont {Fdez.~Galv{\'a}n}\ \emph {et~al.}(2019)\citenamefont {Fdez.~Galv{\'a}n}, \citenamefont {Vacher}, \citenamefont {Alavi}, \citenamefont {Angeli}, \citenamefont {Aquilante}, \citenamefont {Autschbach}, \citenamefont {Bao}, \citenamefont {Bokarev}, \citenamefont {Bogdanov}, \citenamefont {Carlson}, \citenamefont {Chibotaru}, \citenamefont {Creutzberg}, \citenamefont {Dattani}, \citenamefont {Delcey}, \citenamefont {Dong}, \citenamefont {Dreuw}, \citenamefont {Freitag}, \citenamefont {Frutos}, \citenamefont {Gagliardi}, \citenamefont {Gendron}, \citenamefont {Giussani}, \citenamefont {Gonz{\'a}lez}, \citenamefont {Grell}, \citenamefont {Guo}, \citenamefont {Hoyer}, \citenamefont {Johansson}, \citenamefont {Keller}, \citenamefont {Knecht}, \citenamefont {Kova{\v{c}}evi{\'{c}}}, \citenamefont {K{\"a}llman}, \citenamefont {Li~Manni}, \citenamefont {Lundberg}, \citenamefont {Ma}, \citenamefont {Mai}, \citenamefont {Malhado}, \citenamefont {Malmqvist}, \citenamefont {Marquetand}, \citenamefont
  {Mewes}, \citenamefont {Norell}, \citenamefont {Olivucci}, \citenamefont {Oppel}, \citenamefont {Phung}, \citenamefont {Pierloot}, \citenamefont {Plasser}, \citenamefont {Reiher}, \citenamefont {Sand}, \citenamefont {Schapiro}, \citenamefont {Sharma}, \citenamefont {Stein}, \citenamefont {S{\o}rensen}, \citenamefont {Truhlar}, \citenamefont {Ugandi}, \citenamefont {Ungur}, \citenamefont {Valentini}, \citenamefont {Vancoillie}, \citenamefont {Veryazov}, \citenamefont {Weser}, \citenamefont {Weso{\l}owski}, \citenamefont {Widmark}, \citenamefont {Wouters}, \citenamefont {Zech}, \citenamefont {Zobel},\ and\ \citenamefont {Lindh}}]{omolcas2019}%
  \BibitemOpen
  \bibfield  {author} {\bibinfo {author} {\bibfnamefont {I.}~\bibnamefont {Fdez.~Galv{\'a}n}}, \bibinfo {author} {\bibfnamefont {M.}~\bibnamefont {Vacher}}, \bibinfo {author} {\bibfnamefont {A.}~\bibnamefont {Alavi}}, \bibinfo {author} {\bibfnamefont {C.}~\bibnamefont {Angeli}}, \bibinfo {author} {\bibfnamefont {F.}~\bibnamefont {Aquilante}}, \bibinfo {author} {\bibfnamefont {J.}~\bibnamefont {Autschbach}}, \bibinfo {author} {\bibfnamefont {J.~J.}\ \bibnamefont {Bao}}, \bibinfo {author} {\bibfnamefont {S.~I.}\ \bibnamefont {Bokarev}}, \bibinfo {author} {\bibfnamefont {N.~A.}\ \bibnamefont {Bogdanov}}, \bibinfo {author} {\bibfnamefont {R.~K.}\ \bibnamefont {Carlson}}, \bibinfo {author} {\bibfnamefont {L.~F.}\ \bibnamefont {Chibotaru}}, \bibinfo {author} {\bibfnamefont {J.}~\bibnamefont {Creutzberg}}, \bibinfo {author} {\bibfnamefont {N.}~\bibnamefont {Dattani}}, \bibinfo {author} {\bibfnamefont {M.~G.}\ \bibnamefont {Delcey}}, \bibinfo {author} {\bibfnamefont {S.~S.}\ \bibnamefont {Dong}}, \bibinfo {author}
  {\bibfnamefont {A.}~\bibnamefont {Dreuw}}, \bibinfo {author} {\bibfnamefont {L.}~\bibnamefont {Freitag}}, \bibinfo {author} {\bibfnamefont {L.~M.}\ \bibnamefont {Frutos}}, \bibinfo {author} {\bibfnamefont {L.}~\bibnamefont {Gagliardi}}, \bibinfo {author} {\bibfnamefont {F.}~\bibnamefont {Gendron}}, \bibinfo {author} {\bibfnamefont {A.}~\bibnamefont {Giussani}}, \bibinfo {author} {\bibfnamefont {L.}~\bibnamefont {Gonz{\'a}lez}}, \bibinfo {author} {\bibfnamefont {G.}~\bibnamefont {Grell}}, \bibinfo {author} {\bibfnamefont {M.}~\bibnamefont {Guo}}, \bibinfo {author} {\bibfnamefont {C.~E.}\ \bibnamefont {Hoyer}}, \bibinfo {author} {\bibfnamefont {M.}~\bibnamefont {Johansson}}, \bibinfo {author} {\bibfnamefont {S.}~\bibnamefont {Keller}}, \bibinfo {author} {\bibfnamefont {S.}~\bibnamefont {Knecht}}, \bibinfo {author} {\bibfnamefont {G.}~\bibnamefont {Kova{\v{c}}evi{\'{c}}}}, \bibinfo {author} {\bibfnamefont {E.}~\bibnamefont {K{\"a}llman}}, \bibinfo {author} {\bibfnamefont {G.}~\bibnamefont {Li~Manni}}, \bibinfo
  {author} {\bibfnamefont {M.}~\bibnamefont {Lundberg}}, \bibinfo {author} {\bibfnamefont {Y.}~\bibnamefont {Ma}}, \bibinfo {author} {\bibfnamefont {S.}~\bibnamefont {Mai}}, \bibinfo {author} {\bibfnamefont {J.~P.}\ \bibnamefont {Malhado}}, \bibinfo {author} {\bibfnamefont {P.~{\AA}.}\ \bibnamefont {Malmqvist}}, \bibinfo {author} {\bibfnamefont {P.}~\bibnamefont {Marquetand}}, \bibinfo {author} {\bibfnamefont {S.~A.}\ \bibnamefont {Mewes}}, \bibinfo {author} {\bibfnamefont {J.}~\bibnamefont {Norell}}, \bibinfo {author} {\bibfnamefont {M.}~\bibnamefont {Olivucci}}, \bibinfo {author} {\bibfnamefont {M.}~\bibnamefont {Oppel}}, \bibinfo {author} {\bibfnamefont {Q.~M.}\ \bibnamefont {Phung}}, \bibinfo {author} {\bibfnamefont {K.}~\bibnamefont {Pierloot}}, \bibinfo {author} {\bibfnamefont {F.}~\bibnamefont {Plasser}}, \bibinfo {author} {\bibfnamefont {M.}~\bibnamefont {Reiher}}, \bibinfo {author} {\bibfnamefont {A.~M.}\ \bibnamefont {Sand}}, \bibinfo {author} {\bibfnamefont {I.}~\bibnamefont {Schapiro}}, \bibinfo
  {author} {\bibfnamefont {P.}~\bibnamefont {Sharma}}, \bibinfo {author} {\bibfnamefont {C.~J.}\ \bibnamefont {Stein}}, \bibinfo {author} {\bibfnamefont {L.~K.}\ \bibnamefont {S{\o}rensen}}, \bibinfo {author} {\bibfnamefont {D.~G.}\ \bibnamefont {Truhlar}}, \bibinfo {author} {\bibfnamefont {M.}~\bibnamefont {Ugandi}}, \bibinfo {author} {\bibfnamefont {L.}~\bibnamefont {Ungur}}, \bibinfo {author} {\bibfnamefont {A.}~\bibnamefont {Valentini}}, \bibinfo {author} {\bibfnamefont {S.}~\bibnamefont {Vancoillie}}, \bibinfo {author} {\bibfnamefont {V.}~\bibnamefont {Veryazov}}, \bibinfo {author} {\bibfnamefont {O.}~\bibnamefont {Weser}}, \bibinfo {author} {\bibfnamefont {T.~A.}\ \bibnamefont {Weso{\l}owski}}, \bibinfo {author} {\bibfnamefont {P.-O.}\ \bibnamefont {Widmark}}, \bibinfo {author} {\bibfnamefont {S.}~\bibnamefont {Wouters}}, \bibinfo {author} {\bibfnamefont {A.}~\bibnamefont {Zech}}, \bibinfo {author} {\bibfnamefont {J.~P.}\ \bibnamefont {Zobel}}, \ and\ \bibinfo {author} {\bibfnamefont {R.}~\bibnamefont
  {Lindh}},\ }\bibfield  {title} {\emph {\bibinfo {title} {Open{M}olcas: From source code to insight},\ }}\href {\doibase 10.1021/acs.jctc.9b00532} {\bibfield  {journal} {\bibinfo  {journal} {Journal of Chemical Theory and Computation}\ }\textbf {\bibinfo {volume} {15}},\ \bibinfo {pages} {5925} (\bibinfo {year} {2019})}\BibitemShut {NoStop}%
\bibitem [{\citenamefont {Breneman}\ and\ \citenamefont {Wiberg}(1990)}]{breneman1990}%
  \BibitemOpen
  \bibfield  {author} {\bibinfo {author} {\bibfnamefont {C.~M.}\ \bibnamefont {Breneman}}\ and\ \bibinfo {author} {\bibfnamefont {K.~B.}\ \bibnamefont {Wiberg}},\ }\bibfield  {title} {\emph {\bibinfo {title} {Determining atom-centered monopoles from molecular electrostatic potentials. {T}he need for high sampling density in formamide conformational analysis},\ }}\href {\doibase 10.1002/jcc.540110311} {\bibfield  {journal} {\bibinfo  {journal} {Journal of Computational Chemistry}\ }\textbf {\bibinfo {volume} {11}},\ \bibinfo {pages} {361} (\bibinfo {year} {1990})}\BibitemShut {NoStop}%
\bibitem [{\citenamefont {Malmqvist}\ and\ \citenamefont {Roos}(1989)}]{malmqvist1989}%
  \BibitemOpen
  \bibfield  {author} {\bibinfo {author} {\bibfnamefont {P.-{\AA}.}\ \bibnamefont {Malmqvist}}\ and\ \bibinfo {author} {\bibfnamefont {B.~O.}\ \bibnamefont {Roos}},\ }\bibfield  {title} {\emph {\bibinfo {title} {The {CASSCF} state interaction method},\ }}\href {\doibase 10.1016/0009-2614(89)85347-3} {\bibfield  {journal} {\bibinfo  {journal} {Chemical Physics Letters}\ }\textbf {\bibinfo {volume} {155}},\ \bibinfo {pages} {189} (\bibinfo {year} {1989})}\BibitemShut {NoStop}%
\bibitem [{\citenamefont {Malmqvist}\ \emph {et~al.}(2002)\citenamefont {Malmqvist}, \citenamefont {Roos},\ and\ \citenamefont {Schimmelpfennig}}]{malmqvist2002}%
  \BibitemOpen
  \bibfield  {author} {\bibinfo {author} {\bibfnamefont {P.-{\AA}.}\ \bibnamefont {Malmqvist}}, \bibinfo {author} {\bibfnamefont {B.~O.}\ \bibnamefont {Roos}}, \ and\ \bibinfo {author} {\bibfnamefont {B.}~\bibnamefont {Schimmelpfennig}},\ }\bibfield  {title} {\emph {\bibinfo {title} {The restricted active space ({RAS}) state interaction approach with spin{\textendash}orbit coupling},\ }}\href {\doibase 10.1016/s0009-2614(02)00498-0} {\bibfield  {journal} {\bibinfo  {journal} {Chemical Physics Letters}\ }\textbf {\bibinfo {volume} {357}},\ \bibinfo {pages} {230} (\bibinfo {year} {2002})}\BibitemShut {NoStop}%
\bibitem [{\citenamefont {Widmark}\ \emph {et~al.}(1990)\citenamefont {Widmark}, \citenamefont {Malmqvist},\ and\ \citenamefont {Roos}}]{widmark1990}%
  \BibitemOpen
  \bibfield  {author} {\bibinfo {author} {\bibfnamefont {P.-O.}\ \bibnamefont {Widmark}}, \bibinfo {author} {\bibfnamefont {P.-{\AA}.}\ \bibnamefont {Malmqvist}}, \ and\ \bibinfo {author} {\bibfnamefont {B.~O.}\ \bibnamefont {Roos}},\ }\bibfield  {title} {\emph {\bibinfo {title} {Density matrix averaged atomic natural orbital ({ANO}) basis sets for correlated molecular wave functions},\ }}\href {\doibase 10.1007/bf01120130} {\bibfield  {journal} {\bibinfo  {journal} {Theoretica Chimica Acta}\ }\textbf {\bibinfo {volume} {77}},\ \bibinfo {pages} {291} (\bibinfo {year} {1990})}\BibitemShut {NoStop}%
\bibitem [{\citenamefont {Aquilante}\ \emph {et~al.}(2007)\citenamefont {Aquilante}, \citenamefont {Lindh},\ and\ \citenamefont {Pedersen}}]{aquilante2007}%
  \BibitemOpen
  \bibfield  {author} {\bibinfo {author} {\bibfnamefont {F.}~\bibnamefont {Aquilante}}, \bibinfo {author} {\bibfnamefont {R.}~\bibnamefont {Lindh}}, \ and\ \bibinfo {author} {\bibfnamefont {T.~B.}\ \bibnamefont {Pedersen}},\ }\bibfield  {title} {\emph {\bibinfo {title} {Unbiased auxiliary basis sets for accurate two-electron integral approximations},\ }}\href {\doibase 10.1063/1.2777146} {\bibfield  {journal} {\bibinfo  {journal} {The Journal of Chemical Physics}\ }\textbf {\bibinfo {volume} {127}},\ \bibinfo {pages} {114107} (\bibinfo {year} {2007})}\BibitemShut {NoStop}%
\bibitem [{dat()}]{data}%
  \BibitemOpen
  \href@noop {} {}\bibinfo {note} {A. Mattioni, J. K. Staab,
  W. J. A. Blackmore,
  D. Reta, J. Iles-Smith, A. Nazir, and N. F. Chilton, {\it Vibronic effects on the quantum tunnelling of magnetisation in {K}ramers single-molecule magnets}, University of Manchester Figshare, doi.org/10.48420/21892887.v1 (2023)}\BibitemShut {NoStop}%
\end{thebibliography}

\begin{thebibliography}{12}%
\makeatletter
\providecommand \@ifxundefined [1]{%
 \@ifx{#1\undefined}
}%
\providecommand \@ifnum [1]{%
 \ifnum #1\expandafter \@firstoftwo
 \else \expandafter \@secondoftwo
 \fi
}%
\providecommand \@ifx [1]{%
 \ifx #1\expandafter \@firstoftwo
 \else \expandafter \@secondoftwo
 \fi
}%
\providecommand \natexlab [1]{#1}%
\providecommand \enquote  [1]{``#1''}%
\providecommand \bibnamefont  [1]{#1}%
\providecommand \bibfnamefont [1]{#1}%
\providecommand \citenamefont [1]{#1}%
\providecommand \href@noop [0]{\@secondoftwo}%
\providecommand \href [0]{\begingroup \@sanitize@url \@href}%
\providecommand \@href[1]{\@@startlink{#1}\@@href}%
\providecommand \@@href[1]{\endgroup#1\@@endlink}%
\providecommand \@sanitize@url [0]{\catcode `\\12\catcode `\$12\catcode `\&12\catcode `\#12\catcode `\^12\catcode `\_12\catcode `\%12\relax}%
\providecommand \@@startlink[1]{}%
\providecommand \@@endlink[0]{}%
\providecommand \url  [0]{\begingroup\@sanitize@url \@url }%
\providecommand \@url [1]{\endgroup\@href {#1}{\urlprefix }}%
\providecommand \urlprefix  [0]{URL }%
\providecommand \Eprint [0]{\href }%
\providecommand \doibase [0]{http://dx.doi.org/}%
\providecommand \selectlanguage [0]{\@gobble}%
\providecommand \bibinfo  [0]{\@secondoftwo}%
\providecommand \bibfield  [0]{\@secondoftwo}%
\providecommand \translation [1]{[#1]}%
\providecommand \BibitemOpen [0]{}%
\providecommand \bibitemStop [0]{}%
\providecommand \bibitemNoStop [0]{.\EOS\space}%
\providecommand \EOS [0]{\spacefactor3000\relax}%
\providecommand \BibitemShut  [1]{\csname bibitem#1\endcsname}%
\let\auto@bib@innerbib\@empty
\bibitem [{\citenamefont {Kragskow}\ \emph {et~al.}(2023)\citenamefont {Kragskow}, \citenamefont {Mattioni}, \citenamefont {Staab}, \citenamefont {Reta}, \citenamefont {Skelton},\ and\ \citenamefont {Chilton}}]{kragskow2023}%
  \BibitemOpen
  \bibfield  {author} {\bibinfo {author} {\bibfnamefont {J.~G.~C.}\ \bibnamefont {Kragskow}}, \bibinfo {author} {\bibfnamefont {A.}~\bibnamefont {Mattioni}}, \bibinfo {author} {\bibfnamefont {J.~K.}\ \bibnamefont {Staab}}, \bibinfo {author} {\bibfnamefont {D.}~\bibnamefont {Reta}}, \bibinfo {author} {\bibfnamefont {J.~M.}\ \bibnamefont {Skelton}}, \ and\ \bibinfo {author} {\bibfnamefont {N.~F.}\ \bibnamefont {Chilton}},\ }\bibfield  {title} {\emph {\bibinfo {title} {Spin-phonon coupling and magnetic relaxation in single-molecule magnets},\ }}\href {\doibase 10.1039/D2CS00705C} {\bibfield  {journal} {\bibinfo  {journal} {Chem. Soc. Rev.}\ }\textbf {\bibinfo {volume} {52}},\ \bibinfo {pages} {4567} (\bibinfo {year} {2023})}\BibitemShut {NoStop}%
\bibitem [{\citenamefont {Chibotaru}\ \emph {et~al.}(2008)\citenamefont {Chibotaru}, \citenamefont {Ceulemans},\ and\ \citenamefont {Bolvin}}]{Chibotaru2008}%
  \BibitemOpen
  \bibfield  {author} {\bibinfo {author} {\bibfnamefont {L.~F.}\ \bibnamefont {Chibotaru}}, \bibinfo {author} {\bibfnamefont {A.}~\bibnamefont {Ceulemans}}, \ and\ \bibinfo {author} {\bibfnamefont {H.}~\bibnamefont {Bolvin}},\ }\bibfield  {title} {\emph {\bibinfo {title} {Unique definition of the {Z}eeman-splitting $g$ tensor of a {K}ramers doublet},\ }}\href {\doibase 10.1103/PhysRevLett.101.033003} {\bibfield  {journal} {\bibinfo  {journal} {Phys. Rev. Lett.}\ }\textbf {\bibinfo {volume} {101}},\ \bibinfo {pages} {033003} (\bibinfo {year} {2008})}\BibitemShut {NoStop}%
\bibitem [{\citenamefont {Landau}(1932{\natexlab{a}})}]{Landau1932i}%
  \BibitemOpen
  \bibfield  {author} {\bibinfo {author} {\bibfnamefont {L.~D.}\ \bibnamefont {Landau}},\ }\bibfield  {title} {\emph {\bibinfo {title} {Zur {T}heorie der {E}nergie{\"u}bertragung},\ }}\href@noop {} {\bibfield  {journal} {\bibinfo  {journal} {Phyz. Z. Sowjetunion}\ }\textbf {\bibinfo {volume} {1}},\ \bibinfo {pages} {88} (\bibinfo {year} {1932}{\natexlab{a}})}\BibitemShut {NoStop}%
\bibitem [{\citenamefont {Landau}(1932{\natexlab{b}})}]{Landau1932ii}%
  \BibitemOpen
  \bibfield  {author} {\bibinfo {author} {\bibfnamefont {L.~D.}\ \bibnamefont {Landau}},\ }\bibfield  {title} {\emph {\bibinfo {title} {Zur {T}heorie der {E}nergie{\"u}bertragung {II}},\ }}\href@noop {} {\bibfield  {journal} {\bibinfo  {journal} {Phyz. Z. Sowjetunion}\ }\textbf {\bibinfo {volume} {2}},\ \bibinfo {pages} {46} (\bibinfo {year} {1932}{\natexlab{b}})}\BibitemShut {NoStop}%
\bibitem [{\citenamefont {Zener}\ and\ \citenamefont {Fowler}(1932)}]{Zener1932}%
  \BibitemOpen
  \bibfield  {author} {\bibinfo {author} {\bibfnamefont {C.}~\bibnamefont {Zener}}\ and\ \bibinfo {author} {\bibfnamefont {R.~H.}\ \bibnamefont {Fowler}},\ }\bibfield  {title} {\emph {\bibinfo {title} {Non-adiabatic crossing of energy levels},\ }}\href {\doibase 10.1098/rspa.1932.0165} {\bibfield  {journal} {\bibinfo  {journal} {Proceedings of the Royal Society of London. Series A}\ }\textbf {\bibinfo {volume} {137}},\ \bibinfo {pages} {696} (\bibinfo {year} {1932})}\BibitemShut {NoStop}%
\bibitem [{\citenamefont {St{\"u}ckelberg}(1932)}]{Stueckelberg1932}%
  \BibitemOpen
  \bibfield  {author} {\bibinfo {author} {\bibfnamefont {E.~C.~G.}\ \bibnamefont {St{\"u}ckelberg}},\ }\bibfield  {title} {\emph {\bibinfo {title} {{T}heorie der unelastischen {S}t{\"o}sse zwischen {A}tomen},\ }}\href@noop {} {\bibfield  {journal} {\bibinfo  {journal} {Helv. Phys. Acta}\ }\textbf {\bibinfo {volume} {5}},\ \bibinfo {pages} {369} (\bibinfo {year} {1932})}\BibitemShut {NoStop}%
\bibitem [{\citenamefont {Majorana}(1932)}]{Majorana1932}%
  \BibitemOpen
  \bibfield  {author} {\bibinfo {author} {\bibfnamefont {E.}~\bibnamefont {Majorana}},\ }\bibfield  {title} {\emph {\bibinfo {title} {Atomi orientati in campo magnetico variabile},\ }}\href {\doibase 10.1007/bf02960953} {\bibfield  {journal} {\bibinfo  {journal} {Il Nuovo Cimento}\ }\textbf {\bibinfo {volume} {9}},\ \bibinfo {pages} {43} (\bibinfo {year} {1932})}\BibitemShut {NoStop}%
\bibitem [{\citenamefont {Ivakhnenko}\ \emph {et~al.}(2023)\citenamefont {Ivakhnenko}, \citenamefont {Shevchenko},\ and\ \citenamefont {Nori}}]{Ivakhnenko2023}%
  \BibitemOpen
  \bibfield  {author} {\bibinfo {author} {\bibfnamefont {O.~V.}\ \bibnamefont {Ivakhnenko}}, \bibinfo {author} {\bibfnamefont {S.~N.}\ \bibnamefont {Shevchenko}}, \ and\ \bibinfo {author} {\bibfnamefont {F.}~\bibnamefont {Nori}},\ }\bibfield  {title} {\emph {\bibinfo {title} {Nonadiabatic {L}andau-{Z}ener-{S}t{\"u}ckelberg-{M}ajorana transitions, dynamics, and interference},\ }}\href {\doibase https://doi.org/10.1016/j.physrep.2022.10.002} {\bibfield  {journal} {\bibinfo  {journal} {Physics Reports}\ }\textbf {\bibinfo {volume} {995}},\ \bibinfo {pages} {1} (\bibinfo {year} {2023})}\BibitemShut {NoStop}%
\bibitem [{\citenamefont {Goodwin}\ \emph {et~al.}(2017)\citenamefont {Goodwin}, \citenamefont {Ortu}, \citenamefont {Reta}, \citenamefont {Chilton},\ and\ \citenamefont {Mills}}]{Goodwin2017}%
  \BibitemOpen
  \bibfield  {author} {\bibinfo {author} {\bibfnamefont {C.~A.~P.}\ \bibnamefont {Goodwin}}, \bibinfo {author} {\bibfnamefont {F.}~\bibnamefont {Ortu}}, \bibinfo {author} {\bibfnamefont {D.}~\bibnamefont {Reta}}, \bibinfo {author} {\bibfnamefont {N.~F.}\ \bibnamefont {Chilton}}, \ and\ \bibinfo {author} {\bibfnamefont {D.~P.}\ \bibnamefont {Mills}},\ }\bibfield  {title} {\emph {\bibinfo {title} {Molecular magnetic hysteresis at 60 kelvin in dysprosocenium},\ }}\href {\doibase 10.1038/nature23447} {\bibfield  {journal} {\bibinfo  {journal} {Nature}\ }\textbf {\bibinfo {volume} {548}},\ \bibinfo {pages} {439} (\bibinfo {year} {2017})}\BibitemShut {NoStop}%
\bibitem [{\citenamefont {Liu}\ \emph {et~al.}(2016)\citenamefont {Liu}, \citenamefont {Chen}, \citenamefont {Liu}, \citenamefont {Vieru}, \citenamefont {Ungur}, \citenamefont {Jia}, \citenamefont {Chibotaru}, \citenamefont {Lan}, \citenamefont {Wernsdorfer}, \citenamefont {Gao}, \citenamefont {Chen},\ and\ \citenamefont {Tong}}]{Liu2016}%
  \BibitemOpen
  \bibfield  {author} {\bibinfo {author} {\bibfnamefont {J.}~\bibnamefont {Liu}}, \bibinfo {author} {\bibfnamefont {Y.-C.}\ \bibnamefont {Chen}}, \bibinfo {author} {\bibfnamefont {J.-L.}\ \bibnamefont {Liu}}, \bibinfo {author} {\bibfnamefont {V.}~\bibnamefont {Vieru}}, \bibinfo {author} {\bibfnamefont {L.}~\bibnamefont {Ungur}}, \bibinfo {author} {\bibfnamefont {J.-H.}\ \bibnamefont {Jia}}, \bibinfo {author} {\bibfnamefont {L.~F.}\ \bibnamefont {Chibotaru}}, \bibinfo {author} {\bibfnamefont {Y.}~\bibnamefont {Lan}}, \bibinfo {author} {\bibfnamefont {W.}~\bibnamefont {Wernsdorfer}}, \bibinfo {author} {\bibfnamefont {S.}~\bibnamefont {Gao}}, \bibinfo {author} {\bibfnamefont {X.-M.}\ \bibnamefont {Chen}}, \ and\ \bibinfo {author} {\bibfnamefont {M.-L.}\ \bibnamefont {Tong}},\ }\bibfield  {title} {\emph {\bibinfo {title} {A stable pentagonal bipyramidal {D}y({III}) single-ion magnet with a record magnetization reversal barrier over 1000 {K}},\ }}\href {\doibase 10.1021/jacs.6b02638} {\bibfield  {journal}
  {\bibinfo  {journal} {Journal of the American Chemical Society}\ }\textbf {\bibinfo {volume} {138}},\ \bibinfo {pages} {5441} (\bibinfo {year} {2016})}\BibitemShut {NoStop}%
\bibitem [{\citenamefont {Wysocki}\ and\ \citenamefont {Park}(2020)}]{Wysocki2020}%
  \BibitemOpen
  \bibfield  {author} {\bibinfo {author} {\bibfnamefont {A.~L.}\ \bibnamefont {Wysocki}}\ and\ \bibinfo {author} {\bibfnamefont {K.}~\bibnamefont {Park}},\ }\bibfield  {title} {\emph {\bibinfo {title} {Hyperfine and quadrupole interactions for {Dy} isotopes in {DyPc}$_2$ molecules},\ }}\href {\doibase 10.1088/1361-648x/ab757b} {\bibfield  {journal} {\bibinfo  {journal} {Journal of Physics: Condensed Matter}\ }\textbf {\bibinfo {volume} {32}},\ \bibinfo {pages} {274002} (\bibinfo {year} {2020})}\BibitemShut {NoStop}%
\bibitem [{\citenamefont {Ferch}\ \emph {et~al.}(1974)\citenamefont {Ferch}, \citenamefont {Dankwort},\ and\ \citenamefont {Gebauer}}]{Ferch1974}%
  \BibitemOpen
  \bibfield  {author} {\bibinfo {author} {\bibfnamefont {J.}~\bibnamefont {Ferch}}, \bibinfo {author} {\bibfnamefont {W.}~\bibnamefont {Dankwort}}, \ and\ \bibinfo {author} {\bibfnamefont {H.}~\bibnamefont {Gebauer}},\ }\bibfield  {title} {\emph {\bibinfo {title} {Hyperfine structure investigations in {D}y{I} with the atomic beam magnetic resonance method},\ }}\href {https://www.sciencedirect.com/science/article/pii/0375960174908147} {\bibfield  {journal} {\bibinfo  {journal} {Physics Letters A}\ }\textbf {\bibinfo {volume} {49}},\ \bibinfo {pages} {287} (\bibinfo {year} {1974})}\BibitemShut {NoStop}%
\end{thebibliography}

%

\end{bibunit}

\end{onecolumngrid}

\end{document}